\documentclass[apj,twocolumn]{header/openjournal}
\usepackage[T1]{fontenc}
\usepackage{lmodern} 
\usepackage{ae,aecompl}

\usepackage{soul}     
\usepackage{needspace} 
\usepackage{graphicx} 
\usepackage[para]{threeparttable} 
\usepackage[most]{tcolorbox} 
\usepackage{tikz}     

\usepackage{amsmath}   
\usepackage{amssymb}   
\usepackage{dsfont}    
\usepackage{esint}     
\usepackage{abraces}   
\usepackage{orcidlink} 
\usepackage{cleveref}  
\usepackage{bm}		   
\usepackage{cancel}    
\usepackage{xfrac}     
\usepackage[a]{esvect} 
\usepackage{tikz}
\usepackage{circledsteps}
\usepackage{pstricks}

\newcommand{\mathbfss}[1]{\textbf{\textsf{#1}}}
\newcommand{\mTensor}[1]{\mathbfss{#1}}
\newcommand{\mVector}[1]{\bm{#1}} 
\newcommand{\mVectorUnit}[1]{\widehat{\mVector{#1}}} 

\newcommand{\mpFrac}[2]{\frac{\partial #1}{\partial #2}}

\newcommand{\rbrac}[1]{\left( #1 \right)}
\newcommand{\sbrac}[1]{\left[ #1 \right]}
\newcommand{\cbrac}[1]{\left\{ #1 \right\}}

\newcommand{\bigO}[1]{\mathcal{O}\rbrac{#1}}
\newcommand{\EvalAtH}[3]{\ensuremath{\!\sbrac{#1,\,#2,\,#3}}}
\newcommand{\EvalAtV}[3]{\ensuremath{\!\begin{bmatrix}#1\\#2\\#3\end{bmatrix}}}
\newcommand{\ie}{\textit{i.e.},\,}
\newcommand{\eg}{\textit{e.g.},\,}
\newcommand{\etc}{\textit{etc.}\,}
\newcommand{\viz}{\textit{viz.}\,}
\newcommand{\quokka}{\textsc{Quokka}}
\newcommand{\nquad}[1][1]{\hspace*{#1em}\ignorespaces}

\newcommand{\sref}[1]{Section~\ref{#1}}
\newcommand{\aref}[1]{Appendix~\ref{#1}}
\crefname{equation}{Equation}{Equations}
\Crefname{equation}{Equation}{Equations}
\crefname{figure}{Figure}{Figures}
\Crefname{figure}{Figure}{Figures}
\crefname{table}{Table}{Tables}
\Crefname{table}{Table}{Tables}

\defcitealias{Wibking22a}{WK22}
\defcitealias{Felker18a}{FS18}
\defcitealias{Balsara25a}{B25a}
\defcitealias{Balsara25b}{B25b}
\defcitealias{Londrillo04a}{LD04}

\begin{document}

\title[Magnetising \quokka]{Magnetising the \quokka~code with fast, second-order accurate, error-correcting schemes for magnetohydrodynamics on GPUs}

\author{
Neco Kriel
\orcidlink{0000-0002-3558-3926}
}
\email{nkriel@uni-bonn.de}
\affiliation{Research School of Astronomy and Astrophysics, Australian National University, Canberra, ACT 2611, Australia}
\affiliation{Argelander-Institut f\"ur Astronomie, Universit\"at Bonn, Auf dem H\"ugel 71, 53121 Bonn, Germany}

\author{
Elizabeth Cole-Kodikara
\orcidlink{0000-0002-9496-0859}
}
\affiliation{Research School of Astronomy and Astrophysics, Australian National University, Canberra, ACT 2611, Australia}

\author{
Benjamin Wibking
\orcidlink{0000-0003-3175-2291}
}
\affiliation{Department of Physics and Astronomy, Michigan State University, East Lansing, MI 48824, USA}

\author{
Mark R. Krumholz
\orcidlink{0000-0003-3893-854X}
}
\affiliation{Research School of Astronomy and Astrophysics, Australian National University, Canberra, ACT 2611, Australia}

\author{
Chong-Chong He
\orcidlink{0000-0002-2332-8178}
}
\affiliation{Research School of Astronomy and Astrophysics, Australian National University, Canberra, ACT 2611, Australia}

\author{
Anish Sarkar
\orcidlink{0009-0006-6879-8751}
}
\affiliation{Research School of Astronomy and Astrophysics, Australian National University, Canberra, ACT 2611, Australia}

\author{
Pak Shing Li
\orcidlink{0000-0001-8077-7095}
}
\affiliation{Shanghai Astronomical Observatory, Chinese Academy of Sciences, Shanghai 200030, People's Republic of China}

\begin{abstract}
    We implement a second-order accurate constrained transport (CT) magnetohydrodynamics (MHD) module with first-order flux correction in the open-source, adaptive mesh refinement (AMR), GPU-accelerated code \quokka, supporting both ideal and resistive (constant Ohmic) regimes. For computing and averaging the electromotive forces (EMFs) at cell edges, the basis of any CT-MHD method, we experiment with a wide range of recent, state-of-the-art schemes, together with different reconstruction schemes. Alongside these, we develop a new EMF compute scheme that requires fewer reconstruction steps than existing approaches. We evaluate these scheme combinations on the basis of accuracy, stability, and GPU throughput, and demonstrate that our new scheme achieves accuracy comparable to the best-in-class existing methods, together with greater stability in reconnection-dominated flows and roughly $25$--$60\%$ higher GPU throughput. Using this scheme, \quokka~achieves excellent results across a wide range of MHD flow regimes, reaching $>50$ million cell updates per GPU per second, with $>70\%$ parallel efficiency out to $>500$ GPUs. Finally, we confirm that our CT implementation preserves divergence-free magnetic fields (to machine precision) under AMR.
\end{abstract}

\keywords{MHD --- methods: numerical}

\maketitle

\section{Introduction}

\quokka~is a new, open-source, GPU-accelerated radiation-hydrodynamics code aimed at modelling radiating, highly compressible flows across a wide range of scales \citep[hereafter \citetalias{Wibking22a}]{Wibking22a}. In this paper we extend \quokka~to include a magnetohydrodynamics (MHD) module, and in the process test a broad combination of Riemann solvers, constrained-transport (CT) schemes, and error-correction strategies in order to identify ones that offer a good combination of accuracy and performance in these complex flow regimes. This comparison of available approaches in the literature will not only inform our design choices for implementing MHD into \quokka, but we anticipate that it will also be directly useful to others implementing MHD in similar GPU-oriented codes. Moreover, we introduce a new EMF compute scheme that is more efficient on GPU with comparable accuracy; paired with our recommended EMF averaging scheme, we also find it to be more stable in reconnection-dominated flows than most of the alternative schemes we test. To this end, the remainder of this Introduction will be dedicated to first situating \quokka~within the broader landscape of astrophysical simulation codes, before we turn to the numerical challenges posed by highly compressible MHD flows, which \quokka~aims to simulate.

\subsection{Why \quokka?}

While a number of mature grid-based codes already provide rich multiphysics capabilities for galaxy formation and the interstellar medium (ISM), including \textsc{Athena++} \citep{Stone20a}, \textsc{Enzo} \citep{Bryan14a}, \textsc{Flash} \citep{Fryxell00a}, \textsc{Orion} \citep{Li21b}, \textsc{Pencil} \citep{Brandenburg02a}, \textsc{Pluto} \citep{Mignone07a}, and \textsc{Ramses} \citep{Teyssier02a}, these codebases are primarily CPU-based, with GPU support still an active area of development. There are also newer codes like \textsc{Cholla} \citep{Schneider15a, Caddy24a} and \textsc{AthenaK} \citep{Stone24a} that were developed from the outset for GPU architectures, and already include MHD solvers; however, they lack many of the multiphysics features required for simulations of galaxies, star formation, and ISM physics more broadly. For example, \textsc{AthenaK} focuses on applications in high-energy astrophysics, and therefore does not provide modules for ISM cooling, star formation, or stellar feedback, and only includes radiative transfer for relativistic flows. Similarly, while \textsc{Cholla} does include ISM-relevant cooling, it does not include either radiative transfer or self-consistent star formation and feedback, and also does not provide adaptive mesh refinement (AMR), a capability crucial for resolving the evolution of collapse under self-gravity. Taken together, this highlights a gap in the current code landscape: there is a lack of GPU-native, AMR-enabled codes that combine non-relativistic radiation-hydrodynamics with the broader multiphysics required for ISM and star-formation applications.

\quokka~attempts to bridge this gap. It is a GPU-native\footnote{
    \quokka~can also be built and run on CPU-only systems.
} radiation-hydrodynamics code built on \textsc{AMReX} \citep{Zhang19b}, designed for high-resolution AMR studies of non-relativistic, radiating, compressible flows. Recently, the codebase has grown to include a broader set of physics modules, including multi-group, non-relativistic radiation transport \citep{He24a, He24b}, complex gas-particle interactions \citep{He25a}, and subgrid models aimed at ISM, circumgalactic, and star-formation applications, which has been successfully deployed in simulations of galactic winds launched by supernovae \citep{Vijayan24a, Vijayan25a, huang2026quokka}. However, what \quokka~has lacked so far is the ability to simulate magnetised flows, so in this paper we add and validate this capability.

\subsection{Design considerations for highly compressible MHD}
\label{sec:intro:goals}

The environments that \quokka~was designed to model (from flows during cosmological structure formation and galaxy assembly, to star-forming interstellar gas, protostellar and protoplanetary discs, and planetary flows) are highly compressible, with dynamics that evolve over a wide range of spatial and temporal scales. In these environments the sonic Mach number $\mathcal{M}$ (flow speed relative to the sound speed), Alfv\'en Mach number (relative to the Alfv\'en speed), and plasma-$\beta$ (gas-to-magnetic pressure ratio) can each vary by many orders of magnitude, with the gas also often at least partially ionised, leading to dynamically important magnetic fields threading these flows. A code that aims to explore these environments must therefore satisfy several competing requirements: it must be (i) robust in the presence of high-$\mathcal{M}$ flows and strong shocks, (ii) accurate enough to capture low-$\mathcal{M}$, instability-driven dynamics, and (iii) scalable to very high resolutions on modern hardware, particularly on GPUs, and often in combination with AMR.

A common approach for satisfying these challenges is to use a Godunov finite-volume scheme \citep{Toro97a}. This is the approach \quokka~adopts, where conserved quantities are stored as cell averages and updated through fluxes across cell interfaces. In such a scheme one collects the evolved fields into a conservative state vector $\mVector{S}_\mathrm{c}$ obeying
\begin{align}
    \mpFrac{\mVector{S}_\mathrm{c}}{t}
        + \nabla\cdot\mVector{F}\sbrac{\mVector{S}_\mathrm{c}}
        &= \mVector{G}\sbrac{\mVector{S}_\mathrm{c}}
    , \label{eqn:godunov:conservation-law}
\end{align}
where $\mVector{F}[\mVector{S}_\mathrm{c}]$ provides the fluxes of conserved quantities, and $\mVector{G}[\mVector{S}_\mathrm{c}]$ collects explicit sources and sinks (\eg~gravity and radiation couplings, or momentum forcing together with the corresponding energy injection it implies). Integrating over cell volumes $\mathcal{V}$, one obtains the semi-discrete evolution equation
\begin{align}
    \mpFrac{}{t} \langle \mVector{S}_\mathrm{c}\rangle_\mathcal{V}
        + \frac{1}{|\mathcal{V}|} \int_{\partial\mathcal{V}}
            \mVector{F} \cdot \mVectorUnit{n}
       \, \mathrm{d} A
        &= \langle \mVector{G} \rangle_\mathcal{V}
    , \label{eqn:godunov:integral}
\end{align}
where
\begin{align}
    \langle \mVector{S}_\mathrm{c}\rangle_\mathcal{V}
        \equiv \frac{1}{|\mathcal{V}|} \int_\mathcal{V}
            \mVector{S}_\mathrm{c}
       \, \mathrm{d} V
\end{align}
is the average value of $\mVector{S}_\mathrm{c}$ over a cell volume (a similar definition follows for $\langle\mVector{G}\rangle_\mathcal{V}$) and $\partial\mathcal{V}$ is the surface of the cell face, with area $A$. The second term on the left-hand side of \cref{eqn:godunov:integral} follows from the divergence theorem, and for cuboid cells (which \quokka~uses) this integral can be simplified to a sum over the area-averaged fluxes across the six faces adjoining each cell and its neighbours. It then follows that, because these interfaces in turn each carry a single flux that leaves one cell and enters its neighbour, conservation is enforced by construction.

Schemes in this family have proven robust at capturing strong shocks, while still able to retain high levels of accuracy in smooth flows \citep{kritsuk2011comparing}, and, because they operate directly on the conserved variables, extend straightforwardly to AMR. The main numerical choices lie in how one approximates the face-averaged fluxes, which typically involves (i) reconstructing interface states from the cell averages and (ii) solving a local Riemann problem to obtain an upwinded flux; different combinations trade off stability, accuracy, and computational cost.

In ideal MHD, considering an isolated system with $\mVector{G} = 0$ (\ie~no external mass, momentum, or energy sources), the conservative system can be written compactly\footnote{
    Here, and throughout this paper, we adopt the usual rationalised-Gaussian unit system, whereby all non-magnetic quantities are expressed in CGS units, and the magnetic field $\mVector{b}$ is expressed in Gauss units divided by $\sqrt{4\pi}$.
    \label{fnote:b-units}
} as
\begin{align}
    \scalebox{0.925}{$\displaystyle
        \mpFrac{}{t}\begin{Bmatrix}
            \rho \\
            \rho\mVector{u} \\
            e_\mathrm{tot} \\
            \mVector{b}
        \end{Bmatrix}
            = -\nabla\cdot\begin{Bmatrix}
                \rho\mVector{u} \\
                \rho(\mVector{u}\otimes\mVector{u})
                    + \rbrac{p + \dfrac{b^2}{2}} \mTensor{I}
                    - \mVector{b}\otimes\mVector{b} \\
                \rbrac{e_\mathrm{int} + p + \dfrac{b^2}{2}}\mVector{u}
                    - (\mVector{u}\cdot\mVector{b})\, \mVector{b} \\
                \mVector{u}\otimes\mVector{b}
                    - \mVector{b}\otimes\mVector{u}
            \end{Bmatrix}
    ,$}
    \label{eqn:mhd:conservation-laws}
\end{align}
since there are no explicit source or sink terms in ideal MHD. Here $\rho$, $\mVector{u}$, and $\mVector{b}$ are the fluid density, fluid velocity, and magnetic field, respectively, with total energy density $e_\mathrm{tot} = e_\mathrm{int} + (1/2)\rbrac{\rho u^2 + b^2}$, and internal energy $e_\mathrm{int}$; here, also, $b^2 \equiv \mVector{b}\cdot\mVector{b}$ and $\otimes$ is the tensor (or outer) product such that, for example, $\mVector{u}\otimes\mVector{b} \equiv u_i b_j$ in index notation.

On the surface, this system of equations might look like a straightforward extension of the Euler system (\ie~inviscid hydrodynamics), which may lead one to naively assume that evolving \cref{eqn:mhd:conservation-laws} in a Godunov framework simply requires the hydro-Riemann solver be replaced with an MHD-aware Riemann solver. However, magnetised flows face a unique constraint that is not automatically satisfied by a standard Godunov scheme: the numerical method must satisfy $\nabla\cdot\mVector{b} = 0$. This is not guaranteed for a purely cell-centred flux update of $\mVector{b}$, which is a well-known numerical problem to which many approaches have been proposed \citep{kritsuk2011comparing}, including the development of CT methods \citep{Evans88a}.

In many ways CT is a natural extension of the Godunov approach to magnetic fields, where, rather than evolving cell-averaged components of $\mVector{b}$ from $\mVector{F}[\mVector{S}_\mathrm{c}]$, one instead evolves face-averaged normal-components of $\mVector{b}$ from edge-centred electromotive forces (EMFs) in a way that preserves discrete $\nabla\cdot\mVector{b}$ by construction. This follows from the central fact that, for ideal MHD, the magnetic field evolves according to the induction equation
\begin{align}
    \mpFrac{\mVector{b}}{t}
        &= -\nabla\times\mVector{\varepsilon}
    , \label{eqn:induction:continuous}
\end{align}
where $\mVector{\varepsilon} = -\mVector{u}\times\mVector{b}$ is the ideal EMF. Integrating the normal component of \cref{eqn:induction:continuous} over a cell face $\partial\mathcal{V}$ with unit normal direction $\mVectorUnit{n}$, gives
\begin{align}
    \int_{\partial\mathcal{V}} \mpFrac{\mVector{b}}{t}\cdot\mVectorUnit{n}\, \mathrm{d} A
        &= -\int_{\partial\mathcal{V}}
            \rbrac{\nabla\times\mVector{\varepsilon}}
            \cdot\mVectorUnit{n}\, \mathrm{d} A
    . \label{eqn:step:integrate-dbdt}
\end{align}
Commuting the time derivative on the left-hand side motivates the definition of the face-averaged normal field,
\begin{align}
    \langle{b_n}\rangle_{\partial\mathcal{V}}
        &\equiv \frac{1}{|\partial\mathcal{V}|}
            \int_{\partial\mathcal{V}}
            \mVector{b}\cdot\mVectorUnit{n}\, \mathrm{d} A
    ,
\end{align}
while Stokes' theorem converts the surface integral of the curl, on the right-hand side of \cref{eqn:step:integrate-dbdt}, into a circulation (\ie~closed line integral) of the EMF around the face boundary $\partial\mathcal{A}$, giving the semi-discrete evolution equation
\begin{align}
    \mpFrac{}{t}\langle{b_n}\rangle_{\partial\mathcal{V}}
        &= -\frac{1}{|\partial\mathcal{V}|}
            \oint_{\partial\mathcal{A}}
            \mVector{\varepsilon}\cdot\mathrm{d}\mVector{\ell}
    . \label{eqn:ct:circulation-update}
\end{align}
CT is then a discrete realisation of \cref{eqn:ct:circulation-update}, in which $\mVector{b}$ is stored directly as face-averaged normal components, and EMFs as edge-averaged components tangent to the boundary edge. From this, the magnetic update is obtained by approximating the line integral (in \cref{eqn:ct:circulation-update}) as a sum of the EMF components at the edges bounding the face. With this staggering, each edge EMF contributes to exactly two neighbouring face updates, with opposite sign, so the discrete divergence (defined by the oriented sum of face-averaged fluxes over the cell boundary) is preserved by construction: if $\nabla\cdot\mVector{b}$ is initially zero on this stencil, it remains so to machine precision thereafter.

With this context, it then follows that enabling ideal MHD in \quokka~amounts to extending its finite-volume flux update in two coupled ways: (i) we replace the purely hydrodynamic flux function with an MHD-aware Riemann solver, so we can compute upwinded face-centred fluxes from reconstructed left and right interface states, and (ii) we compute the magnetic update using CT. In the sections that follow we describe our choice of MHD-aware Riemann solver and set of CT schemes.

\subsection{Outline of this paper}

The remainder of this paper is organised as follows. In \sref{sec:method} we present the MHD module added to \quokka, describing the set of (Riemann and CT) solvers, reconstruction schemes, and our flux correction strategy. In \sref{sec:scheme-comparisons} we compare the accuracy and performance of different schemes, before recommending our preferred scheme combination for \quokka. In \sref{sec:stress-tests} we demonstrate the correctness, robustness, and broader capability of our recommended scheme across the remainder of our validation suite, and then finally summarise our findings and discuss future extensions of MHD for \quokka~in \sref{sec:conclusions}.

\section{Numerical implementation}
\label{sec:method}

Here we introduce the different parts of our MHD module. We first describe the mesh we use to discretise the equations in \sref{sec:methods:mesh}. We then describe the method we use to compute the fluxes of cell-centred variables in \sref{sec:methods:flux}, and to compute EMFs for the face-centred magnetic fields in \sref{sec:methods:emf}. In \sref{sec:methods:time-stepping} we explain how we chain these calculations together within our overall time-stepping strategy, including our implementation of flux correction, and conclude in \sref{sec:methods:amr} with a discussion of how the module was extended to AMR. Because we track a large number of quantities, with many different types of indices (indicating grid position, time, direction, \etc), we collect all the symbols we use in \cref{table:symbols}, with the notational conventions discussed in \aref{sec:notation}. We only work through the update in the $x_0$-direction, with the updates in other directions following by cyclic permutation of the indices.

\begin{table*}
\centering
\caption{Symbols used throughout this paper. Units are quoted in the rationalised-Gaussian system (\ie~CGS for all non-magnetic quantities, with the magnetic field measured in Gauss divided by $\sqrt{4\pi}$; see \cref{fnote:b-units}). A dash indicates a dimensionless quantity, and ``mixed'' a collection whose elements carry different units.
\label{table:symbols}}
\small
\renewcommand{\arraystretch}{0.95}
\setlength{\tabcolsep}{4pt}
\begin{tabular}{l p{0.55\textwidth} l l}
\hline\hline
Symbol
    & Description
    & Units
    & Introduced in \\
\hline
\multicolumn{4}{l}{\textit{Domain variables}} \\
\hline
$[i, j, k]$
    & Mesh indices; integers denote cell centres, half-integers faces or edges
    & --
    & \cref{eqn:state:conserved} \\
$x_d$
    & Coordinate axis direction, $d \in \{0, 1, 2\}$, following the right-hand rule
    & $\mathrm{cm}$
    & \cref{fig:schematic:grid} \\
$L$
    & Length of the simulation along an axis of the domain
    & $\mathrm{cm}$
    & \sref{sec:scheme-comparison:waves} \\
$N$
    & Number of cells along an axis of the domain
    & --
    & \sref{sec:scheme-comparison:waves} \\
$\Delta{x_d}$
    & Cell width along the $x_d$-direction
    & $\mathrm{cm}$
    & \cref{eqn:dbdt:discrete} \\
$\Delta{x}_\mathrm{min}$
    & Smallest cell width on a given AMR level
    & $\mathrm{cm}$
    & \cref{eqn:resistivity:cfl} \\
$\ell$
    & Characteristic length scales
    & $\mathrm{cm}$
    & \sref{sec:scheme-comparison:dissipation} \\
$\mathcal{V}$
    & Cell volume, with measure $|\mathcal{V}|$
    & $\mathrm{cm^{3}}$
    & \cref{eqn:godunov:integral} \\
$\partial\mathcal{V}$
    & Cell face, with area $|\partial\mathcal{V}|$ and area element $A$
    & $\mathrm{cm^{2}}$
    & \cref{eqn:godunov:integral} \\
$\partial\mathcal{A}$
    & Closed circuit traced by the four edges bounding a face
    & --
    & \cref{eqn:ct:circulation-update} \\
$\mVectorUnit{n}$
    & Unit vector normal to a face
    & --
    & \cref{eqn:godunov:integral} \\
$\mathrm{d}\mVector{\ell}$
    & Line element directed along a face boundary
    & $\mathrm{cm}$
    & \cref{eqn:ct:circulation-update} \\
\hline
\multicolumn{4}{l}{\textit{Time stepping}} \\
\hline
$t$
    & Time value
    & $\mathrm{s}$
    & \cref{eqn:godunov:conservation-law} \\
$\Delta t$
    & Time step
    & $\mathrm{s}$
    & \cref{eqn:resistivity:cfl} \\
$[n]$
    & Old time level; \eg~$[n + 1/2]$ is the intermediate time
    & --
    & \sref{sec:methods:time-stepping} \\
$\mathrm{CFL}$
    & Courant-Friedrichs-Lewy number
    & --
    & \cref{eqn:resistivity:cfl} \\
HO, FO
    & Acronyms for high-order and first-order estimates of fluxes and EMFs
    & --
    & \cref{fig:schematic:flowchart} \\
\hline
\multicolumn{4}{l}{\textit{Reconstructed-state labels}} \\
\hline
$S_\mathrm{I}$
    & States either side of an $x_0$-face, $S_\mathrm{I} \in \{L, R\}$ (left, right)
    & --
    & \cref{fig:schematic:b-common} \\
$S_\mathrm{II}$
    & States either side of an $x_1$-face, $S_\mathrm{II} \in \{B, T\}$ (bottom, top)
    & --
    & \cref{fig:schematic:b-common} \\
$Q_\mathrm{I}$
    & Four sides at an edge, $Q_\mathrm{I} \in \{L, T, R, B\}$
    & --
    & \cref{fig:schematic:q26} \\
$Q_\mathrm{II}$
    & Four corners at an edge, $Q_\mathrm{II} \in \{LT, RT, RB, LB\}$
    & --
    & \cref{fig:schematic:fs18} \\
$[x_0 x_1]$
    & Sequence of 1D reconstructions; $[x_1 x_0]$ is the reverse
    & --
    & \cref{eqn:emf:fs18-average} \\
\hline
\multicolumn{4}{l}{\textit{State vectors, fluxes, and operators}} \\
\hline
$\mVector{S}_\mathrm{c}$
    & Vector of conserved quantities
    & mixed
    & \cref{eqn:state:conserved} \\
$\mVector{S}_\mathrm{p}$
    & Vector of primitive quantities
    & mixed
    & \cref{eqn:state:primitive} \\
$\mVector{S}$
    & Vector of quantities from $\mVector{S}_\mathrm{c}$ and $\mVector{S}_\mathrm{p}$, used to define a problem setup
    & mixed
    & \cref{eqn:wave:background} \\
$\mVector{F}$
    & Flux of the conserved quantities across a face
    & mixed
    & \cref{eqn:godunov:conservation-law} \\
$\mVector{G}$
    & Explicit sources and sinks of the conserved quantities
    & mixed
    & \cref{eqn:godunov:conservation-law} \\
$\mTensor{I}$
    & Identity tensor
    & --
    & \cref{eqn:mhd:conservation-laws} \\
$\otimes$
    & Tensor (outer) product, $\mVector{u}\otimes\mVector{b} \equiv u_i b_j$
    & --
    & \cref{eqn:mhd:conservation-laws} \\
$D^{[m]}$
    & Update increment evaluated at time step $[m]$
    & set by argument
    & \cref{eqn:rk:increment-state} \\
$\langle\,\cdot\,\rangle_\mathcal{V}$
    & Volume-average over a cell; $\langle\,\cdot\,\rangle_{\partial\mathcal{V}}$ denotes the corresponding face-average
    & set by argument
    & \cref{eqn:godunov:integral} \\
\hline
\multicolumn{4}{l}{\textit{Field variables}} \\
\hline
$\rho$
    & Mass density
    & $\mathrm{g\,cm^{-3}}$
    & \cref{eqn:state:conserved} \\
$\mVector{u}$
    & Fluid velocity
    & $\mathrm{cm\,s^{-1}}$
    & \cref{eqn:mhd:conservation-laws} \\
$p$
    & Gas (thermal) pressure
    & $\mathrm{erg\,cm^{-3}}$
    & \cref{eqn:state:primitive} \\
$\mVector{b}$
    & Magnetic field
    & $\mathrm{G}/\sqrt{4\pi}$
    & \cref{eqn:mhd:conservation-laws} \\
$e_\mathrm{int}$
    & Internal energy density
    & $\mathrm{erg\,cm^{-3}}$
    & \cref{eqn:mhd:conservation-laws} \\
$e_\mathrm{tot}$
    & Total energy density, $e_\mathrm{int} + (\rho u^2 + b^2)/2$
    & $\mathrm{erg\,cm^{-3}}$
    & \cref{eqn:mhd:conservation-laws} \\
$\mVector{\varepsilon}$
    & Electromotive force (EMF), $\mVector{u}\times\mVector{b}$
    & $\mathrm{G\,cm\,s^{-1}}/\sqrt{4\pi}$
    & \cref{eqn:induction:continuous} \\
$\mVector{j}$
    & Current density, $\nabla\times\mVector{b}$
    & $\mathrm{G\,cm^{-1}}/\sqrt{4\pi}$
    & \cref{eqn:resistivity:ohms-law} \\
$\eta$
    & Ohmic resistivity, constant in space and time
    & $\mathrm{cm^{2}\,s^{-1}}$
    & \cref{eqn:resistivity:ohms-law} \\
$\gamma$
    & Adiabatic index
    & --
    & \sref{sec:scheme-comparison:waves} \\
$\mathcal{M}$
    & Sonic Mach number, $u / c_\mathrm{s}$
    & --
    & \sref{sec:intro:goals} \\
\hline
\multicolumn{4}{l}{\textit{Wave speeds}} \\
\hline
$c_\mathrm{s}$
    & Adiabatic sound speed, $\sqrt{\gamma p / \rho}$
    & $\mathrm{cm\,s^{-1}}$
    & \sref{sec:scheme-comparison:waves:fast} \\
$c_\mathrm{A}$
    & Alfv\'en speed, $b / \sqrt{\rho}$
    & $\mathrm{cm\,s^{-1}}$
    & \sref{sec:scheme-comparison:waves:alfven-linear} \\
$c_\mathrm{fm}$
    & Fast magnetosonic speed
    & $\mathrm{cm\,s^{-1}}$
    & \cref{eqn:waves:fast-speed} \\
$c_\mathrm{sm}$
    & Slow magnetosonic speed
    & $\mathrm{cm\,s^{-1}}$
    & \cref{eqn:waves:slow-speed} \\
$\alpha_{d}^{[\pm]}$
    & Non-negative bounding wavespeed magnitudes on the $x_d$-face
    & $\mathrm{cm\,s^{-1}}$
    & \cref{eqn:emf:alpha-pm} \\
\hline
\hline
\end{tabular}
\end{table*}

\subsection{Mesh}
\label{sec:methods:mesh}

\begin{figure}
    \centering
    \includegraphics[width=0.9\linewidth]{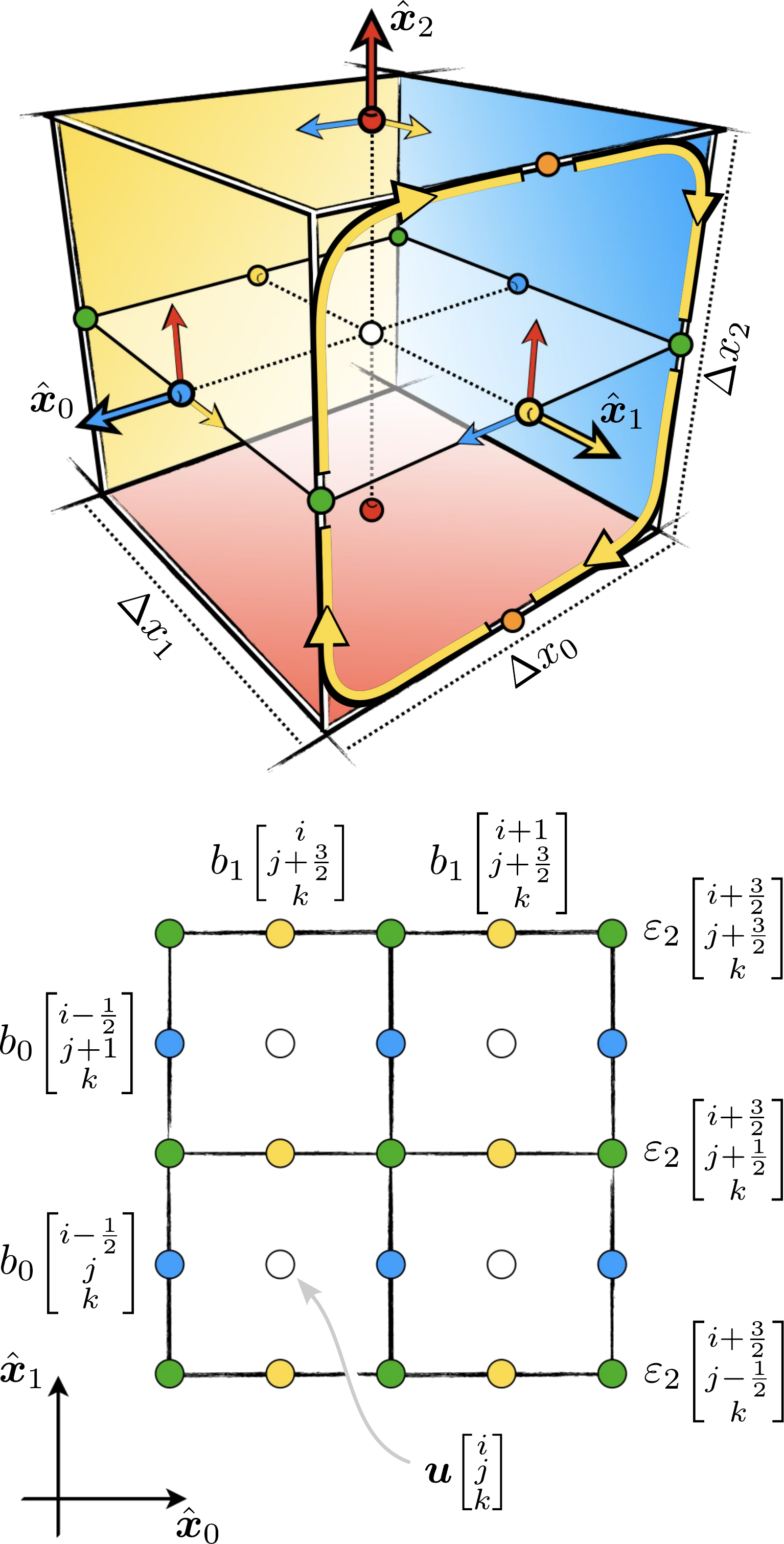}
    \caption{The \quokka~mesh shown in perspective view (left) and a slice in the $(x_0,x_1)$-plane (right). Cell-centred quantities (white points) are shown alongside face-centred magnetic field components (blue, yellow, and red points, for the $x_0$, $x_1$, and $x_2$ faces respectively) and edge-centred EMF components (green points).}
    \label{fig:schematic:grid}
\end{figure}

Our implementation adopts the staggered mesh scheme originally developed for computational electromagnetics by \citet{Yee66a}, later applied to magnetospheric MHD simulations \citep{Brecht81a}, and first used for astrophysical MHD by \citet{Evans88a}; today, this approach is standard in essentially all astrophysical CT-MHD implementations \citep[\eg][]{Fromang06a, Gardiner08a, Li12a, Benitez-Llambay16a, Caddy24a, Stone24a}. Consistent with this staggering, we store all conserved quantities, except for magnetic fields, at cell centres (illustrated in \cref{fig:schematic:grid}),\footnote{
    In addition to the MHD variables described in \cref{eqn:state:conserved}, \quokka~also supports passive scalars and a dual-energy formalism, where the internal energy is evolved alongside the total energy. These quantities extend $\mVector{S}_\mathrm{c}$ with extra cell-centred quantities, but, for simplicity, in this paper we restrict the discussion to the conserved variables relevant for MHD.
    \label{fnote:dual-energy}
}
and thus our vector of conserved cell-centred quantities is
\begin{align}
    \mVector{S}_\mathrm{c}\EvalAtV{i}{j}{k}
        = \begin{Bmatrix}
            \rho\EvalAtH{i}{j}{k} \\
            \rho\EvalAtH{i}{j}{k} \mVector{u}\EvalAtH{i}{j}{k} \\
            e_\mathrm{tot}\EvalAtH{i}{j}{k}
        \end{Bmatrix}
    , \label{eqn:state:conserved}
\end{align}
where we use integer indices $i$, $j$, and $k$ to denote cell-centred quantities. Each of these should be understood as an average over cell $\EvalAtH{i}{j}{k}$, but from this point forward we will omit the angle bracket notation introduced in \cref{eqn:godunov:integral}. We store the magnetic field on cell faces, with the component normal to a face stored at the face centre (see \cref{fig:schematic:grid}), so that a face-averaged quantity carries one half-integer index and two integer indices, \viz~$b_0$ is indexed as $\,\EvalAtH{i + 1/2}{j}{k}$, $b_1$ as $\,\EvalAtH{i}{j + 1/2}{k}$, and $b_2$ as $\,\EvalAtH{i}{j}{k + 1/2}$. Edge-averaged quantities carry two half-integer indices, \eg~$\EvalAtH{i + 1/2}{j + 1/2}{k}$, which sits on the $x_2$-edge.

\subsection{Flux calculation}
\label{sec:methods:flux}

\sref{sec:intro:goals} introduced the Godunov approach that \quokka~adopts, where $\mVector{F}[\mVector{S}_\mathrm{c}]$ is computed in two parts: (i) reconstructing the primitive states on either side of the interface, and (ii) solving a Riemann problem to obtain an upwinded flux. Our reconstruction closely follows the approach \quokka~already uses for pure hydrodynamics, as described in \citetalias{Wibking22a}. We therefore only summarise it here, and highlight the differences introduced by the inclusion of magnetic fields.

We first convert our vector of conserved variables to a vector of cell-centred primitive variables that includes the magnetic field:
\begin{align}
    \mVector{S}_\mathrm{p}\EvalAtV{i}{j}{k}
        = \begin{Bmatrix}
            \rho\EvalAtH{i}{j}{k} \\
            \mVector{u}\EvalAtH{i}{j}{k} \\
            p\EvalAtH{i}{j}{k} \\
            \mVector{b}\EvalAtH{i}{j}{k}
        \end{Bmatrix}
    , \label{eqn:state:primitive}
\end{align}
where $p$ is the gas pressure. Here we compute the cell-averaged magnetic field from the adjacent face-centred values, \ie
\begin{align}
    b_0\EvalAtV{i}{j}{k}
        = \frac{1}{2} \rbrac{
            b_0\EvalAtV{i + 1/2}{j}{k} + b_0\EvalAtV{i - 1/2}{j}{k}
        }
    .
\end{align}

Next, we reconstruct $\mVector{S}_\mathrm{p}$ at faces from cell centres using 1D polynomial reconstruction along each coordinate direction, obtaining left and right states at each interface. \quokka~supports piecewise-constant reconstruction (\ie~we take the primitive state vector at the face of each cell to be equal to the cell-centred primitive state), a piecewise linear method (PLM) using either a monotonized-central or minmod slope limiter \citep{van-Leer74a}, a piecewise parabolic method (PPM) similar to that of \citet{Colella84a} but with a modified approach to monotonicity preservation (see \citetalias{Wibking22a} for details), and the extremum-preserving PPM method (PPM-EP) described by \citet{Rider07a}. We will test several of these choices in \sref{sec:scheme-comparisons}. Note that, for the magnetic fields, we reconstruct only the components transverse to the reconstruction direction, since the normal component is already available at the face. For example, if we reconstruct in the $x_0$-direction, then we reconstruct only $b_1$ and $b_2$ at the face, where $b_0$ is stored natively.

The second part of the flux calculation is to solve the Riemann problem. Our default method for this is the HLLD Riemann solver \citep{Miyoshi05a}, which returns (i) the face-centred fluxes $\mVector{F}$ that enter the finite-volume updates of $\rho$, $\rho\mVector{u}$, and $e_\mathrm{tot}$, and (ii) a set of resolved interface states, including the normal and transverse components of $\mVector{u}$ and $\mVector{b}$ associated with the intermediate waves in the Riemann fan. In our CT formulation we do not use the magnetic part of the HLLD flux to advance $\mVector{b}$; instead, we use the upwinded interface states when constructing edge-centred EMFs for the magnetic update, as described in \sref{sec:methods:emf}. We therefore store these interface states for use in the next step.

To improve robustness in challenging flows, we also apply two local corrections to the numerical fluxes returned by HLLD. The first is a carbuncle fix for grid-aligned strong shocks, based on a pressure-sensitive blending of transverse dissipation. The second is a low-Mach pressure rescaling that restores the correct $\bigO{\mathcal{M}^2}$ scaling of pressure perturbations in the limit $\mathcal{M} \rightarrow 0$. For both these fixes we follow the method of \citet{Minoshima21a}.

In addition to the HLLD Riemann solver, we also implement a Local Lax-Friedrichs (LLF) solver \citep{Rusanov62a}, which returns the same quantities but is much more dissipative. We use this more dissipative flux in our flux correction strategy, described in full in \sref{sec:methods:time-stepping}.

\subsection{EMF calculation}
\label{sec:methods:emf}

The second part of advancing the MHD equations in time is to compute the EMFs required to update the magnetic field. As with the flux calculation, this happens in two parts: (i) computing EMF estimates at the cell edges, and (ii) averaging these into the single EMF used to update the magnetic field. In what follows, we only work through an update in the $x_0$-direction, for which the face has edges oriented in the $x_1$- and $x_2$-directions. Thus our goal for cell $\EvalAtH{i}{j}{k}$ is to arrive at estimates for the four average EMFs on the edges surrounding face $\EvalAtH{i + 1/2}{j}{k}$, so that we can write the semi-discrete evolution equation (\cref{eqn:ct:circulation-update}) as
\begin{align}
    \scalebox{0.925}{$\displaystyle
    \begin{aligned}
        &\mpFrac{b_0}{t}\EvalAtV{i + 1/2}{j}{k}
            =  \frac{1}{\Delta{x_1}\, \Delta{x_2}} \Bigg(
        \\
        &\nquad[3]
            \Delta{x_2} \, \varepsilon_2\EvalAtV{i + 1/2}{j - 1/2}{k}
            + \Delta{x_1} \, \varepsilon_1\EvalAtV{i + 1/2}{j}{k + 1/2}
        \\
        &\nquad[4]
            - \Delta{x_2} \, \varepsilon_2\EvalAtV{i + 1/2}{j + 1/2}{k}
            - \Delta{x_1} \, \varepsilon_1\EvalAtV{i + 1/2}{j}{k - 1/2}
        \Bigg)
    ,
    \end{aligned}
    $}
    \label{eqn:dbdt:discrete}
\end{align}
where $\Delta{x_1}$ and $\Delta{x_2}$ are the cell widths in the $x_1$- and $x_2$-directions. Of these four edges we only work through the $x_2$-edge, $\varepsilon_2\EvalAtH{i + 1/2}{j \pm 1/2}{k}$.

\subsubsection{The EMF compute scheme}
\label{sec:methods:emf:compute}

\begin{figure}
    \centering
    \includegraphics[width=\linewidth]{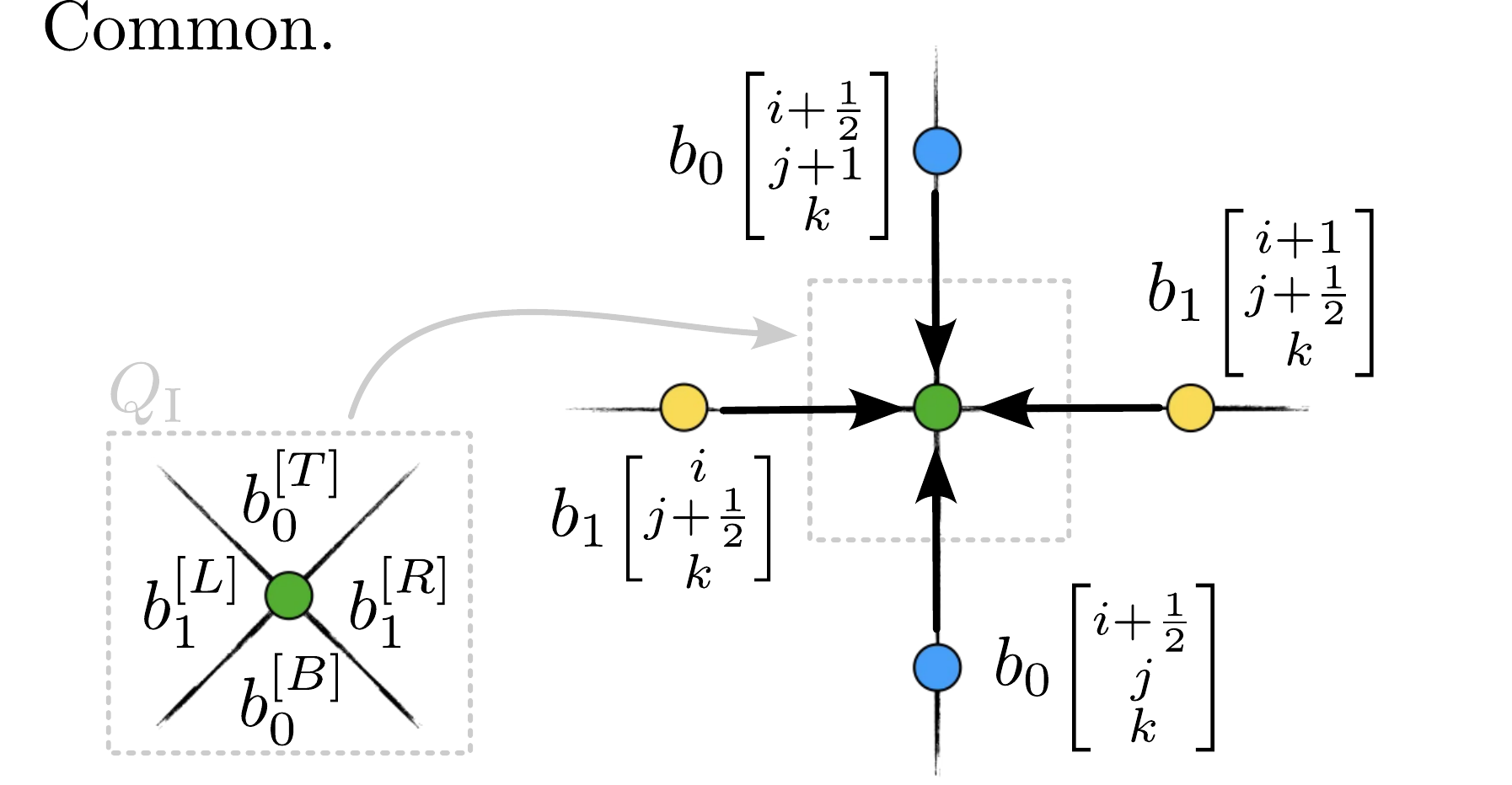}
    \caption{Reconstruction of face-centred magnetic fields to the cell edge (green point), common to all EMF compute schemes: $b_0$ (blue points, $x_0$ faces) yields $b_0^{[B]}$ and $b_0^{[T]}$; $b_1$ (yellow points, $x_1$ faces) yields $b_1^{[L]}$ and $b_1^{[R]}$.}
    \label{fig:schematic:b-common}
\end{figure}

Here we consider three schemes, of which two are implemented as published, and the third is a modified version of an existing approach. All three start in the same way (illustrated in \cref{fig:schematic:b-common}), by first reconstructing the face-centred magnetic field component $b_0\EvalAtH{i \pm 1/2}{j}{k}$ in the $x_1$-direction to yield $b_{0}^{[S_\mathrm{II}]}\EvalAtH{i \pm 1/2}{j \pm 1/2}{k}$, and reconstructing the face-centred magnetic field component $b_1\EvalAtH{i}{j \pm 1/2}{k}$ in the $x_0$-direction to yield $b_{1}^{[S_\mathrm{I}]}\EvalAtH{i \pm 1/2}{j \pm 1/2}{k}$. Here the superscripts $S_\mathrm{I} \in \{L, R\}$ indicate values on the left and right side of an $x_0$-face, and similarly $S_\mathrm{II} \in \{B, T\}$ indicate the values on the bottom and top of an $x_1$-face;\footnote{
    While ``left'' and ``bottom'' both denote the reconstructed value on the lower side of an interface relative to the reconstruction direction, and ``right'' and ``top'' the higher side, we use separate words to signal which reconstruction direction the state is associated with; ``left'' and ``right'' with the $x_0$-direction, and ``bottom'' and ``top'' with the $x_1$-direction.
} the $L$ and $R$ values, and the $B$ and $T$ values, may be different due to limiting in the reconstruction. This reconstruction, and all the other reconstruction steps we will describe below, can be done using any of the PC, PLM, PPM, or PPM-EP reconstruction methods described in \sref{sec:methods:flux}; the choices of EMF compute scheme and reconstruction method are independent, so in \sref{sec:scheme-comparisons} we compare different pairings.

Once the magnetic fields are reconstructed at the cell edges, the remaining steps differ depending on the scheme.

\begin{figure}
    \centering
    \includegraphics[width=\linewidth]{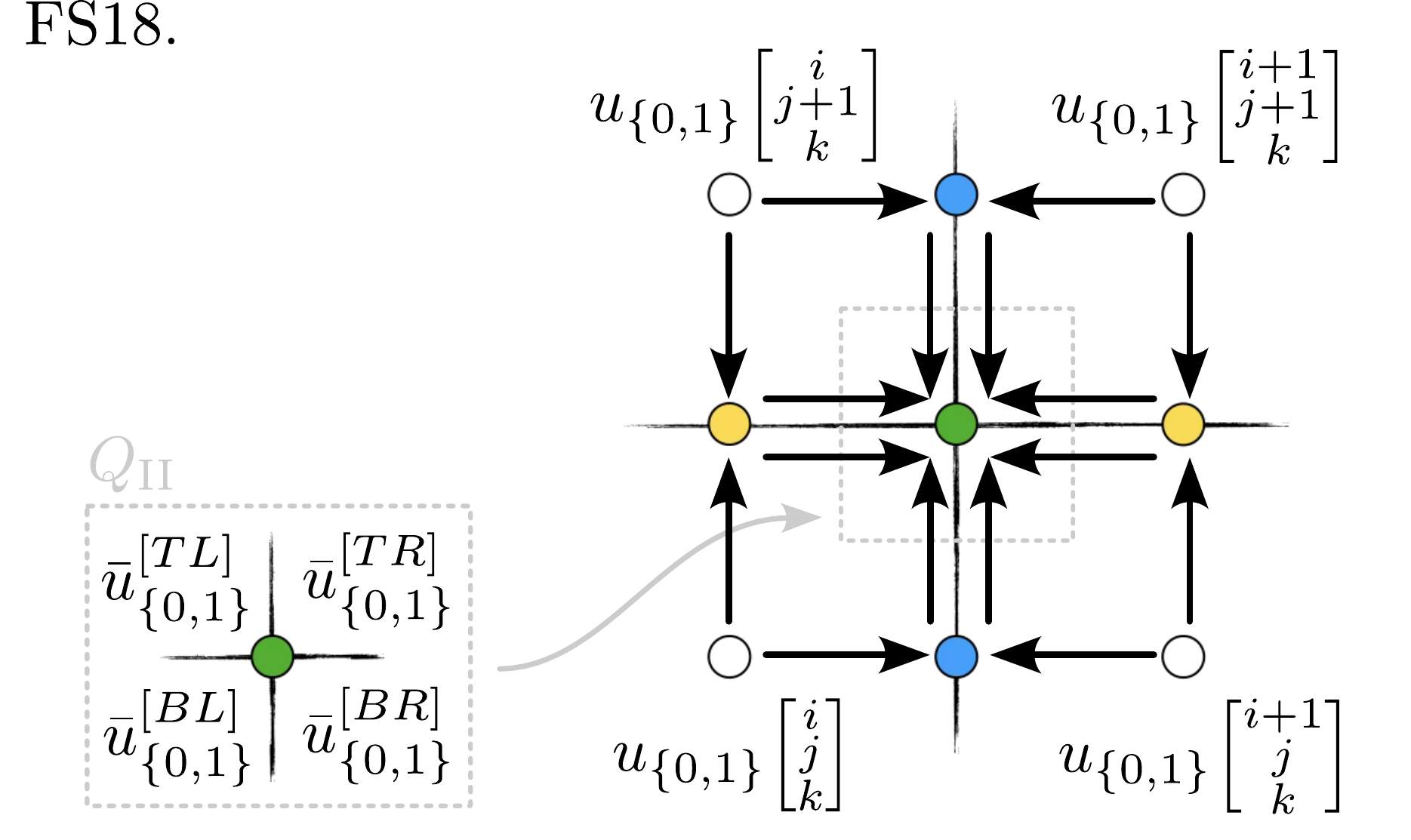}
    \caption{Velocity reconstruction for the \citetalias{Felker18a} compute scheme: cell-centred $u_0$ and $u_1$ (white point) are reconstructed to cell faces (blue and yellow points) and then to the cell edge (green point), yielding four estimates $\bar{u}_{\cbrac{0,1}}^{[Q_\mathrm{II}]}$ (\cref{eqn:emf:fs18-average}), which combine with the magnetic field estimates (\cref{fig:schematic:b-common}) to give the EMF (\cref{eqn:emf:fs18}).}
    \label{fig:schematic:fs18}
\end{figure}

Following the \citet[][hereafter \citetalias{Felker18a}; illustrated in \cref{fig:schematic:fs18}]{Felker18a} scheme, we begin from the cell-centred velocity components $u_0\EvalAtH{i}{j}{k}$ and $u_1\EvalAtH{i}{j}{k}$, the only two velocity components that contribute to $\varepsilon_2$, which we denote jointly as $\mVector{u}_{\cbrac{0,1}}\EvalAtH{i}{j}{k}$. We first reconstruct $\mVector{u}_{\cbrac{0,1}}\EvalAtH{i}{j}{k}$ in the $x_0$- and $x_1$-directions to obtain the face-centred velocities $\mVector{u}_{\cbrac{0,1}}^{[S_\mathrm{I}: x_0]}\EvalAtH{i \pm 1/2}{j}{k}$ and $\mVector{u}_{\cbrac{0,1}}^{[S_\mathrm{II}: x_1]}\EvalAtH{i}{j \pm 1/2}{k}$, where the superscript $x_0$ or $x_1$ indicates the direction in which reconstruction was performed. We then reconstruct these face-centred velocities at cell edges, reconstructing the $x_0$-face-centred velocities in the $x_1$-direction and the $x_1$-face-centred velocities in the $x_0$-direction. This yields eight quantities at each edge, which we denote $\mVector{u}_{\cbrac{0,1}}^{[Q_\mathrm{II}: x_0 x_1]}\EvalAtH{i \pm 1/2}{j \pm 1/2}{k}$ and $\mVector{u}_{\cbrac{0,1}}^{[Q_\mathrm{II}: x_1 x_0]}\EvalAtH{i \pm 1/2}{j \pm 1/2}{k}$, where the corner subscripts $Q_\mathrm{II} \in \{LB, RB, LT, RT\}$ indicate the four possible corners adjacent to the edge, and the superscripts $x_0 x_1$ and $x_1 x_0$ indicate the reconstruction order: $x_0 x_1$ denotes a quantity produced by reconstructing first in the $x_0$-direction at the $x_0$-face and then in the $x_1$-direction, to arrive at the edge joining the $x_0$- and $x_1$-faces; $x_1 x_0$ indicates reconstructing in the opposite order. Since reconstruction is not a commutative operation (due to limiting), these two orderings in general yield distinct values, giving two estimates for each corner quantity $Q_\mathrm{II} \in \{LB, RB, LT, RT\}$: one from reconstructing in $x_0$ then $x_1$, and one from the reverse order. We average these two estimates,
\begin{align}
    \scalebox{0.925}{$\displaystyle
    \begin{aligned}
        \bar{\mVector{u}}_{\cbrac{0,1}}^{[Q_\mathrm{II}]}\EvalAtV{i \pm 1/2}{j \pm 1/2}{k}
            &= \frac{1}{2} \Bigg(
                \mVector{u}_{\cbrac{0,1}}^{[Q_\mathrm{II}: x_0 x_1]}\EvalAtV{i \pm 1/2}{j \pm 1/2}{k}
            \\
            &\nquad[4]
                + \mVector{u}_{\cbrac{0,1}}^{[Q_\mathrm{II}: x_1 x_0]}\EvalAtV{i \pm 1/2}{j \pm 1/2}{k}
            \Bigg)
    ,
    \end{aligned}
    $}
    \label{eqn:emf:fs18-average}
\end{align}
reducing the number of distinct velocity estimates at each cell edge from eight to four.

Finally, with four velocity estimates and two magnetic field estimates now available at each edge, we combine these to produce four EMF estimates at each edge:
\begin{align}
    \scalebox{0.925}{$\displaystyle
    \begin{aligned}
        \varepsilon_{2}^{[Q_\mathrm{II}]}\EvalAtV{i \pm 1/2}{j \pm 1/2}{k}
            &= -\bar{u}_{0}^{[Q_\mathrm{II}]}\EvalAtV{i \pm 1/2}{j \pm 1/2}{k}
                b_{1}^{[S_\mathrm{I}]}\EvalAtV{i \pm 1/2}{j \pm 1/2}{k}
            \\
            &\nquad[1]
                + \bar{u}_{1}^{[Q_\mathrm{II}]}\EvalAtV{i \pm 1/2}{j \pm 1/2}{k}
                b_{0}^{[S_\mathrm{II}]}\EvalAtV{i \pm 1/2}{j \pm 1/2}{k}
    .
    \end{aligned}
    $}
    \label{eqn:emf:fs18}
\end{align}

\begin{figure}
    \centering
    \includegraphics[width=\linewidth]{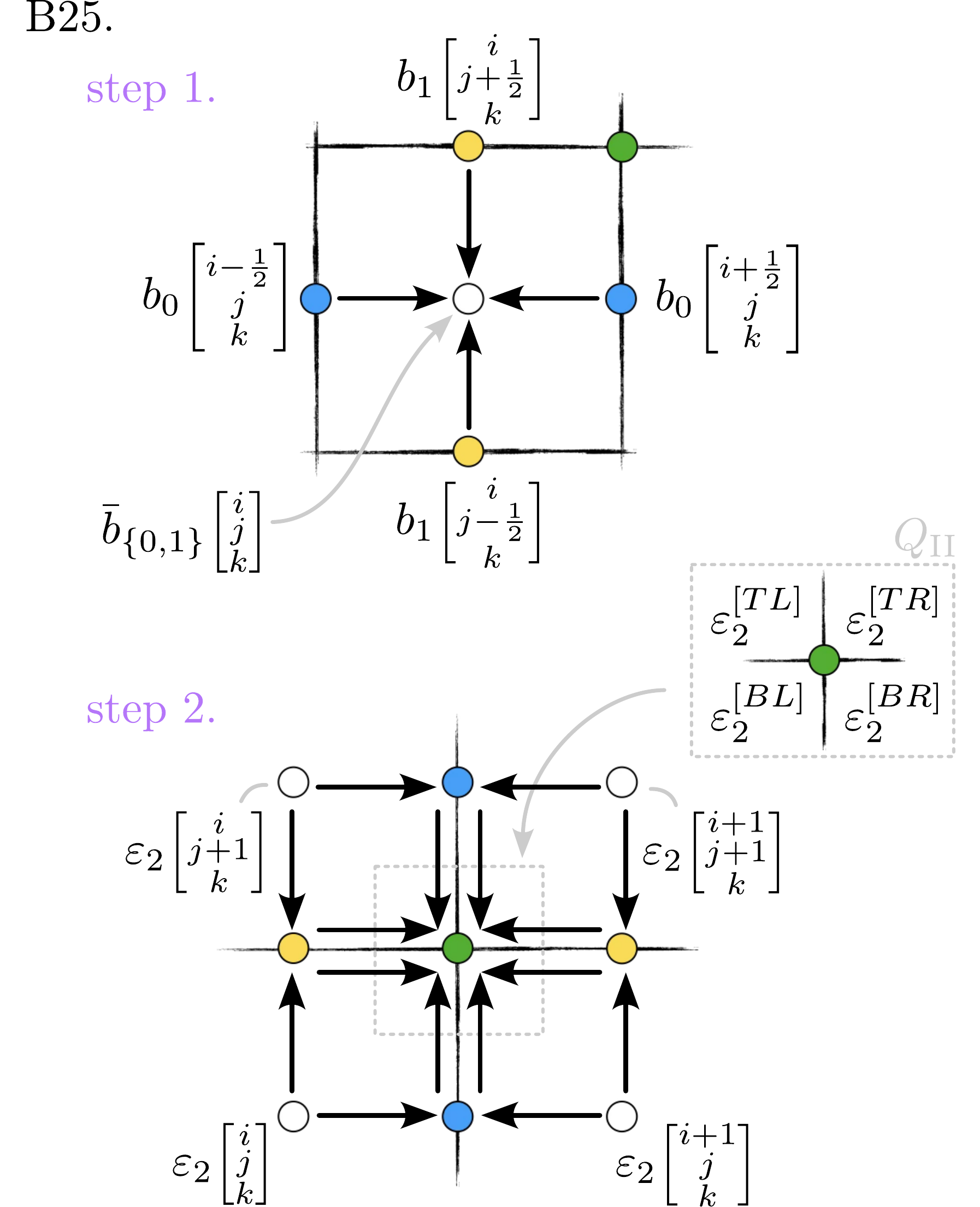}
    \caption{Two-step EMF construction for the \citetalias{Balsara25a} compute scheme: $b_0$ and $b_1$ (blue and yellow points) are averaged to the cell centre (white point) to compute $\varepsilon_2$ there (\cref{eqn:emf:b25a-b-avg,eqn:emf:b25a-emf-centre}), and then reconstructed to the cell edge (green point), yielding four estimates $\bar{\varepsilon}_2^{[Q_\mathrm{II}]}$ (\cref{eqn:emf:b25a-average}).}
    \label{fig:schematic:b25}
\end{figure}

Following the \citeauthor{Balsara25a} (2025, hereafter \citetalias{Balsara25a}; illustrated in \cref{fig:schematic:b25}) scheme, we first average the face-centred magnetic fields to cell centres,
\begin{align}
    b_0\EvalAtV{i}{j}{k}
        = \frac{1}{2}\rbrac{
                b_0\EvalAtV{i + 1/2}{j}{k}
                + b_0\EvalAtV{i - 1/2}{j}{k}
            }
    , \label{eqn:emf:b25a-b-avg}
\end{align}
and similarly for $b_1\EvalAtH{i}{j}{k}$. We then evaluate the $x_2$-component of the cell-centred EMF,
\begin{align}
    \varepsilon_2\EvalAtV{i}{j}{k}
        = -u_0\EvalAtV{i}{j}{k} b_1\EvalAtV{i}{j}{k} + u_1\EvalAtV{i}{j}{k} b_0\EvalAtV{i}{j}{k}
    , \label{eqn:emf:b25a-emf-centre}
\end{align}

We next reconstruct these cell-centred EMFs at the cell faces, yielding $\varepsilon_{2}^{[S_\mathrm{I}: x_0]}\EvalAtH{i \pm 1/2}{j}{k}$ and $\varepsilon_{2}^{[S_\mathrm{II}: x_1]}\EvalAtH{i}{j \pm 1/2}{k}$. We then reconstruct these face-centred EMFs at the cell edges, in the $x_1$-direction for the $x_0$-face-centred EMF and in the $x_0$-direction for the $x_1$-face-centred EMF, yielding eight estimates of each EMF at the edge, which we denote $\varepsilon_{2}^{[Q_\mathrm{II}: x_0 x_1]}\EvalAtH{i \pm 1/2}{j \pm 1/2}{k}$ and $\varepsilon_{2}^{[Q_\mathrm{II}: x_1 x_0]}\EvalAtH{i \pm 1/2}{j \pm 1/2}{k}$, where, as before, the superscripts $x_0 x_1$ and $x_1 x_0$ indicate the reconstruction order.

As in the \citetalias{Felker18a} method, we have two estimates for each of the four corners LB, RB, LT, and RT, resulting from the two possible orderings of the 1D reconstructions, which we average,
\begin{align}
\begin{aligned}
    \varepsilon_{2}^{[Q_\mathrm{II}]}\EvalAtV{i \pm 1/2}{j \pm 1/2}{k}
        &= \frac{1}{2}\Bigg(
            \varepsilon_{2}^{[Q_\mathrm{II}: x_0 x_1]}\EvalAtV{i \pm 1/2}{j \pm 1/2}{k}
        \\
        &\nquad[4]
            + \varepsilon_{2}^{[Q_\mathrm{II}: x_1 x_0]}\EvalAtV{i \pm 1/2}{j \pm 1/2}{k}
        \Bigg)
    .
\end{aligned}
    \label{eqn:emf:b25a-average}
\end{align}

\begin{figure}
    \centering
    \includegraphics[width=\linewidth]{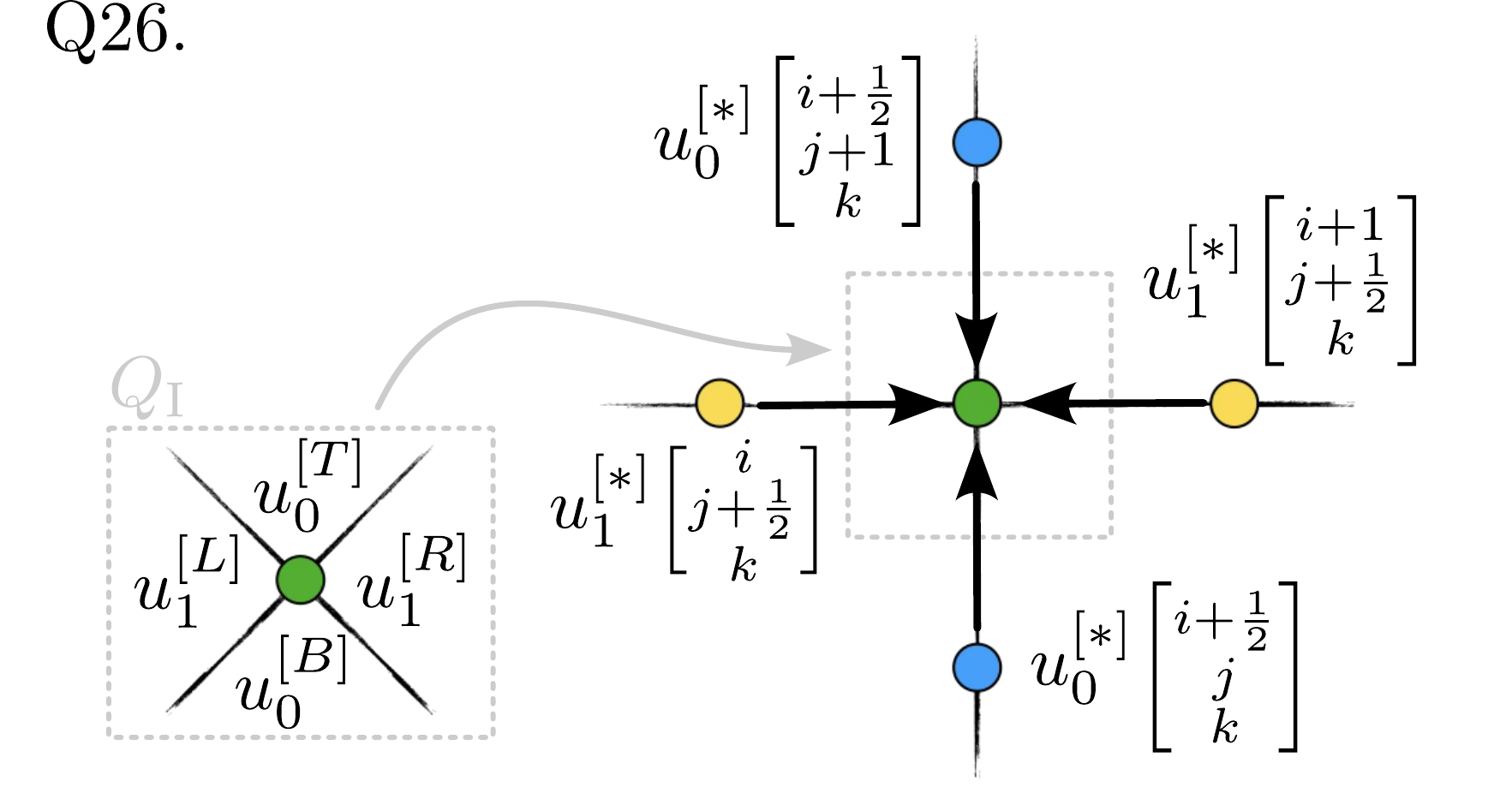}
    \caption{Velocity reconstruction for the Q26 compute scheme. Starting from face-normal velocities $u_0^{[*]}$ and $u_1^{[*]}$ (\cref{eqn:emf:q26-facevel}) at the $x_0$ and $x_1$ cell faces (blue and yellow points), these are reconstructed at the cell edge (green point) to yield $u_0^{[B]}$ and $u_0^{[T]}$, and $u_1^{[L]}$ and $u_1^{[R]}$. These combine with the magnetic fields reconstructed at the edge (\cref{fig:schematic:b-common}) to yield the EMF (\cref{eqn:emf:q26}).}
    \label{fig:schematic:q26}
\end{figure}

We refer to the final scheme that we consider as the Quokka 2026 (Q26) scheme (illustrated in \cref{fig:schematic:q26}), which modifies the upwind CT approach of \citet{DelZanna07a} \citep[see also][]{Mignone21a}. As with the \citetalias{Felker18a} method, we independently reconstruct the velocity and magnetic field at the cell edge, and combine them there to compute the EMF, but with an important difference: rather than reconstructing velocities at the face from the cell centre and then at the edge from the face, we instead begin from face-normal velocities constructed from a wavespeed-weighted average of the reconstructed left and right states at the face (\cref{eqn:emf:q26-facevel}), using the same bounding wavespeeds as the Riemann solver (\sref{sec:methods:flux}) rather than its resolved fluxes or interface states, and therefore require only a single reconstruction operation for them to reach the relevant cell edges.\footnote{
    We originally attempted to compute the face velocity by dividing the mass flux returned by the Riemann solver by the upwind-state density, but found that this was unstable for the complex magnetic field geometries resolved in high-resolution, turbulent systems like the Orszag-Tang vortex and the current sheet tested later in this paper.
} We estimate this face-normal velocity, which we denote $u_{0}^{[*]}\EvalAtH{i \pm 1/2}{j}{k}$ on the $x_0$-faces (and similarly $u_1^{[*]}\EvalAtH{i}{j \pm 1/2}{k}$ on the $x_1$-faces, \etc), via a fast-magnetosonic-wavespeed-weighted average of the reconstructed left and right normal-velocity states at the face,
\begin{align}
    u_{0}^{[*]}
        = \frac{\alpha_0^{[+]} u_{0}^{[L]} + \alpha_0^{[-]} u_{0}^{[R]}}{\alpha_0^{[+]} + \alpha_0^{[-]}}
    , \label{eqn:emf:q26-facevel}
\end{align}
where every quantity is evaluated at the shared face $\EvalAtH{i + 1/2}{j}{k}$. Following \citet{Miyoshi05a} (their eqn. 67), we estimate the non-negative bounding wavespeeds of the Riemann fan at the $x_0$-face as
\begin{align}
    \alpha_0^{[+]}
        &\scalebox{0.925}{$\displaystyle{}= \max\sbrac{0,\ \max\sbrac{u_0^{[L]}, u_0^{[R]}} + \max\sbrac{c_\mathrm{fm}^{[L]}, c_\mathrm{fm}^{[R]}}}$}
    \\
    \alpha_0^{[-]}
        &\scalebox{0.925}{$\displaystyle{}= \max\sbrac{0,\ \max\sbrac{c_\mathrm{fm}^{[L]}, c_\mathrm{fm}^{[R]}} - \min\sbrac{u_0^{[L]}, u_0^{[R]}}},$}
    \label{eqn:emf:alpha-pm}
\end{align}
where $c_\mathrm{fm}^{[L]}$ and $c_\mathrm{fm}^{[R]}$ are the fast-magnetosonic speed (\cref{eqn:waves:fast-speed}) evaluated at the reconstructed left and right states. The wavespeed averaging we use is the same as equation 57 of \citet{DelZanna07a}, but here we apply it to the normal component of the velocity rather than to the transverse components, so that each face carries a single velocity component instead of two, and correspondingly fewer fields need reconstructing to the cell edges.

Once we have computed the face velocities $u_0^{[*]}$ and $u_1^{[*]}$, we reconstruct $u_{0}^{[*]}$ in the $x_1$-direction to yield $u_{0}^{[*;S_\mathrm{II}]}\EvalAtH{i \pm 1/2}{j \pm 1/2}{k}$, and reconstruct $u_{1}^{[*]}$ in the $x_0$-direction to yield $u_{1}^{[*;S_\mathrm{I}]}\EvalAtH{i \pm 1/2}{j \pm 1/2}{k}$. For the $x_2$-edges, we can then compute four estimates of the EMF as
\begin{align}
    \scalebox{0.925}{$\displaystyle
    \begin{aligned}
        \varepsilon_{2}^{[Q_\mathrm{II}]}\EvalAtV{i \pm 1/2}{j \pm 1/2}{k}
            &= -u_{0}^{[*;S_\mathrm{II}]}\EvalAtV{i \pm 1/2}{j \pm 1/2}{k}
                b_{1}^{[S_\mathrm{I}]}\EvalAtV{i \pm 1/2}{j \pm 1/2}{k}
            \\
            &\nquad[1]
                + u_{1}^{[*;S_\mathrm{I}]}\EvalAtV{i \pm 1/2}{j \pm 1/2}{k}
                b_{0}^{[S_\mathrm{II}]}\EvalAtV{i \pm 1/2}{j \pm 1/2}{k}
    .
    \end{aligned}
    $}
    \label{eqn:emf:q26}
\end{align}

In comparison to the \citetalias{Felker18a} and \citetalias{Balsara25a} schemes, it is clear that the Q26 scheme involves significantly fewer reconstructions and fewer temporary variables. However, this is partially offset by an increase in the cost of the flux calculation associated with ghost zones. For reasons we discuss in \sref{sec:methods:emf:averaging}, our EMF averaging schemes require that we calculate the fluxes not just in the real cells of the domain, but also for at least one ghost zone. The Q26 scheme in general requires more, at least if we wish to use a higher-order reconstruction method: we must be able to reconstruct $u_{0}^{[*]}$ in the $x_1$-direction, which requires knowledge of $u_{0}^{[*]}$ not just at the one ghost face adjacent to the real cells, but in two ghost zones for PLM reconstruction, and three for PPM or PPM-EP. We are therefore required to solve the Riemann problem in this many more ghost zones. We will test the performance impact of this trade-off (fewer reconstruction steps but more ghost zones) in \sref{sec:scheme-comparisons}.

\subsubsection{The EMF averaging scheme}
\label{sec:methods:emf:averaging}

Each of the EMF compute schemes described in \sref{sec:methods:emf:compute} yields four quadrant-estimates of the EMF around each corner of the cell edge; for the $x_2$-edges, these are denoted $\varepsilon_{2}^{[Q_\mathrm{II}]}\EvalAtH{i \pm 1/2}{j \pm 1/2}{k}$. The final step is to average these into a single edge-centred EMF, $\varepsilon_{2}\EvalAtH{i \pm 1/2}{j \pm 1/2}{k}$, for which we consider two different schemes. Both also require the magnetic fields reconstructed at the cell edges: $b_{0}^{[S_\mathrm{II}]}\EvalAtH{i \pm 1/2}{j \pm 1/2}{k}$ and $b_{1}^{[S_\mathrm{I}]}\EvalAtH{i \pm 1/2}{j \pm 1/2}{k}$ (see \cref{fig:schematic:b-common}), and the wavespeed magnitudes there, taken from the face-centred values (\cref{eqn:emf:alpha-pm}):
\begin{align}
    \scalebox{0.875}{$\displaystyle
        \alpha_0^{[\pm]}\EvalAtV{i + 1/2}{j + 1/2}{k}
            = \max\sbrac{
                \alpha_0^{[\pm]}\EvalAtV{i + 1/2}{j}{k},
                \alpha_0^{[\pm]}\EvalAtV{i + 1/2}{j + 1}{k}
            }
    .$}
    \label{eqn:emf:alpha-edge}
\end{align}
$\alpha_1^{[\pm]}$ is computed analogously after cyclic permutation. Note that computing these wavespeeds requires that we solve the Riemann problem not just over the real cells of the domain, but for one ghost zone as well.

Our first scheme, following the \citet[hereafter \citetalias{Londrillo04a}]{Londrillo04a} approach, combines the four corner EMFs using the solution to a 2D HLL Riemann solver. Our implementation follows equation 41 of \citetalias{Felker18a},
\begin{align}
\begin{aligned}
    \varepsilon_{2}\EvalAtV{i \pm 1/2}{j \pm 1/2}{k}
        &= \frac{
            \alpha_0^{[+]}\alpha_1^{[+]} \varepsilon_2^{[LB]}
                + \alpha_0^{[-]}\alpha_1^{[+]} \varepsilon_2^{[RB]}
        }{
            \rbrac{\alpha_0^{[+]}+\alpha_0^{[-]}} \rbrac{\alpha_1^{[+]}+\alpha_1^{[-]}}
        }
        \\
        &\nquad[1]
            + \frac{
                \alpha_0^{[+]}\alpha_1^{[-]} \varepsilon_2^{[LT]}
                + \alpha_0^{[-]}\alpha_1^{[-]} \varepsilon_2^{[RT]}
            }{
                \rbrac{\alpha_0^{[+]}+\alpha_0^{[-]}} \rbrac{\alpha_1^{[+]}+\alpha_1^{[-]}}
            }
        \\
        &\nquad[1]
            - \frac{\alpha_1^{[+]}\alpha_1^{[-]}}{\alpha_1^{[+]}+\alpha_1^{[-]}} \rbrac{b_0^{[T]} - b_0^{[B]}}
        \\
        &\nquad[1]
            + \frac{\alpha_0^{[+]}\alpha_0^{[-]}}{\alpha_0^{[+]}+\alpha_0^{[-]}} \rbrac{b_1^{[R]} - b_1^{[L]}}
    ,
\end{aligned}
    \label{eqn:emf:ld04}
\end{align}
where every quantity in this equation is evaluated at the shared edge $\EvalAtH{i \pm 1/2}{j \pm 1/2}{k}$. This yields a single upwinded EMF estimate $\varepsilon_2\EvalAtH{i \pm 1/2}{j \pm 1/2}{k}$ at each cell edge.

Our second scheme uses the new 2D Riemann solver developed by \citeauthor{Balsara25b} (2025, hereafter \citetalias{Balsara25b}) to combine the EMFs. Unlike \cref{eqn:emf:ld04}, this cannot be expressed as a single average of the four corner EMFs: instead, depending on the signs of the one-sided wavespeeds bounding the edge, it selects among a raw corner EMF, a one-sided (single-direction) resolved state, or a fully 2D resolved state analogous to \cref{eqn:emf:ld04}. We do not reproduce this full matrix here, and instead refer the reader to equations 3.2-3.10 of \citetalias{Balsara25a} for the complete definition and discussion of this method.

Note, the EMF compute scheme (\citetalias{Balsara25a}) and EMF averaging scheme (\citetalias{Balsara25b}) are independent choices in our framework, so we can pair the \citetalias{Balsara25b} Riemann solver with any combination of four corner EMFs, regardless of the EMF compute scheme used to obtain them; conversely, we are free to use the \citetalias{Londrillo04a} averaging scheme rather than \citetalias{Balsara25b} on corner EMFs even if they were derived using the \citetalias{Balsara25a} compute scheme. Below in \sref{sec:scheme-comparisons} we will test every such combination.

\subsection{Constant Ohmic resistivity}
\label{sec:methods:resistivity}

Up to this point we have described our scheme for the ideal MHD equations. We also support resistive MHD flows with a constant (in space and time) Ohmic resistivity $\eta$, through an extension of the ideal EMF (\cref{eqn:induction:continuous}) using the generalised Ohm's law:
\begin{align}
    \mVector{\varepsilon}
        &= -\mVector{u}\times\mVector{b} + \eta\, \mVector{j}
    , \label{eqn:resistivity:ohms-law}
\end{align}
where $\mVector{j} = \nabla\times\mVector{b}$ is the current density in rationalised-Gaussian units. Conveniently, since $\eta\, \mVector{j}$ is constructed on the same stencil (\ie~on cell edges) as the EMFs that are used to evaluate \cref{eqn:ct:circulation-update}, the resistive term is added as an explicit magnetic-diffusion source during the existing EMF calculation in \sref{sec:methods:emf}. This reuses the kernel launches and memory footprint of the existing EMF calculation, and thus does not impact the CT update, so $\nabla\cdot\mVector{b} = 0$ remains preserved by construction.

\begin{figure}
    \centering
    \includegraphics[width=\linewidth]{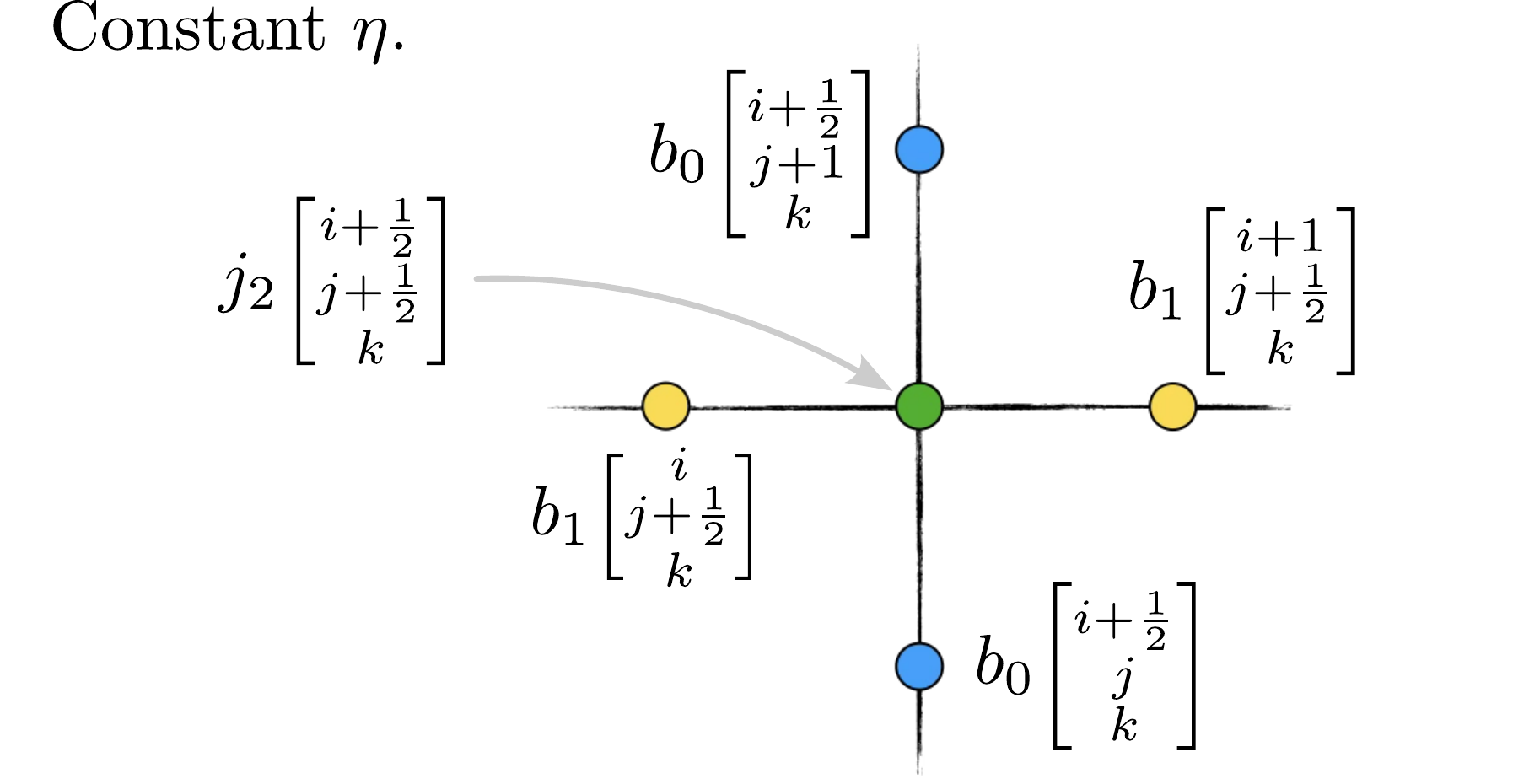}
    \caption{The constant-$\eta$ stencil for the Ohmic contribution to the edge-centred EMF $\varepsilon_2$ (green point), constructed from the four adjacent face-centred magnetic field values: $b_0$ (blue points) and $b_1$ (yellow points) (\cref{eqn:resistivity:current-stencil}). The corresponding stencils for $j_0$ and $j_1$ follow by cyclic permutation.}
    \label{fig:schematic:eta-constant}
\end{figure}

For scalar $\eta$, only the component of $\mVector{j}$ aligned with each cell edge enters the EMF correction. Without loss of generality, the component along the $x_2$-edge, centred at $\EvalAtH{i + 1/2}{j + 1/2}{k}$ (see \cref{fig:schematic:eta-constant}), is
\begin{align}
    j_2
        &= \mpFrac{b_1}{x_0} - \mpFrac{b_0}{x_1}
    .
\end{align}
The staggered mesh places each partial derivative exactly at the edge, so a centred difference of the adjacent face-centred values requires no reconstruction; doing so gives
\begin{align}
    \scalebox{0.925}{$\displaystyle
    \begin{aligned}
        &j_2\EvalAtV{i + 1/2}{j + 1/2}{k}
            = \frac{1}{\Delta{x_0}} \rbrac{
                    b_1\EvalAtV{i + 1}{j + 1/2}{k}
                    - b_1\EvalAtV{i}{j + 1/2}{k}
                } \\
            &\nquad[3]
                - \frac{1}{\Delta{x_1}} \rbrac{
                    b_0\EvalAtV{i + 1/2}{j + 1}{k}
                    - b_0\EvalAtV{i + 1/2}{j}{k}
                }
    ,
    \end{aligned}
    $}
        \label{eqn:resistivity:current-stencil}
\end{align}
and corresponding expressions for $j_0$ and $j_1$ follow by cyclic permutation of the indices. The resistive contribution is then added to the EMF computed in \sref{sec:methods:emf:averaging},
\begin{align}
    \scalebox{0.925}{$\displaystyle
        \varepsilon_2\EvalAtV{i + 1/2}{j + 1/2}{k}
            = -(\mVector{u}\times\mVector{b})_2\EvalAtV{i + 1/2}{j + 1/2}{k}
                + \eta\, j_2\EvalAtV{i + 1/2}{j + 1/2}{k}
    ,$}
    \label{eqn:resistivity:emf-update}
\end{align}
so that the resistive term enters the same line-integral update of the face-centred magnetic field as the ideal EMF (\cref{eqn:ct:circulation-update}). This treatment is explicit in time, and reduces exactly to the ideal scheme when $\eta = 0$.

Note that, since the energy flux associated with the induction equation is linear in the EMF, adding the resistive term $\eta\,\mVector{j}$ to $\mVector{\varepsilon}$ (\cref{eqn:resistivity:ohms-law}) adds a corresponding $\eta\,\mVector{j}\times\mVector{b}$ term to the total-energy flux of \cref{eqn:mhd:conservation-laws}, on top of its existing ideal-MHD terms. Without this addition, energy would still be conserved by construction, but the flux across each face would be wrong: the magnetic energy that \cref{eqn:resistivity:emf-update} removes from $\mVector{b}$ would not be correctly transferred into internal energy, so energy would be split incorrectly between the two cells bordering that face. With this term, the dissipated magnetic energy is instead correctly converted into internal energy (Joule heating) at the rate $\eta\, j^2$ implied by \cref{eqn:resistivity:current-stencil}. We add this contribution to the flux on each cell face by gathering the resistive EMFs already computed on its four bounding edges.

Treating the resistive term explicitly comes with a diffusive (parabolic) time-step constraint, where stability requires
\begin{align}
    \Delta t
        &\leq \frac{(\Delta{x}_\mathrm{min})^2}{6\, \eta}
    , \label{eqn:resistivity:cfl}
\end{align}
with $\Delta{x}_\mathrm{min}$ the smallest cell width on the AMR-level and $\mathrm{CFL}$ the Courant-Friedrichs-Lewy (CFL) number. We incorporate this constraint into the existing CFL machinery by treating $2\, \eta / \Delta{x}_\mathrm{min}$ as an effective signal speed contributed to the per-cell maximum alongside the MHD wave speeds, so that the hyperbolic and parabolic constraints share the same user-controlled CFL number. Note that, as for the hyperbolic constraint, the dimensional factor is absorbed into the CFL number: satisfying \cref{eqn:resistivity:cfl} in 3D therefore requires $\mathrm{CFL} < 1/3$. We currently disable AMR subcycling-in-time when $\eta \ne 0$ to keep the resistive constraint consistent across refinement levels; relaxing this restriction to allow a per-level resistive time step is left as future work.

\subsection{Time-stepping with flux correction}
\label{sec:methods:time-stepping}

\begin{figure}
    \centering
    \includegraphics[width=\linewidth]{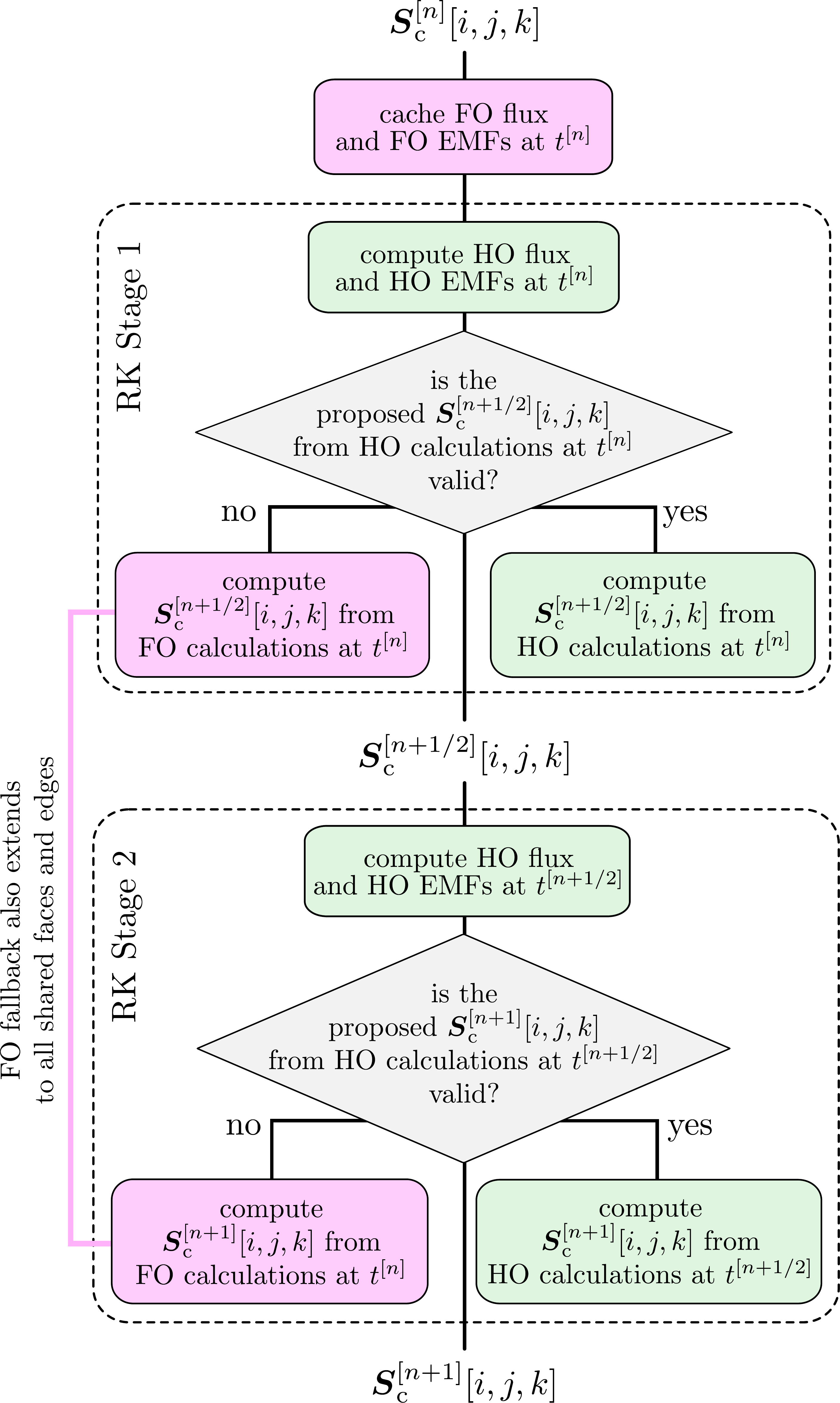}
    \caption{Flow chart of the \quokka~time-stepping strategy combining RK2-SSP with first-order flux correction. Here quantities labelled as high-order (HO) are computed using either PLM, PPM, or PPM-EP reconstruction and the HLLD Riemann solver, while those labelled first-order (FO) are computed using PC reconstruction and the LLF Riemann solver.}
    \label{fig:schematic:flowchart}
\end{figure}

By default, and as in \citetalias{Wibking22a}, \quokka~uses a strong-stability-preserving, second-order-accurate Runge-Kutta (RK2-SSP) time-stepping strategy\footnote{
    In problems that include the radiation submodule, we change to the asymptotic-preserving IMEX PD-ARS method following \citet{Chu19a}, implemented as described by \citet{He24a}. However, this method reduces to RK2-SSP for the non-radiation parts of the update, including MHD, and we therefore will not distinguish the two schemes here.
} \citep{Shu88a} that is robust under the standard hyperbolic CFL restriction. Nonetheless, for strong shocks or high-Mach-number flows, this scheme can become unstable even while time steps obey the CFL condition; indeed, there is no known method for guaranteeing stability in multidimensional schemes with accuracies greater than first order \citep{Godunov59a}. To mitigate this problem, we introduce a time-stepping strategy featuring a dynamic first-order flux correction (FOFC) that maintains stability even in difficult-to-integrate flows. We describe this in detail below (and summarise it in \cref{fig:schematic:flowchart}), drawing attention to the different spatial stencils that the cell-centred hydrodynamic quantities and face-centred magnetic fields each depend on.

Starting at time $t^{[n]}$ with a set of cell-centred conserved quantities $\mVector{S}_\mathrm{c}^{[n]}\EvalAtH{i}{j}{k}$ and face-centred magnetic fields $b_0^{[n]}\EvalAtH{i + 1/2}{j}{k}$, we advance to time $t^{[n + 1]} = t^{[n]} + \Delta t$ as follows.\footnote{
    For simplicity we omit the extra operations involved if we are using the \quokka~dual energy formalism. If dual energy is in use, we apply the dual energy correction described in \citetalias{Wibking22a} following each calculation of a cell-centred state.
}
At the start of each time step (see the first, pink box in \cref{fig:schematic:flowchart}), we begin by computing a set of first-order, highly-dissipative fluxes of conserved cell-centred quantities and EMFs, for each cell face and edge respectively, using the old-time state $\mVector{S}_\mathrm{c}^{[n]}\EvalAtH{i}{j}{k}$ and $b_0^{[n]}\EvalAtH{i + 1/2}{j}{k}$. We carry out this flux calculation as described in \sref{sec:methods:flux}, using PC reconstruction and the LLF Riemann solver, and then use the resulting wavespeeds to compute first-order EMFs as described in \sref{sec:methods:emf}. We denote the resulting first-order fluxes of conserved cell-centred quantities as $\mVector{F}_\mathrm{FO}^{[n]}\EvalAtH{i \pm 1/2}{j}{k}$ and the corresponding EMFs as $\mVector{\varepsilon}_\mathrm{FO}^{[n]}\EvalAtH{i \pm 1/2}{j \pm 1/2}{k}$; we cache these first-order estimates, to fall back on later in the update, should the high-order scheme produce an invalid state.

For the first RK stage (see \cref{fig:schematic:flowchart}), we compute a set of high-order, low-dissipation fluxes and EMFs as described in \sref{sec:methods:flux} and \sref{sec:methods:emf}; this calculation is identical to the one used to compute $\mVector{F}_\mathrm{FO}^{[n]}$ and $\mVector{\varepsilon}_\mathrm{FO}^{[n]}$ above, except that we use a higher-order reconstruction scheme (PLM, PPM, or PPM-EP as chosen by the user) and the HLLD Riemann solver, rather than PC reconstruction coupled with the LLF Riemann solver; we denote the resulting fluxes and EMFs as $\mVector{F}_\mathrm{HO}^{[n]}\EvalAtH{i \pm 1/2}{j}{k}$ and $\mVector{\varepsilon}_\mathrm{HO}^{[n]}\EvalAtH{i \pm 1/2}{j \pm 1/2}{k}$. From a generic set of fluxes and EMFs at time step $[m]$, we define the update increment for the conserved cell-centred quantities,
\begin{align}
    \scalebox{0.925}{$\displaystyle
    \begin{aligned}
        &D^{[m]}\sbrac{\mVector{S}_\mathrm{c}}\EvalAtV{i}{j}{k}
            = \Delta t \Bigg(
            \\
            &\nquad[3]
                \, \frac{1}{\Delta{x_0}} \rbrac{
                    \mVector{F}^{[m]}\EvalAtV{i - 1/2}{j}{k}
                    - \mVector{F}^{[m]}\EvalAtV{i + 1/2}{j}{k}
                }
            \\
            &\nquad[2]
                + \frac{1}{\Delta{x_1}} \rbrac{
                    \mVector{F}^{[m]}\EvalAtV{i}{j - 1/2}{k}
                    - \mVector{F}^{[m]}\EvalAtV{i}{j + 1/2}{k}
                }
            \\
            &\nquad[2]
                + \frac{1}{\Delta{x_2}} \rbrac{
                    \mVector{F}^{[m]}\EvalAtV{i}{j}{k - 1/2}
                    - \mVector{F}^{[m]}\EvalAtV{i}{j}{k + 1/2}
                }
            \Bigg)
        ,
    \end{aligned}
    $}
    \label{eqn:rk:increment-state}
\end{align}
and, analogously, the update increment for the face-centred magnetic fields from the EMFs at the same time step
\begin{align}
    \scalebox{0.925}{$\displaystyle
    \begin{aligned}
        &D^{[m]}\sbrac{b_0}\EvalAtV{i + 1/2}{j}{k}
            = \frac{\Delta t}{\Delta{x_1} \, \Delta{x_2}} \Bigg(
            \\
            &\nquad[3]
                \, \Delta{x_2} \, \varepsilon_2^{[m]}\EvalAtV{i + 1/2}{j - 1/2}{k}
                + \Delta{x_1} \, \varepsilon_1^{[m]}\EvalAtV{i + 1/2}{j}{k + 1/2}
            \\
            &\nquad[2]
                - \Delta{x_2} \, \varepsilon_2^{[m]}\EvalAtV{i + 1/2}{j + 1/2}{k}
                - \Delta{x_1} \, \varepsilon_1^{[m]}\EvalAtV{i + 1/2}{j}{k - 1/2}
        \Bigg)
        .
    \end{aligned}
    $}
    \label{eqn:rk:increment-bfield}
\end{align}
The proposed intermediate state of the first RK stage then follows directly,
\begin{align}
    \mVector{S}_\mathrm{c}^{[n + 1/2]}\EvalAtV{i}{j}{k}
        &= \mVector{S}_\mathrm{c}^{[n]}\EvalAtV{i}{j}{k}
            + D^{[n]}\sbrac{\mVector{S}_\mathrm{c}}\EvalAtV{i}{j}{k}
    ,
\end{align}
and similarly for the face-centred magnetic fields,
\begin{align}
    \scalebox{0.925}{$\displaystyle
    \begin{aligned}
        b_0^{[n + 1/2]}\EvalAtV{i + 1/2}{j}{k}
            &= b_0^{[n]}\EvalAtV{i + 1/2}{j}{k}
            \\
            &\nquad[3]
            + D^{[n]}\sbrac{b_0}\EvalAtV{i + 1/2}{j}{k}
        .
    \end{aligned}
    $}
    \label{eqn:rk1:update}
\end{align}

\begin{figure}
    \centering
    \includegraphics[width=0.8\linewidth]{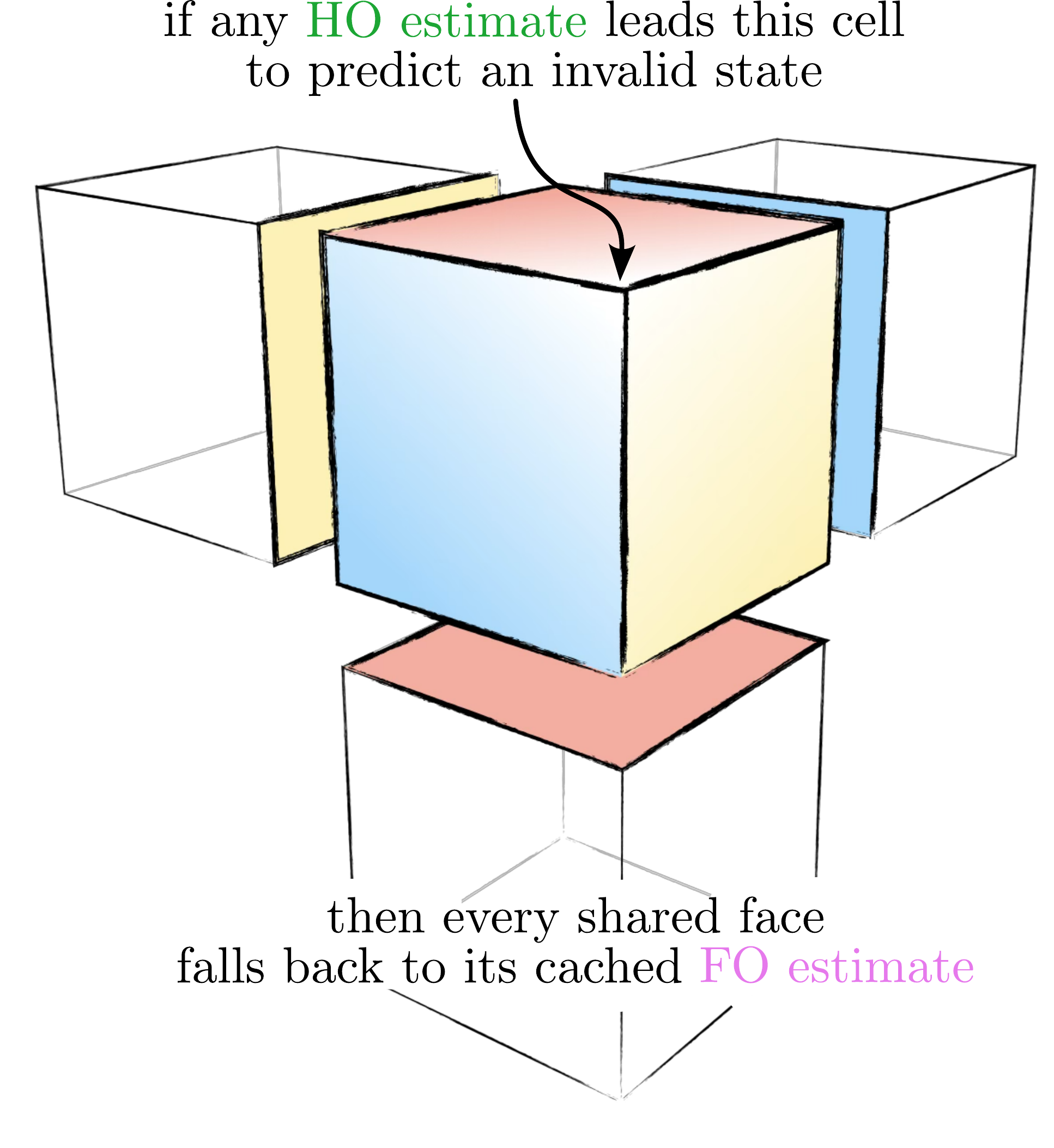}
    \caption{When the state of a cell is flagged as invalid after a higher-order update, first-order flux correction (FOFC) replaces all of its fluxes with first-order estimates; this includes those shared with neighbouring cells. Note that faces are shaded by their face normal direction, using the same colours as in \cref{fig:schematic:b-common}.}
    \label{fig:schematic:fofc}
\end{figure}

We then check whether any of the intermediate state or magnetic quantities are physically invalid; at a minimum this means flagging cell centre states with a negative density, but we also provide a means for users to flag other, more restrictive criteria, \eg~velocities that are unphysically large. For any cell $\EvalAtH{i}{j}{k}$ flagged as invalid (see \cref{fig:schematic:fofc} for an illustration), we replace the HO fluxes and EMFs for all of its faces and edges with their cached FO values, $\mVector{F}_\mathrm{FO}^{[n]}\EvalAtH{i \pm 1/2}{j}{k}$ and $\mVector{\varepsilon}_\mathrm{FO}^{[n]}\EvalAtH{i \pm 1/2}{j \pm 1/2}{k}$, and recompute the state and magnetic fields for that cell by again applying \cref{eqn:rk1:update} with the replaced fluxes. We also recalculate the intermediate state vector and magnetic fields for any neighbouring cell sharing a face with the flagged cell. This ensures that fluxes and EMFs are applied consistently for the update step, which is a necessary condition for conservation and maintenance of $\nabla\cdot\mVector{b} = 0$.

For the second RK stage (see \cref{fig:schematic:flowchart}), we compute a set of high-order fluxes and EMFs exactly as we did to compute $\mVector{F}_\mathrm{HO}^{[n]}$ and $\mVector{\varepsilon}_\mathrm{HO}^{[n]}$ above, but we now use the intermediate-time states $\mVector{S}_\mathrm{c}^{[n + 1/2]}\EvalAtH{i}{j}{k}$ and $b_0^{[n + 1/2]}\EvalAtH{i + 1/2}{j}{k}$ as inputs. We denote these fluxes and EMFs as $\mVector{F}_\mathrm{HO}^{[n + 1/2]}\EvalAtH{i \pm 1/2}{j}{k}$ and $\mVector{\varepsilon}_\mathrm{HO}^{[n + 1/2]}\EvalAtH{i \pm 1/2}{j \pm 1/2}{k}$, giving access to the update increments $D^{[n + 1/2]}\sbrac{\mVector{S}_\mathrm{c}}$ and $D^{[n + 1/2]}\sbrac{b_0}$ following \cref{eqn:rk:increment-state,eqn:rk:increment-bfield}. Together with the increments at time step $[n]$, determined during the first RK stage (using the HO fluxes and EMFs, unless they were replaced by their cached FO equivalents), the proposed final state of the second RK stage is simply the average of the increments from both time steps,
\begin{align}
    \scalebox{0.925}{$\displaystyle
    \begin{aligned}
        \mVector{S}_\mathrm{c}^{[n + 1]}\EvalAtV{i}{j}{k}
            = \mVector{S}_\mathrm{c}^{[n]}\EvalAtV{i}{j}{k}
                &+ \frac{1}{2}\Bigg(
                D^{[n]}\sbrac{\mVector{S}_\mathrm{c}}\EvalAtV{i}{j}{k}
            \\
            &\nquad[3]
                + D^{[n + 1/2]}\sbrac{\mVector{S}_\mathrm{c}}\EvalAtV{i}{j}{k}
            \Bigg)
    ,
    \end{aligned}
    $}
\end{align}
and similarly for the face-centred magnetic fields,
\begin{align}
    \scalebox{0.925}{$\displaystyle
    \begin{aligned}
        &b_0^{[n + 1]}\EvalAtV{i + 1/2}{j}{k}
            = b_0^{[n]}\EvalAtV{i + 1/2}{j}{k}
                + \frac{1}{2}\Bigg(
            \\
            &\nquad[3]
                D^{[n]}\sbrac{b_0}\EvalAtV{i + 1/2}{j}{k}
                + D^{[n + 1/2]}\sbrac{b_0}\EvalAtV{i + 1/2}{j}{k}
            \Bigg)
    ,
    \end{aligned}
    $}
    \label{eqn:rk2:update}
\end{align}

We again check for physically invalid state or magnetic quantities in any cells, and if one is found, we replace the RK-averaged fluxes and EMFs on the faces and edges of that cell (\cref{eqn:rk2:update}) outright with their cached FO values, $\mVector{F}_\mathrm{FO}^{[n]}\EvalAtH{i \pm 1/2}{j}{k}$ and $\mVector{\varepsilon}_\mathrm{FO}^{[n]}\EvalAtH{i \pm 1/2}{j \pm 1/2}{k}$. We then recompute $\mVector{S}_\mathrm{c}^{[n + 1]}\EvalAtH{i}{j}{k}$ and $b_0^{[n + 1]}\EvalAtH{i + 1/2}{j}{k}$ in both the flagged cell and every cell sharing a face with it, using the replaced fluxes.

This completes the dynamic update cycle, which reduces to a standard RK2-SSP update if no flux replacements take place in either RK stage, and to a forward Euler (first-order in time) update if the HO fluxes are replaced in both stages. Importantly, the correction only affects the cells that require it, so regions that remain well behaved can still achieve an HO update. Intermediate levels of approximation are also possible, namely when only the first RK stage falls back to the FO estimate. By contrast, a cell that falls back at the second stage always reverts to an FO update for that step, irrespective of what its first stage used.

\subsection{AMR considerations}
\label{sec:methods:amr}

\quokka~uses \citet{Berger84a} block-structured AMR with \citet{Berger89a} adaptive time-stepping, as implemented in the \textsc{AMReX} library \citep{Zhang19b}. Since the cell-centred update couples to other refinement levels only through ghost zones, it is straightforward to extend the existing pure hydrodynamic AMR implementation from \citetalias{Wibking22a} to CT-MHD, and to support subcycled time-stepping across AMR levels. For the cell-centred quantities, the existing \quokka~flux-register machinery already ensures that coarse and fine levels use the same fluxes across shared faces, so conservation is maintained across coarse-fine boundaries.

Supporting the face-centred magnetic field under AMR, however, requires two additional ingredients specific to CT: (i) how to refine or coarsen the face-centred field when the resolution changes, and (ii) how to correct for EMF-driven flux mismatches across coarse-fine boundaries. For refinement and coarsening, \quokka~uses the divergence-preserving interpolator introduced by \citet{Vanella10a}, and implemented as part of \textsc{AMReX}. This guarantees that $\nabla\cdot\mVector{b} = 0$ still holds when we add or remove levels, or when we fill ghost zones. For these mismatches, we implement the EMF-correction scheme described by \citet{Balsara01a}, which is also used in the \textsc{Orion2} code \citep{Li12a}. This scheme corrects the magnetic field estimate on coarse cell faces that lie at coarse-fine boundaries to ensure that they are updated using the same EMFs as were used to update the fine cell faces, which maintains $\nabla\cdot\mVector{b} = 0$ across coarse-fine interfaces.

\section{Comparison of schemes}
\label{sec:scheme-comparisons}

\sref{sec:methods:emf} introduced three EMF compute schemes (\citetalias{Felker18a}, \citetalias{Balsara25a}, and Q26) and two EMF averaging schemes (\citetalias{Londrillo04a} and \citetalias{Balsara25b}), each of which can independently be paired with PLM, PPM, or PPM-EP reconstruction, giving 18 distinct scheme combinations in total. We run all 18 in \sref{sec:scheme-comparison:waves} and \sref{sec:scheme-comparison:orszag-tang}, where we compare the two higher-order schemes against the more diffusive PLM solutions, and then only use PPM and PPM-EP for the remaining tests. In this section we have two aims: (i) to validate our implementation against a reduced set of MHD test problems\footnote{
    \quokka~is a 3D code, so 1D and 2D problems are run in a 3D domain, extended uniformly along the trivial axes with the minimum number of cells needed for ghost-zone exchange. When discussing simulation runs, we only quote the resolutions along the non-trivial axes.
    \label{note:quokka:3d}
}, deferring the remainder of our validation suite to \sref{sec:stress-tests}; and (ii) to compare the 18 combinations directly, both in their accuracy and in their GPU throughput. Unless stated otherwise, all tests use $\mathrm{CFL} = 0.3$.

To this end, we first test the numerical convergence of each scheme combination in \sref{sec:scheme-comparison:waves}, isolating the accuracy of the reconstruction scheme in smooth flow regimes; we then test how well each scheme captures and resolves shocks in \sref{sec:scheme-comparison:shocks}, where discontinuities stress the Riemann solver; and finally, we test all 18 combinations against two more challenging, multidimensional non-linear problems in \sref{sec:scheme-comparison:orszag-tang} and \sref{sec:scheme-comparison:reconnection}. We then measure the relative throughput of each scheme in \sref{sec:scheme-comparison:speed}, before recommending in \sref{sec:scheme-comparison:recommendation} which scheme to use in \quokka.

\subsection{Convergence of magnetosonic waves}
\label{sec:scheme-comparison:waves}

For all four waves we test, we start by considering a uniform background state. Here we write $\mVector{S}$ (without a $\mathrm{c}$ or $\mathrm{p}$ subscript) for a vector of quantities from $\mVector{S}_\mathrm{c}$ and $\mVector{S}_\mathrm{p}$, used to define a problem setup.
\begin{align}
    \mVector{S}^{[\mathrm{bg}]}
        = \begin{Bmatrix}
                \rho \\
                \rho \mVector{u} \\
                \mVector{b} \\
                e_\mathrm{tot}
            \end{Bmatrix}
        = \begin{Bmatrix}
                \rho_\mathrm{bg} \\
                \mVector{0} \\
                b_\mathrm{bg} \mVectorUnit{n}_\mathrm{bg} \\
                \dfrac{p_\mathrm{bg}}{\gamma-1} + \dfrac{b_\mathrm{bg}^2}{2}
            \end{Bmatrix}
    , \label{eqn:wave:background}
\end{align}
where $\rho_\mathrm{bg}$, $p_\mathrm{bg}$, $b_\mathrm{bg}$, and $\mVectorUnit{n}_\mathrm{bg}$ are the background density, pressure, magnetic field magnitude, and magnetic field direction, respectively, and $\gamma = 5/3$ is the adiabatic index. In each wave test, we perturb this background state with a plane wave travelling along $\mVectorUnit{n}_k = \mVector{k}/k$, where $\mVector{k}$ is the wave vector, in a triply-periodic, cubic domain of side length $L = 1$. We always choose every component of $\mVector{k}L/2\pi$ to be an integer, so the wave is continuous across the periodic boundaries. Since this section compares all 18 EMF scheme combinations, we further restrict to grid-aligned wave propagation, $\mVector{k}L/2\pi = \{1,0,0\}$; later, in \sref{sec:stress-tests:misaligned-waves}, we verify that waves misaligned with the grid also propagate correctly.

\subsubsection{Linearly polarised Alfv\'en wave}
\label{sec:scheme-comparison:waves:alfven-linear}

This wave is the simplest, since only $\mVector{u}$ and $\mVector{b}$ are perturbed:
\begin{align}
    \mVector{S}^{[\delta\mathrm{AL}]}
        = \delta_\mathrm{wave} \cos\sbrac{
                \omega t - \mVector{k}\cdot\mVector{x}
            }
        \begin{Bmatrix}
            0 \\
            -c_\mathrm{A} \mVectorUnit{n}_\perp \\
            b_\mathrm{bg} \mVectorUnit{n}_\perp \\
            \delta_\mathrm{wave} c_\mathrm{A}^2
        \end{Bmatrix}
    ,
\end{align}
with $\delta_\mathrm{wave} = 10^{-6}$ a dimensionless number parameterising the perturbation amplitude, $c_\mathrm{A} = b_\mathrm{bg} / \sqrt{\rho_\mathrm{bg}}$ the Alfv\'en speed, $\mVectorUnit{n}_\perp$ a unit vector orthogonal to $\mVectorUnit{n}_{k}$, $\omega = c_\mathrm{A} k \cos\theta$ the angular frequency, and $\theta = \cos^{-1}\sbrac{\mVectorUnit{n}_\mathrm{bg} \cdot \mVectorUnit{n}_{k}}$ the angle between the background magnetic field and wave vector. For this test we use $\mVectorUnit{n}_\mathrm{bg} = \{1,0,0\}$ and $\mVectorUnit{n}_\perp = \{0,0,1\}$, so that $\theta = 0$. Since the restoring force is purely magnetic tension, $\mVector{u}$ and $\mVector{b}$ oscillate in-phase along the same transverse direction $\mVectorUnit{n}_\perp$.

\subsubsection{Circularly polarised Alfv\'en wave}
\label{sec:scheme-comparison:waves:alfven-circular}

Unlike the linearly-polarised Alfv\'en wave, here $\mVector{u}$ and $\mVector{b}$ are perturbed along two transverse directions at once, with a $90^\circ$ phase shift between them. The restoring force is again purely magnetic tension, so $\rho$ and $p$ remain unperturbed:
\begin{align}
    \mVector{S}^{[\delta\mathrm{AC}]}
        = \begin{Bmatrix}
            0 \\
            -c_\mathrm{A} \delta_\mathrm{wave} \rbrac{
                    \sin\phi\, \mVectorUnit{n}_{\perp,1}
                    + \cos\phi\, \mVectorUnit{n}_{\perp,2}
                } \\
            b_\mathrm{bg} \delta_\mathrm{wave} \rbrac{
                    \sin\phi\, \mVectorUnit{n}_{\perp,1}
                    + \cos\phi\, \mVectorUnit{n}_{\perp,2}
                } \\
            \delta_\mathrm{wave}^2 b_\mathrm{bg}^2
        \end{Bmatrix}
    ,
\end{align}
with $\phi = \omega t - \mVector{k}\cdot\mVector{x}$, $\omega = c_\mathrm{A} k$, and $\mVectorUnit{n}_{\perp,1}$ and $\mVectorUnit{n}_{\perp,2}$ the two mutually-orthogonal directions transverse to $\mVectorUnit{n}_{k}$.

This wave shares the same grid-aligned geometry as the linearly-polarised Alfv\'en wave ($\theta = 0$, so that $\mVectorUnit{n}_\mathrm{bg} \parallel \mVectorUnit{n}_{k}$); for this test we use $\mVectorUnit{n}_{\perp,1} = \{0,1,0\}$ and $\mVectorUnit{n}_{\perp,2} = \{0,0,1\}$. Since the transverse displacement of the perturbation has constant magnitude at every phase, unlike the linearly-polarised Alfv\'en, fast, or slow waves, this perturbation is an exact solution of the full non-linear MHD equations, regardless of amplitude $\delta_\mathrm{wave}$.

\subsubsection{Fast magnetosonic wave}
\label{sec:scheme-comparison:waves:fast}

Unlike both Alfv\'en waves, this wave also perturbs $\rho$ and $p$:
\begin{align}
\begin{aligned}
    &\mVector{S}^{[\delta\mathrm{fm}]}
        = \delta_\mathrm{wave} \cos\sbrac{
                \omega t - \mVector{k}\cdot\mVector{x}
            }
        \\
        &
        \scalebox{0.85}{$\displaystyle
            \begin{Bmatrix}
                \phi_1 \rho_\mathrm{bg} \\
                c_\mathrm{fm} (
                        \phi_1 \mVectorUnit{n}_{k} - \phi_2 \mVectorUnit{n}_\perp
                    ) \\
                b_\mathrm{bg} \mVectorUnit{n}_\perp \\
                \dfrac{\phi_1 p_\mathrm{bg} \gamma}{\gamma-1}
                    + b_\mathrm{bg}^2\sin\theta
                    + \delta_\mathrm{wave} \rbrac{
                        \dfrac{b_\mathrm{bg}^2}{2}
                        + \rho_\mathrm{bg} c_\mathrm{fm}^2 \rbrac{\phi_1^2 + \phi_2^2}
                    }
        \end{Bmatrix}
        ,$}
\end{aligned}
\end{align}
where $\mVectorUnit{n}_\perp$ is a unit vector orthogonal to $\mVectorUnit{n}_{k}$, lying in the plane formed by $\mVectorUnit{n}_{k}$ and $\mVectorUnit{n}_\mathrm{bg}$, with
\begin{align}
    \scalebox{0.925}{$\displaystyle
        c_\mathrm{fm}^2
            = \frac{1}{2} \rbrac{
                c_\mathrm{s}^2 + c_\mathrm{A}^2 + \rbrac{
                    \rbrac{
                        c_\mathrm{s}^2 + c_\mathrm{A}^2
                    }^2 - 4 c_\mathrm{s}^2 c_\mathrm{A}^2 \cos^2\theta
                }^{1/2}
            }
    ,$}
    \label{eqn:waves:fast-speed}
\end{align}
the fast magnetosonic speed, $c_\mathrm{s} = \sqrt{\gamma p_\mathrm{bg} / \rho_\mathrm{bg}}$ the sound speed, $\omega = c_\mathrm{fm} k$ the angular frequency, and the geometry factors $\phi_1$ and $\phi_2$ given by
\begin{align}
    \phi_1
        &= \frac{c_\mathrm{fm}^2 - c_\mathrm{A}^2 \cos^2\theta}{c_\mathrm{fm}^2 \sin\theta} \\
    \phi_2
        &= \rbrac{\frac{c_\mathrm{A}}{c_\mathrm{fm}}}^2 \cos\theta
    .
\end{align}
For this test we use $\mVectorUnit{n}_\mathrm{bg} = \{0,1,0\}$, so that $\theta = 90^\circ$ and $\mVectorUnit{n}_\perp = \{0,1,0\}$.

\subsubsection{Slow magnetosonic wave}
\label{sec:scheme-comparison:waves:slow}

Like the fast magnetosonic wave, the same combined magnetic and gas pressure restoring force acts here, with $\rho$ and $p$ perturbed alongside $\mVector{u}$ and $\mVector{b}$:
\begin{align}
\begin{aligned}
    &\mVector{S}^{[\delta\mathrm{sm}]}
        = \delta_\mathrm{wave} \cos\sbrac{
                \omega t - \mVector{k}\cdot\mVector{x}
            }
        \\
        &
        \scalebox{0.85}{$\displaystyle
            \begin{Bmatrix}
                \phi_1 \rho_\mathrm{bg} \\
                c_\mathrm{sm} (
                        -\phi_1 \mVectorUnit{n}_{k}
                        + \phi_2 \mVectorUnit{n}_\perp
                    ) \\
                b_\mathrm{bg} \mVectorUnit{n}_\perp \\
                \dfrac{\phi_1 p_\mathrm{bg} \gamma}{\gamma-1}
                    + b_\mathrm{bg}^2\sin\theta
                    + \delta_\mathrm{wave} \rbrac{
                            \dfrac{b_\mathrm{bg}^2}{2}
                            + \rho_\mathrm{bg} c_\mathrm{sm}^2 (\phi_1^2 + \phi_2^2)
                        }
            \end{Bmatrix}
        ,$}
\end{aligned}
\end{align}
where $\mVectorUnit{n}_\perp$ is the same as for the fast magnetosonic wave, with
\begin{align}
    \scalebox{0.925}{$\displaystyle
        c_\mathrm{sm}^2
            = \frac{1}{2}\rbrac{
                c_\mathrm{s}^2
                + c_\mathrm{A}^2
                - \rbrac{
                        \rbrac{c_\mathrm{s}^2
                        + c_\mathrm{A}^2}^2
                        - 4 c_\mathrm{s}^2 c_\mathrm{A}^2 \cos^2\theta
                    }^{1/2}
            }
    ,$}
    \label{eqn:waves:slow-speed}
\end{align}
the slow magnetosonic wave speed, $\omega = c_\mathrm{sm} k$ the angular frequency, and
\begin{align}
    \phi_1
        &= \frac{
            c_\mathrm{sm}^2 - c_\mathrm{A}^2\cos^2\theta
        }{
            c_\mathrm{sm}^2 \sin\theta
        } \\
    \phi_2
        &= \rbrac{\frac{c_\mathrm{A}}{c_\mathrm{sm}}}^2 \cos\theta
    .
\end{align}
For this test we use $\mVectorUnit{n}_\mathrm{bg} = \{1,1,0\}/\sqrt{2}$, so that $\theta = 45^\circ$ and $\mVectorUnit{n}_\perp = \{0,1,0\}$.

\subsubsection{Analysis of convergence}

For all four waves we adopt $\rho_\mathrm{bg} = b_\mathrm{bg} = 1$ with $p_\mathrm{bg} = 1/\gamma$ so $c_\mathrm{s} = 1$, and initialise each wave test\footnote{
    Initialising the magnetic field requires some care to ensure \mbox{$\nabla\cdot\mVector{b} = 0$} to machine precision under the \quokka~stencil. Rather than set the magnetic field directly, we use the vector potential corresponding to the magnetic field at the centres of cell edges (the same locations at which we store EMFs), and then compute the magnetic field at the face of each cell by taking the curl of the magnetic vector potential using the same stencil.
} by applying its perturbation to the background state (\cref{eqn:wave:background}). We then evolve each test for a single wave period, $t = 2\pi/\omega$, before comparing the evolved wave profile with the initial one. Across the convergence study, we test resolutions in doublings from $N = 16$ to $2048$ cells along the wave propagation direction, repeating this process for each wave, across all 18 scheme combinations.

\begin{figure}
    \centering
    \includegraphics[width=\linewidth]{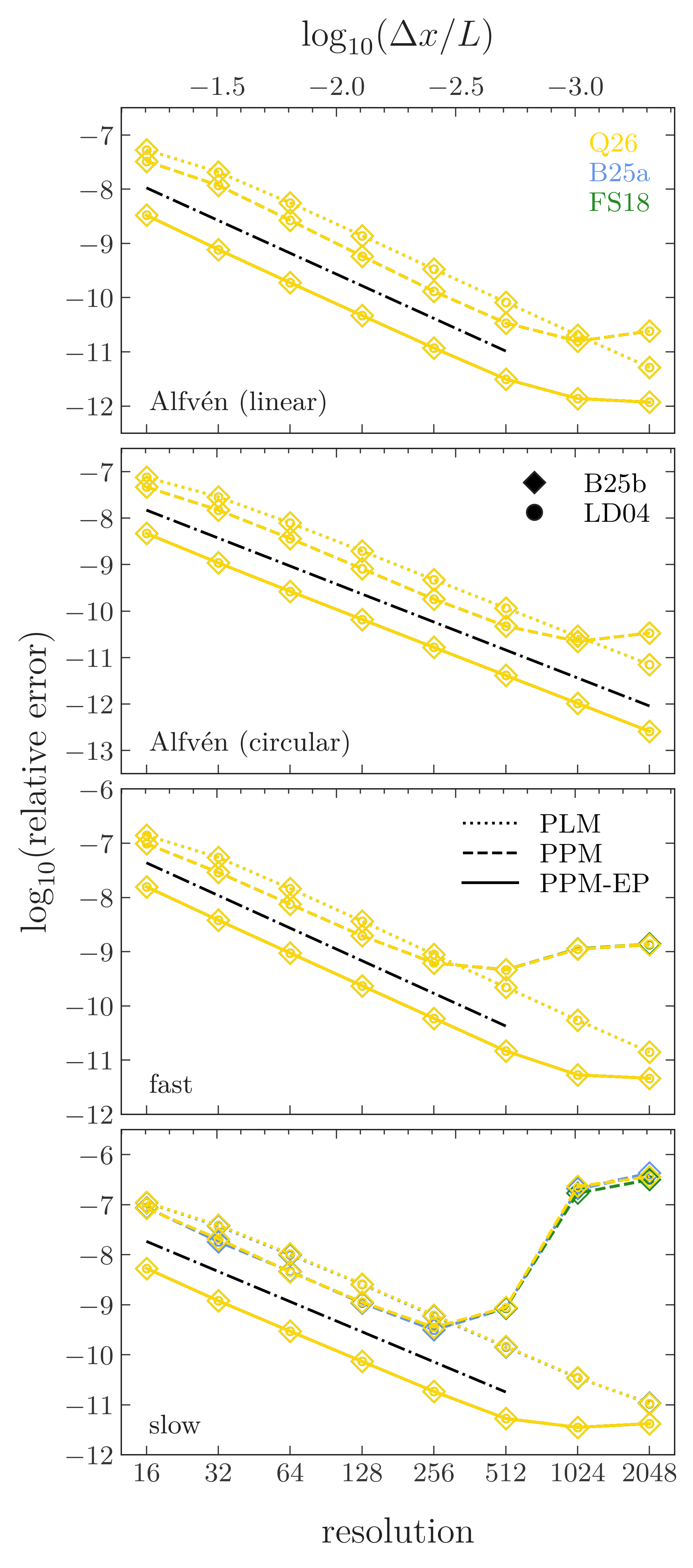}
    \caption{Convergence of all 18 scheme combinations: EMF compute (indicated by colours; \sref{sec:methods:emf:compute}), EMF averaging (indicated by markers; \sref{sec:methods:emf:averaging}), and reconstruction (indicated by line style; \sref{sec:methods:flux}), against four wave-propagation scenarios: linearly-polarised Alfv\'en (first panel; \sref{sec:scheme-comparison:waves:alfven-linear}), circularly-polarised Alfv\'en (second panel; \sref{sec:scheme-comparison:waves:alfven-circular}), fast (third panel; \sref{sec:scheme-comparison:waves:fast}), and slow (last panel; \sref{sec:scheme-comparison:waves:slow}) waves. The relative error (\cref{eqn:error:relative}) for each scheme combination is plotted against resolution (bottom axis), with the corresponding cell width shown on the top axis. A black dash-dotted line marks second-order convergence.}
    \label{fig:waves:convergence}
\end{figure}

For each simulation we measure the relative mean absolute error between the evolved and initial wave profiles, which we compute as follows: first we compute the mean absolute difference for each element $s$ of the state vector $\mVector{S}$ as
\begin{align}
    \mbox{error}_s
        = \frac{1}{N_s} \sum_{\forall i, j, k} \big|
            s_\mathrm{exact}\EvalAtH{i}{j}{k}
            - s_\mathrm{sim}\EvalAtH{i}{j}{k}
        \big|
    , \label{eqn:error:absolute}
\end{align}
where $s_\mathrm{sim}\EvalAtH{i}{j}{k}$ and $s_\mathrm{exact}\EvalAtH{i}{j}{k}$ are the simulated and analytically-exact values at the coordinate $\EvalAtH{i}{j}{k}$ where $s$ is stored (cell centre or face centre), respectively, $N_s$ is the number of cell centres or faces for that element, and the sums run over all centres or faces. We then similarly compute the normalisation for that element as
\begin{align}
    \mbox{norm}_s
        = \frac{1}{N_s} \sum_{\forall i, j, k} \big|
            s_\mathrm{exact}\EvalAtH{i}{j}{k}
        \big|
    , \label{eqn:error:normalisation}
\end{align}
and combine \cref{eqn:error:absolute} and \cref{eqn:error:normalisation} to compute the relative error across all elements of $\mVector{S}$ as
\begin{align}
    \mbox{relative error}
        = \rbrac{\frac{
                    \sum_{\forall s} \mbox{error}_s^2 N_s^2
                }{
                    \sum_{\forall s} \mbox{norm}_s^2 N_s^2
            }}^{1/2}
    . \label{eqn:error:relative}
\end{align}

\Cref{fig:waves:convergence} plots this relative error for all 18 scheme combinations as a function of resolution (lower x-axis) and cell width (upper x-axis), and broadly shows second-order convergence across all four waves. In some panels, curves for different scheme combinations overlap almost exactly, so not every colour remains visible. This is expected, since all three reconstruction schemes are second-order in space, or higher, while our time-stepping strategy is second-order in time (see \cref{fig:schematic:flowchart}), which restricts the overall convergence to second-order even for PPM and PPM-EP.

By $N = 2048$ we find that PPM-EP converges cleanly to the numerical-precision floor, achieving a similar relative error to PLM at roughly $4\times$ lower resolution. The more diffusive PLM converges equally cleanly, but has yet to reach the precision floor. PPM converges between them up to moderate resolution, reaching a minimum relative error of $10^{-11}$ to $10^{-9}$ depending on the wave. Beyond that it stops decreasing and instead grows, completely departing from the second-order trend they maintain.

This effect is mild for both linearly and circularly polarised Alfv\'en waves, slightly more apparent in the fast magnetosonic wave, and most pronounced in the slow magnetosonic wave. It is not a numerical instability, though, but instead an artefact of the Colella-Woodward monotonicity limiter \citep{Colella84a}, which cannot differentiate between a well-resolved, smooth extremum and a discontinuity, so it clips both alike; this is the same clipping-induced dissipation highlighted by \citet{Felker18a}.

The extremum-preserving version of PPM \citep[PPM-EP;][]{Rider07a} does make this distinction, which is why it avoids the artefact entirely. In any case, the artefact in plain PPM is specific to smooth, advection-dominated flows, where clipping the wave extrema is counterproductive. Real astrophysical flows are rarely as pristine as single-frequency waves: discontinuities and sharp gradients are exactly what the limiter exists to stabilise against. Indeed, of all the tests in this paper, this convergence study is the only one where it appears; see \sref{sec:scheme-comparison:shocks} and \sref{sec:scheme-comparison:orszag-tang} for more detailed comparison of plain PPM and PPM-EP.

\subsection{Resolving magnetised shocks}
\label{sec:scheme-comparison:shocks}

Next we compare our schemes against two 1D MHD shock tube problems, which together provide complementary validation of the Riemann solver at moderate resolutions. The \citet{RyuJones95a} test checks that all seven MHD wave families are resolved, and the \citet{Brio88a} test includes a compound wave that is absent from the Ryu-Jones test. In both tests we consider adiabatic flows in a domain spanning $x_0 \in [-0.5, 0.5]$, with an initial discontinuity at $x_\mathrm{shock} = 0$; since the resulting evolution is self-similar (the speed of every wave is constant in time), we are free to compare our schemes at any snapshot, as long as it is before the fastest wave reaches the domain boundary. We therefore choose the final snapshot in our simulations that meets this criterion.

We test all six combinations of EMF compute and averaging schemes, coupled with PPM-EP reconstruction. We also validate our LLF Riemann solver (which is used in our FOFC strategy; see \sref{sec:methods:time-stepping}) by substituting it in place of HLLD, coupled with the Q26 compute, \citetalias{Balsara25b} averaging, and PPM-EP reconstruction schemes. Finally, we check whether PPM degrades any of the wave families resolved in these tests, using the same EMF scheme combination (Q26 with \citetalias{Balsara25b}), this time coupled with PPM (and the default HLLD solver). We discuss the results of both tests, for all of these scheme combinations, together in \sref{sec:scheme-comparison:shocks:discussion}.

\subsubsection{The Ryu \& Jones (1995) shock tube}
\label{sec:scheme-comparison:shocks:ryu-jones}

For this test we set $\gamma=5/3$ and initialise the left half of the domain, $x_0 \leq x_\mathrm{shock}$, with the state:
\begin{align}
    \mVector{S}_\mathrm{p}^{[L]}
        = \begin{Bmatrix}
                \rho \\
                u_0 \\
                u_1 \\
                u_2 \\
                p \\
                b_0 \\
                b_1 \\
                b_2
            \end{Bmatrix}
        = \begin{Bmatrix}
                1.08 \\
                1.2 \\
                0.01 \\
                0.5 \\
                0.95 \\
                2/\sqrt{4\pi} \\
                3.6/\sqrt{4\pi} \\
                2/\sqrt{4\pi}
            \end{Bmatrix}
    ,
\end{align}
and the right half, $x_0 > x_\mathrm{shock}$, with the state:
\begin{align}
    \mVector{S}_\mathrm{p}^{[R]}
        = \begin{Bmatrix}
                \rho \\
                u_0 \\
                u_1 \\
                u_2 \\
                p \\
                b_0 \\
                b_1 \\
                b_2
            \end{Bmatrix}
        = \begin{Bmatrix}
                1 \\
                0 \\
                0 \\
                0 \\
                1 \\
                2/\sqrt{4\pi} \\
                4/\sqrt{4\pi} \\
                2/\sqrt{4\pi}
            \end{Bmatrix}
    .
\end{align}
We run this problem with $512$ cells until $t=0.2$; we use higher resolution than the Brio-Wu test below, since this test has far richer wave structures, and the differences between schemes only become visually obvious with more cells to resolve them. We then compare every scheme against the exact solution at each cell position, computed with the solver by \citet{Kriel26a}.

\subsubsection{The Brio \& Wu (1988) shock tube}
\label{sec:scheme-comparison:shocks:brio-wu}

For this test we set $\gamma=2$ and initialise the left half of the domain, $x_0 \leq x_\mathrm{shock}$, with the state:
\begin{align}
    \mVector{S}_\mathrm{p}^{[L]}
        = \begin{Bmatrix}
                \rho \\
                u_0 \\
                u_1 \\
                u_2 \\
                p \\
                b_0 \\
                b_1 \\
                b_2
            \end{Bmatrix}
        = \begin{Bmatrix}
                1 \\
                0 \\
                0 \\
                0 \\
                1 \\
                0.75 \\
                1 \\
                0
            \end{Bmatrix}
    ,
\end{align}
and the right half, $x_0 > x_\mathrm{shock}$, with the state:
\begin{align}
    \mVector{S}_\mathrm{p}^{[R]}
        = \begin{Bmatrix}
                \rho \\
                u_0 \\
                u_1 \\
                u_2 \\
                p \\
                b_0 \\
                b_1 \\
                b_2
            \end{Bmatrix}
        = \begin{Bmatrix}
                0.125 \\
                0 \\
                0 \\
                0 \\
                0.1 \\
                0.75 \\
                -1 \\
                0
            \end{Bmatrix}
    .
\end{align}
We run this problem with $256$ cells until $t = 0.1$. Unlike the Ryu-Jones test, the transverse magnetic field for this test is coplanar (\ie~$b_2 = 0$ on both sides of the initial discontinuity), which means the rotational discontinuities in the exact 7-wave solution are degenerate. A compound wave forms in their place: a slow shock attached to a slow rarefaction. Since we do not have an exact solution available, we instead use, as our reference solution, a simulation using Q26 with \citetalias{Balsara25b} and PPM-EP, run at a much higher resolution of $8192$ cells.

\subsubsection{Analysis of shock-tube results}
\label{sec:scheme-comparison:shocks:discussion}

\begin{figure*}
    \centering
    \includegraphics[width=0.85\linewidth]{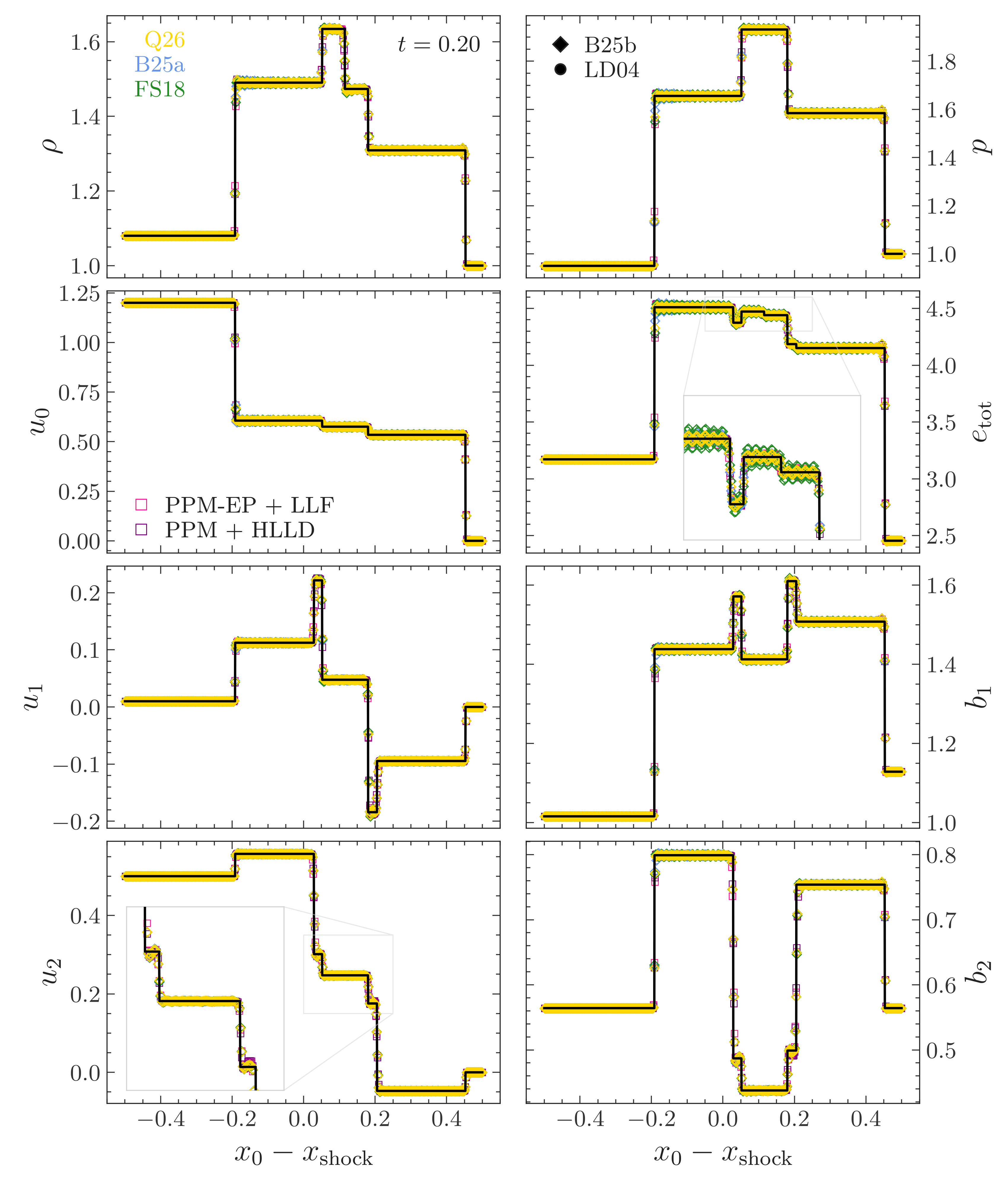}
    \caption{The \citet{RyuJones95a} shock tube test (\sref{sec:scheme-comparison:shocks:ryu-jones}) compared across 6 EMF scheme combinations run at $N = 512$ with PPM-EP reconstruction and our HLLD Riemann solver: EMF compute (indicated by colour; \sref{sec:methods:emf:compute}) and EMF averaging (indicated by marker; \sref{sec:methods:emf:averaging}). We also consider two more scheme combinations, both paired with Q26 compute and \citetalias{Balsara25b} averaging: our LLF Riemann solver with PPM-EP reconstruction (pink squares), and HLLD with plain PPM reconstruction (purple squares). Each panel shows the profile of key quantities at $t = 0.20$ along $x_0$, with exact Riemann solutions (overlayed black line) computed with the solver by \citet{Kriel26a}.}
    \label{fig:ryu-jones-profiles}
\end{figure*}

\begin{figure*}
    \centering
    \includegraphics[width=0.85\linewidth]{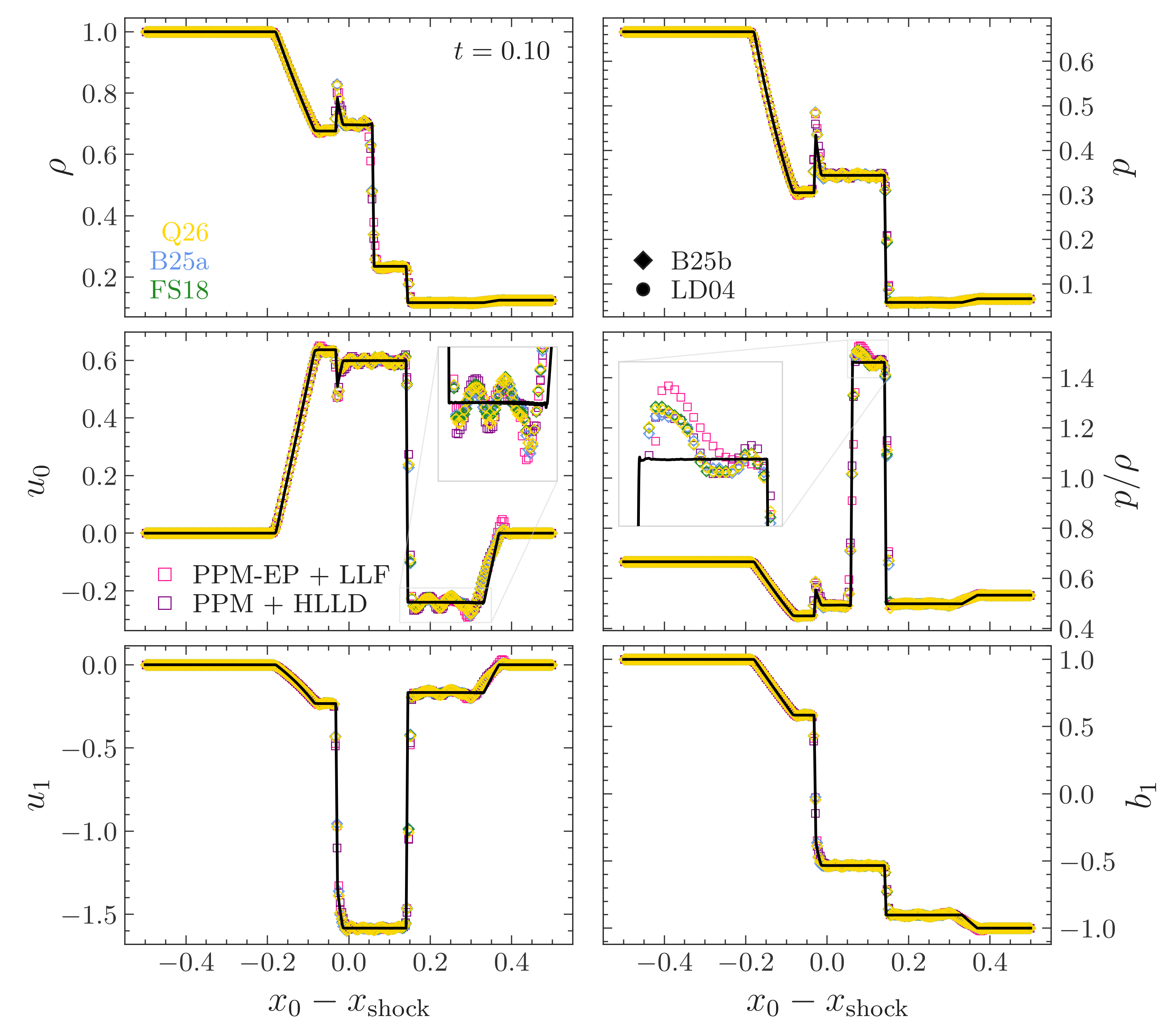}
    \caption{The \citet{Brio88a} shock tube test (\sref{sec:scheme-comparison:shocks:brio-wu}) run at $N = 256$, across the same scheme combinations as \cref{fig:ryu-jones-profiles}, showing profiles of key quantities at $t = 0.1$. A reference solution (black line) is found by running the problem with Q26 compute, \citetalias{Balsara25b} averaging, and PPM-EP reconstruction at $N = 8192$.}
    \label{fig:brio-wu-profiles}
\end{figure*}

As in \sref{sec:scheme-comparison:waves}, every scheme combination we test resolves the same wave structures in both shock tubes (\cref{fig:ryu-jones-profiles,fig:brio-wu-profiles}), closely matching the reference solution of either test. Between the combinations themselves, though, the Riemann solver drives the largest differences: LLF (pink squares) does not resolve the contact discontinuity as well as HLLD (all other points), and as a result shows more ringing in the Brio-Wu $p/\rho$ profile (a quantity proportional to plasma temperature). The EMF compute schemes differ less, with Q26 showing the least ringing of the three, which is most visible in the $e_\mathrm{tot}$ profile of the Ryu-Jones test; the reconstruction order matters least of all: PPM and PPM-EP agree closely across all quantities in both tests, apart from small ringing-related differences in the $u_0$ profile of the Brio-Wu test.

\subsection{Resolving turbulent structures}
\label{sec:scheme-comparison:orszag-tang}

\subsubsection{The Orszag-Tang vortex}
\label{sec:scheme-comparison:orszag-tang:setup}

We next test each of our scheme combinations on the Orszag-Tang vortex \citep{OrszagTang1979}, a 2D non-linear MHD problem, which we initialise with the state
\begin{align}
    \rho(\mVector{x})
        &= \frac{25}{36\pi} \\
    \mVector{u}(\mVector{x})
        &= \sin\sbrac{2\pi x_1}\, \mVectorUnit{x}_0
        - \sin\sbrac{2\pi x_0}\, \mVectorUnit{x}_1 \\
    p(\mVector{x})
        &= \frac{5}{12\pi} \\
    \mVector{b}(\mVector{x})
        &= \frac{1}{\sqrt{4\pi}} \sin\sbrac{2\pi x_1}\, \mVectorUnit{x}_0
        + \frac{1}{\sqrt{4\pi}} \sin\sbrac{4\pi x_0}\, \mVectorUnit{x}_1
    .
\end{align}
This test is widely used to exercise all aspects of an MHD solver, since the smooth initial state evolves to form propagating shocks and thin current sheets along the vortex arms, all surrounded by a flow that remains smooth. The amount of structure resolved in these sheets and shocks is set by the dissipation, so we use this test to compare the numerical dissipation of each scheme combination, simulating this flow with $\gamma = 5/3$ on a periodic domain $x_0, x_1 \in [-0.5, 0.5]$, using $1024^2$ cells and $\mathrm{CFL} = 0.2$. We also compare PPM and PPM-EP reconstruction at $4096^2$ cells, run with Q26 compute and \citetalias{Balsara25b} averaging, to confirm that our conclusions about clipping hold at higher resolution.

The first series of shocks collides at around $t \approx 0.5$, with mixing and shock generation peaking around $t \approx 0.85$; we therefore run every simulation to $t = 1$, to confirm that it completes without failure, and then compare all scheme combinations at $t = 0.85$.

\subsubsection{Analysis of resolved structures}
\label{sec:scheme-comparison:orszag-tang:discussion}

\begin{figure*}
    \centering
    \includegraphics[width=0.95\linewidth]{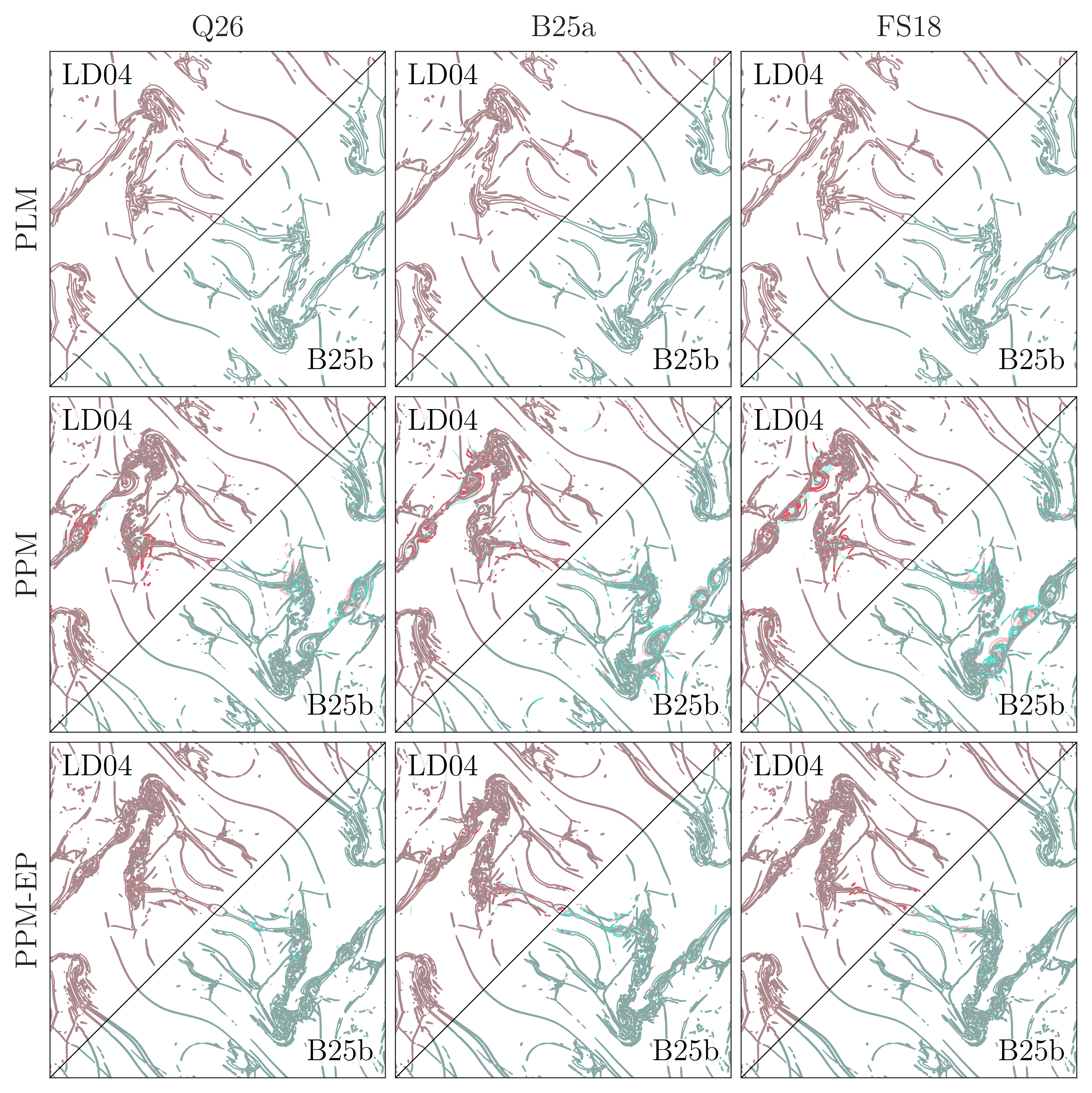}
    \caption{Comparison of all 18 scheme combinations for the Orszag-Tang vortex (\sref{sec:scheme-comparison:orszag-tang:setup}) run at $1024^2$: reconstruction varies across rows (\sref{sec:methods:flux}), EMF compute across columns (\sref{sec:methods:emf:compute}), and both EMF averaging schemes (\sref{sec:methods:emf:averaging}) are compared in each panel. In each panel we compare snapshots at $t = 0.85$ of \citetalias{Londrillo04a} (red) in the top-left corner and \citetalias{Balsara25b} (blue) in the bottom-right corner, with contours tracing $\log_{10}\rbrac{\Delta x\, |\nabla\times\mVector{b}|} = -1.6$.}
    \label{fig:orszag-tang:emf-schemes}
\end{figure*}

\begin{figure}
    \centering
    \includegraphics[width=0.95\linewidth]{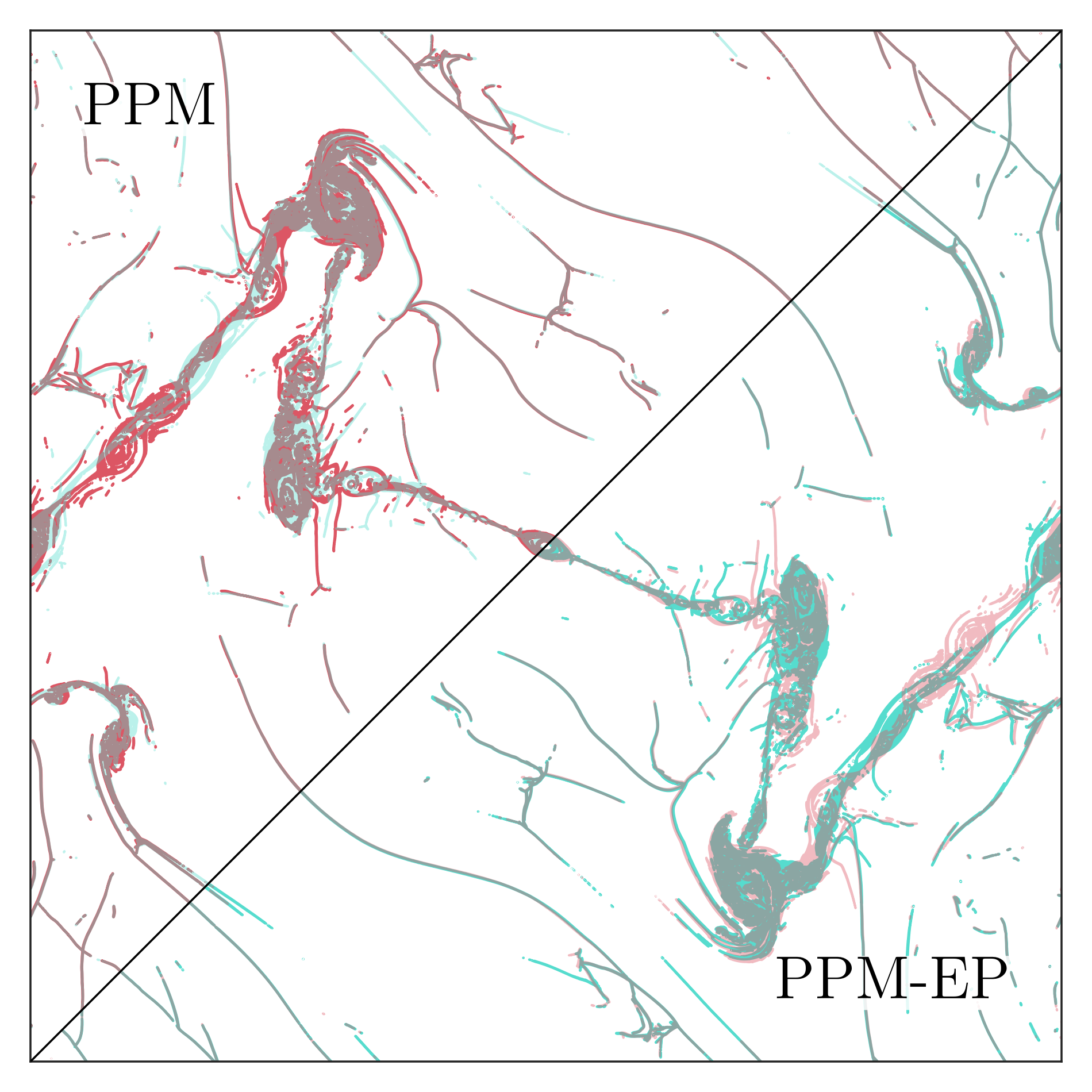}
    \caption{Comparison of PPM and PPM-EP reconstruction, using our recommended EMF combination (Q26 compute and \citetalias{Balsara25b} averaging), for the Orszag-Tang vortex run at $4096^2$, following the same format as \cref{fig:orszag-tang:emf-schemes}.}
    \label{fig:orszag-tang:reconstruction-schemes}
\end{figure}

In this section we exploit the fact that our implementation of all our CT-MHD schemes preserves point-reflection symmetry to bit-level precision.\footnote{
    Achieving this across every Riemann solver, EMF compute scheme, EMF averaging scheme, and reconstruction order required us to take considerable care with how floating-point operations were grouped throughout the solver. We found the symmetry-preserving best-practice guidelines of \citet{Fleischmann19a} especially useful in doing so.
} Thus all differences between schemes we identify reflect differences in how each scheme combination resolves the flow, and not an artefact of numerical symmetry breaking amplified by the non-linear evolution. With this in mind, we compare two schemes by overlaying their resolved current-density structure at $t = 0.85$. We do this by drawing the contour $\log_{10}\rbrac{\Delta x\, |\nabla\times\mVector{b}|} = -1.6$ for both schemes on a panel, opaque in one half of the domain (split along the off-diagonal) and faint in the other; differences between the two schemes are then highlighted by regions where both colours are visible. That is, if two schemes produced exactly identical results we would see only one colour on each side of the diagonal, so any deviation from this is evidence of differences between the schemes.

\Cref{fig:orszag-tang:emf-schemes} uses this plotting technique to compare all 18 scheme combinations run at $1024^2$, arranged in a $3\times 3$ grid, with EMF compute scheme varying across columns, reconstruction order across rows, and EMF averaging scheme within each panel. Across all these comparisons, the schemes broadly resolve the same underlying structures, with the clearest differences arising from reconstruction order: PLM resolves substantially less turbulent structure than either PPM or PPM-EP, regardless of the EMF scheme combination. This mirrors the wave-convergence results (\sref{sec:scheme-comparison:waves}), where PLM is the most diffusive and requires roughly $4\times$ higher resolution to achieve the same accuracy as PPM-EP.

In contrast to \sref{sec:scheme-comparison:waves}, the differences between PPM and PPM-EP are much smaller: both reconstruction schemes resolve the same level of structure across every EMF scheme combination. But, at $1024^2$ it is important to note that the flow remains in a dissipation-regulated regime, and so numerical differences between the schemes are partly suppressed at this resolution. This is because the effective Lundquist number, $c_\mathrm{A}L/\eta$ (where $\eta$ is the effective resistivity set by the combined resolution and EMF scheme), is too small for current sheets to reliably fragment into plasmoid chains, though PPM and PPM-EP already begin to resolve isolated plasmoids along the vortex arms at this resolution, whereas the stronger diffusion of PLM keeps its sheets laminar throughout \citep{Landi15a}. We therefore repeat this comparison with $4096^2$ cells in \cref{fig:orszag-tang:reconstruction-schemes}, where, in both solutions, the lower effective numerical dissipation allows thin current sheets along the vortex arms to become tearing unstable and fragment into plasmoids. These plasmoids are advected along the current sheets, with some coalescing into larger magnetic islands; the precise locations of these islands at $t = 0.85$ differ between the two solutions.

These differences in placement, however, do not provide a measure of relative accuracy: as \citet{Lecoanet16a} demonstrate for the Kelvin-Helmholtz instability, without explicit dissipation, this class of problem (where secondary instabilities are present) has no unique solution as the resolution or numerical scheme changes. The relevant comparison is therefore not the placement of individual plasmoids, but instead whether both schemes recover the same tearing-mediated dynamics, and since both do, we conclude that PPM performs as well as PPM-EP, despite its clipping-induced dissipation.

Finally, among the three EMF compute schemes tested here, the structures resolved in \cref{fig:orszag-tang:emf-schemes} remain consistent. EMF averaging scheme makes little difference with PLM or PPM-EP reconstruction, but plays a larger role with plain PPM in resolving the larger plasmoids that begin forming along the vortex arms at this resolution. Even so, all six EMF scheme combinations coupled with both PPM and PPM-EP reconstruction produce comparable results in this demanding non-linear problem, with every run successfully reaching $t = 1$. We therefore turn to numerical stability, and then computational performance, to distinguish between these schemes.

\subsection{Stability under tearing-mediated reconnection}
\label{sec:scheme-comparison:reconnection}

Unlike the previous tests that compared the accuracy and level of resolved structure between different scheme combinations, here we instead test their numerical stability using the \citet{HawleyStone95a} and \citet{Stone08a} current sheet problem.\footnote{
    We show the full time evolution of this problem in \sref{sec:stress-tests:current-sheet}.
} This test features a strong magnetic field reversal that, when evolved for a long time in low-dissipation flows, grows tearing unstable and produces plasmoids that merge into increasingly larger ones. Each merger leaves behind sharp, transient gradients, which are numerically challenging for any scheme: the time step must become small enough to resolve each merger, and then recover between mergers so the simulation can make progress, while the flux and EMF updates need to resolve those gradients without producing a physically invalid state.

We set this test up with $\gamma = 5/3$ on a periodic domain $x_0, x_1 \in [-0.5, 0.5]$, with the background state,
\begin{align}
    \mVector{S}^{[\mathrm{bg}]}
        = \begin{Bmatrix}
                \rho \\
                \mVector{u} \\
                \mVector{b} \\
                e_\mathrm{tot}
            \end{Bmatrix}
        = \begin{Bmatrix}
                1 \\
                \mVector{0} \\
                b_\mathrm{bg} \, \mathrm{sgn}\sbrac{|x_0| - 0.25} \, \mVectorUnit{x}_1 \\
                \dfrac{p_\mathrm{bg}}{\gamma-1} + \dfrac{b_\mathrm{bg}^2}{2}
            \end{Bmatrix}
    ,
\end{align}
where $b_\mathrm{bg} = 1$ and $p_\mathrm{bg} = 0.05$; these states are perturbed with,
\begin{align}
    \mVector{S}^{[\delta]}
        = \begin{Bmatrix}
                \delta\rho \\
                \delta \mVector{u} \\
                \delta \mVector{b} \\
                \delta e_\mathrm{tot}
            \end{Bmatrix}
        = \begin{Bmatrix}
                0 \\
                \delta u \sin\sbrac{2\pi x_1} \, \mVectorUnit{x}_0 \\
                \mVector{0} \\
                \dfrac{1}{2} (\delta u)^2 \sin^2\sbrac{2\pi x_1}
            \end{Bmatrix},
\end{align}
where $\delta u = 0.1$. We run this with $1024^2$ cells and $\mathrm{CFL} = 0.2$, through to $t = 10$, for all six EMF compute and averaging scheme combinations, coupled with both PPM and PPM-EP reconstruction. We run without explicit resistivity, so reconnection is entirely mediated by numerical dissipation, which at $1024^2$ is comparable to that in the previous Orszag-Tang runs in \cref{fig:orszag-tang:emf-schemes}. Unlike the Orszag-Tang test, though, current sheets here span the full domain, so the effective Lundquist number is larger, and tearing is already excited at $1024^2$, whereas there it requires $4096^2$ (see \cref{fig:orszag-tang:reconstruction-schemes}).

We find that all scheme combinations with plain PPM reconstruction resolve the plasmoid mergers stably all the way through to $t = 10$. However, only two scheme combinations with PPM-EP reconstruction complete successfully: Q26 compute with \citetalias{Balsara25b} averaging, and \citetalias{Balsara25a} compute with \citetalias{Balsara25b} averaging; every other scheme combination becomes numerically unstable during the early period $t \lesssim 1$ when plasmoids are still growing quickly. For PPM-EP, then, the EMF pairing strongly affects stability in this reconnection-dominated regime, whereas plain PPM, with its more aggressive limiter, remains stable regardless.

\subsection{Estimating numerical dissipation}
\label{sec:scheme-comparison:dissipation}

\begin{figure}
    \centering
    \includegraphics[width=\linewidth]{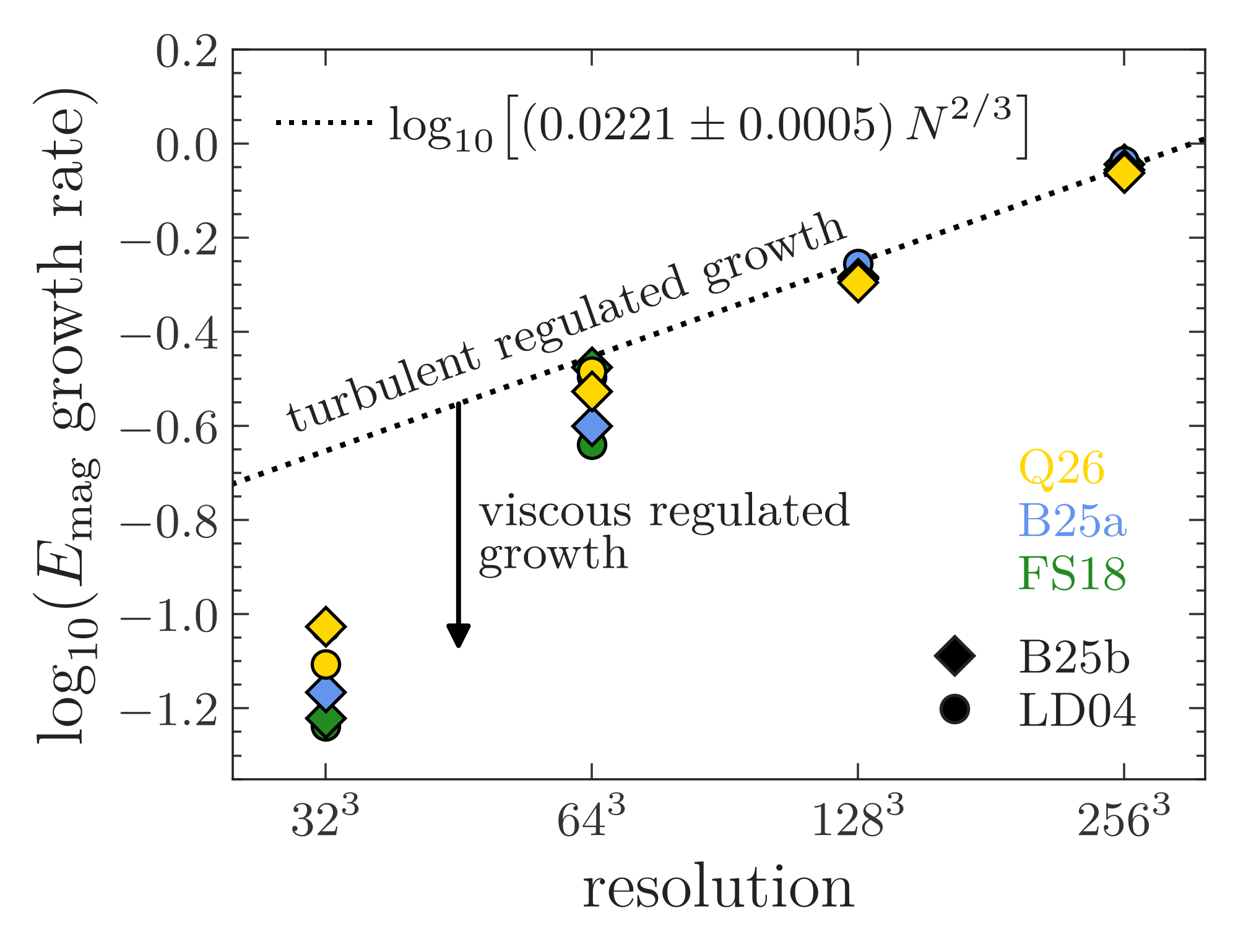}
    \caption{Growth rate of magnetic energy, $E_\mathrm{mag}$, measured during the kinematic phase of small-scale dynamo (SSD) simulations (\sref{sec:scheme-comparison:dissipation}). We run all six combinations of EMF compute (\citetalias{Felker18a}, \citetalias{Balsara25a}, Q26) and averaging (\citetalias{Londrillo04a}, \citetalias{Balsara25b}) schemes, each paired with PPM-EP reconstruction, across a range of resolutions $N$. The dotted line shows the expected $N^{2/3}$ scaling of incompressible turbulent SSD growth regulated by numerical dissipation; this reference line is anchored to the average growth rate measured at $N = 256$.}
    \label{fig:ssd-growth-rate}
\end{figure}

\begin{figure*}
    \centering
    \includegraphics[width=0.95\linewidth]{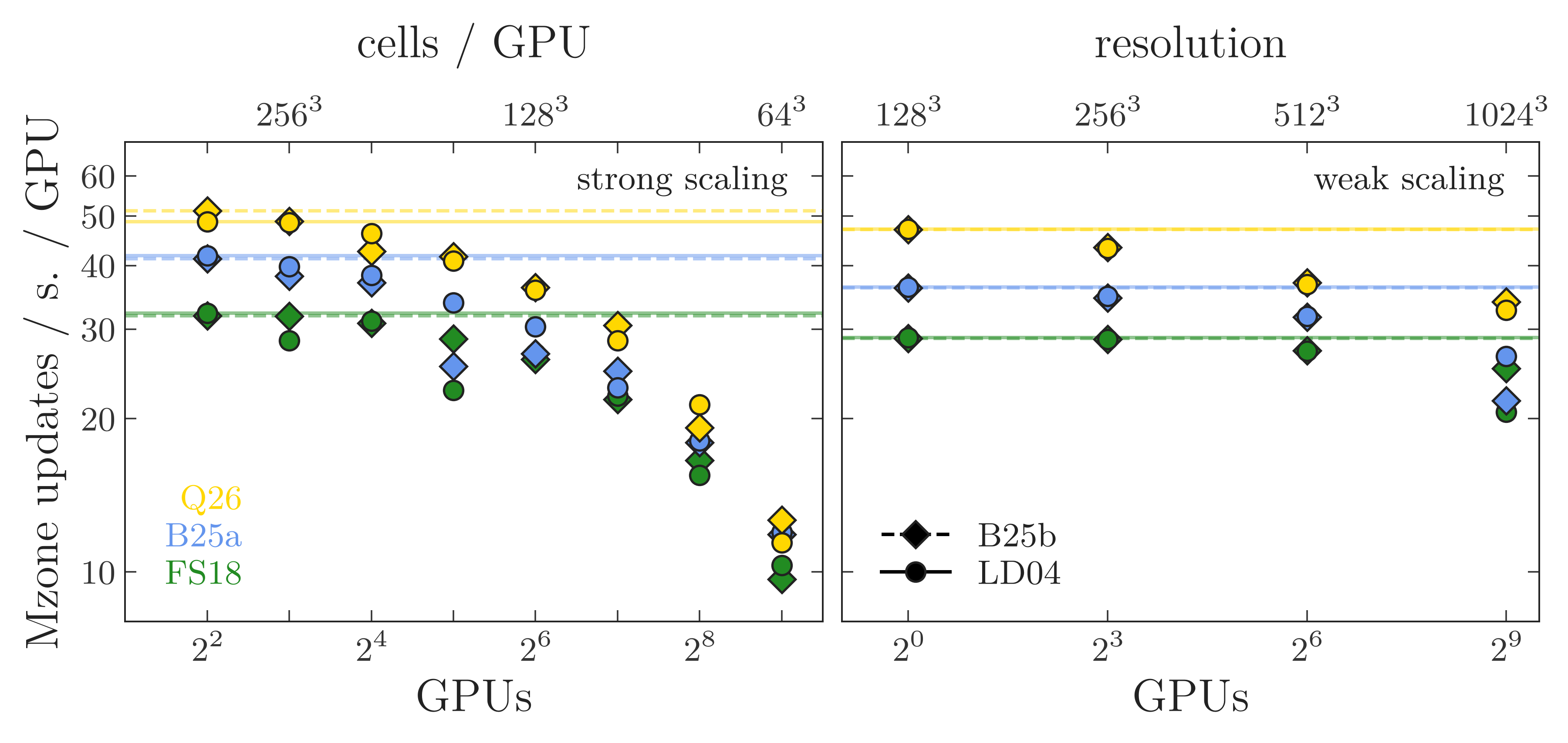}
    \caption{Comparison of 6 EMF scheme combinations for GPU throughput and scaling (\sref{sec:scheme-comparison:speed}), all paired with PPM-EP reconstruction: EMF compute (indicated by colour; \sref{sec:methods:emf:compute}) and EMF averaging (indicated by marker; \sref{sec:methods:emf:averaging}). Cell updates per second per GPU are shown for strong scaling at a fixed problem size of $512^3$ cells (left panel) and weak scaling at $128^3$ cells per GPU (right panel); lines indicate perfect scaling for each scheme combination.}
    \label{fig:gpu-scaling}
\end{figure*}

As a final comparison of how schemes resolve turbulent flow, we simulate the small-scale dynamo (SSD) and measure the growth rate of magnetic energy, $E_\mathrm{mag}$, during the kinematic phase, since it is directly regulated by dissipation and also straightforward to measure. We run without either explicit viscosity or resistivity (ideal MHD), so this growth is regulated entirely by numerical dissipation, thus allowing us to use it to quantify the relative dissipation introduced by each scheme. We restrict our attention to the incompressible flow regime, starting from a flow at rest with an ABC-type, maximally helical seed magnetic field \citep{Arnold65a, Childress70a} that is $10$ orders of magnitude weaker than the steady-state kinetic energy. We drive this flow on $\ell_0 = L / 2$ to a $\mathcal{M} \approx 0.5$ state via a purely solenoidal Ornstein-Uhlenbeck forcing field \citep{Federrath2022_TurbGen} on a periodic, unit-length domain. We run all six EMF compute and averaging scheme combinations, paired with PPM-EP reconstruction, at four resolutions from $N = 32$ to $256$.

For incompressible turbulence, the growth rate of $E_\mathrm{mag}$ is expected to scale as $\mathrm{Re}^{1/2}$ \citep[e.g.,][]{kriel2026universal}, since growth is driven by stretching at the viscous scale, $\ell_\nu$. Assuming the dissipation scale sits at the grid scale, $\ell_\nu \sim \ell_0 / N$, the Kolmogorov relation $\ell_\nu / \ell_0 \sim \mathrm{Re}^{-3/4}$ \citep[\eg][]{schekochihin2004simulations, kriel2022fundamental} gives $\mathrm{Re} \sim N^{4/3}$, and hence the growth rate is expected to scale as $N^{2/3}$. In \cref{fig:ssd-growth-rate} we compare this expectation with measurements from our simulations at different resolutions.

We find that all six EMF scheme combinations converge to the expected scaling by $N \geq 128$, but deviate from it, and from each other, at lower resolutions. At $N = 32$ the energy injection scale and the grid (dissipation) scale are barely a decade apart, and thus energy is dissipated before any cascade can form. At this resolution every EMF scheme combination falls below the expected turbulent growth rate. At $N = 64$ we find that the scale separation $\ell_\nu / \ell_0$ resolved by some schemes, grows sufficiently large for the flow to transition into turbulence-regulated growth; Q26 with \citetalias{Londrillo04a} and \citetalias{Felker18a} with \citetalias{Balsara25b} appear to have already resolved this growth, while the remaining combinations remain suppressed by numerical dissipation. This viscous dissipation-dominated flow regime (not the turbulent flow regime resolved by higher resolutions) most directly exposes the impact of these EMF schemes, since the dissipation footprint of each scheme is roughly the same at different resolutions, but the footprint makes up a larger fraction of the resolved $\ell_\nu$ at low $N$; this becomes less significant as $\ell_0 / \ell_\nu$ grows with $N$.

At $N = 32$, every scheme combination is in this viscous-dominated regime, but Q26 achieves the fastest growth rate (and thus resolves the smallest effective $\ell_\nu$) of the three compute schemes, closely followed by B25a, and then FS18. This ordering largely holds at $N = 64$, even as some scheme combinations begin to resolve turbulence-regulated growth. Averaging scheme has some impact as well, though inconsistently across compute schemes and between $N = 32$ and $64$. At $N \geq 128$, all schemes resolve amplification consistent with incompressible turbulent flows, and differences between schemes at these resolutions are small enough to be consistent with stochastic variation in the turbulence, though a single simulation realization does not let us quantify this directly.

\needspace{5\baselineskip}
\subsection{Performance and scaling}
\label{sec:scheme-comparison:speed}

Next we measure the relative throughput and scaling of all six combinations of the EMF compute and averaging schemes, each paired with PPM-EP. To do so, we first use strong scaling to determine how the throughput of each scheme depends on the GPU decomposition. We then use weak scaling, at a common local problem size, to test how well this throughput is maintained for larger problems.

Since the choice of problem for strong and weak scaling is arbitrary, we use the linearly-polarised Alfv\'en wave defined in \sref{sec:scheme-comparison:waves:alfven-linear}, with $\mVector{k} L / 2\pi = \{1, 2, 3\}$ and $\theta = \pi/4$. We evolve each simulation for 100 time steps on the Setonix machine at the Pawsey Supercomputing Centre in Western Australia\footnote{
    Each compute node carries four AMD Instinct MI250X GPU cards (two GPUs per card), with nodes linked by HPE Slingshot interconnects (allowing direct GPU-to-GPU communication).
}, excluding startup and I/O overheads from the measured time. For strong scaling we hold the problem size fixed at $512^3$ cells and increase the GPU count from $4$ to $512$ in doublings, while for weak scaling we use a common local problem size of $128^3$ cells per GPU, increasing $N_\mathrm{GPU}$ through $\{1, 8, 64, 512\}$.

\Cref{fig:gpu-scaling} shows that throughput is primarily determined by the EMF compute scheme: the choice of averaging scheme has very little effect, since the averaging step does not change the memory footprint or the number of GPU kernel launches. Among the compute schemes, Q26 is consistently the fastest, reaching a throughput of $>50$ million cell updates per second per GPU at $N_\mathrm{GPU} = 4$, which is roughly $25\%$ higher than \citetalias{Balsara25a} and $60\%$ higher than \citetalias{Felker18a}. This speed-up comes from Q26 requiring fewer reconstruction steps and temporary variables, but is partially offset by needing to solve the Riemann problem over a wider shell of ghost zones to evaluate the face velocity in \cref{eqn:emf:q26-facevel} (see \sref{sec:methods:emf:compute} for details). As the subdomain on each GPU becomes smaller, the fraction of it that is made up of ghost zones increases, so the throughput advantage narrows at the largest GPU counts; Q26 remains the fastest throughout our testing, though. Moreover, despite the differences in absolute throughput, most scheme combinations achieve better than $70\%$ weak-scaling efficiency from $1$ to $512$ GPUs, and better than $70\%$ strong-scaling efficiency when each GPU is assigned at least $128^3$ cells.

\subsection{Summary and recommendations}
\label{sec:scheme-comparison:recommendation}

Across all tests, our new Q26 compute scheme (\sref{sec:methods:emf:compute}) produces results in both linear and non-linear problems that are comparable to those of the existing \citetalias{Felker18a} and \citetalias{Balsara25a} compute schemes, while achieving substantially higher throughput (\sref{sec:scheme-comparison:speed}); it also introduces the least numerical dissipation of the three compute schemes at low resolution (\sref{sec:scheme-comparison:dissipation}). Moreover, \citetalias{Balsara25b} averaging proved the most stable in reconnection-dominated flows (\sref{sec:scheme-comparison:reconnection}). We therefore adopt Q26 compute with \citetalias{Balsara25b} averaging as the default EMF scheme combination in \quokka.

As for reconstruction order, we broadly find that both PPM and PPM-EP resolve structures in non-linear flows comparably well (\cref{fig:orszag-tang:reconstruction-schemes}); however, the extrema-preserving limiter in PPM-EP avoids the clipping-induced degradation we found in plain PPM for some smooth waves (\sref{sec:scheme-comparison:waves}). This accuracy can come with a stability cost, though: we find that PPM-EP remains stable with only two of the six EMF scheme combinations in reconnection-dominated flows, compared to all six with plain PPM (\sref{sec:scheme-comparison:reconnection}). Since our recommended EMF pairing (Q26 compute with \citetalias{Balsara25b} averaging) is one of these two, we recommend PPM-EP as the default reconstruction scheme when it is coupled with this pairing (or with \citetalias{Balsara25a} compute and \citetalias{Balsara25b} averaging). If a user instead adopts a different EMF combination, plain PPM is the more stable choice, since its stability does not depend on which EMF combination it is paired with.

\section{Validating our recommended scheme}
\label{sec:stress-tests}

Following the systematic comparison of all three EMF compute schemes (\sref{sec:methods:emf:compute}), both EMF averaging schemes (\sref{sec:methods:emf:averaging}), and three reconstruction schemes (PLM, PPM, PPM-EP) presented in \sref{sec:scheme-comparisons}, we arrive at a recommended combination: Q26 compute with \citetalias{Balsara25b} averaging and PPM-EP reconstruction. Because the comparison that supported this conclusion spanned a large number of scheme combinations, it necessarily relied on a reduced subset of tests, each chosen to probe differences between schemes rather than to exhaustively validate correctness; this section presents the remainder of our validation suite, which we run against our recommended scheme combination.

\subsection{Directional bias in grid-misaligned waves}
\label{sec:stress-tests:misaligned-waves}

To verify that our solver does not introduce directional bias when waves propagate misaligned, rather than along the grid, we simulate the same slow magnetosonic wave introduced in \sref{sec:scheme-comparison:waves}. We choose this wave because its perturbations to $\mVector{u}$ and $\mVector{b}$ are coupled and compressive, which makes it the most stringent test of directional bias among the four waves we test. To probe this, we run it propagating obliquely at two resolutions, checking whether any directional-bias artefact increases with resolution.

We choose $\mVector{k}L/2\pi = \{1, 2, 3\}$ and $\theta \equiv \cos^{-1}\sbrac{\mVectorUnit{n}_\mathrm{bg}\cdot\mVectorUnit{n}_k} = 45^\circ$, so that neither $\mVectorUnit{n}_k$ nor $\mVectorUnit{n}_\mathrm{bg}$ is aligned with any axis of the grid, and simulate until $t = 4\pi/\omega$ (\ie~two wave periods $T = 2\pi/\omega$) in domains of $128^3$ and $512^3$ cells. All other parameters are identical to those used in \sref{sec:scheme-comparison:waves}. For this configuration, $\mVectorUnit{n}_\perp = \{-2,1,0\}/\sqrt{5}$, which is orthogonal to both $\mVectorUnit{n}_k$ and the $x_2$-axis, so the perturbation has no $x_2$ component by construction. With this setup, the magnetic perturbation lies entirely along $\mVectorUnit{n}_\perp$, so $b_0$ carries most of the field signal, and $b_2$ should remain unperturbed from its background value. In practice, however, this is not guaranteed numerically, so we use growth in $b_2$ as a metric for directional bias, which is seeded by reconstructing the wave along grid axes misaligned with its propagation direction. What matters is not the absolute growth of $b_2$, but its growth relative to the real wave signal, carried by $b_0$.

\begin{figure}
    \centering
    \includegraphics[width=\linewidth]{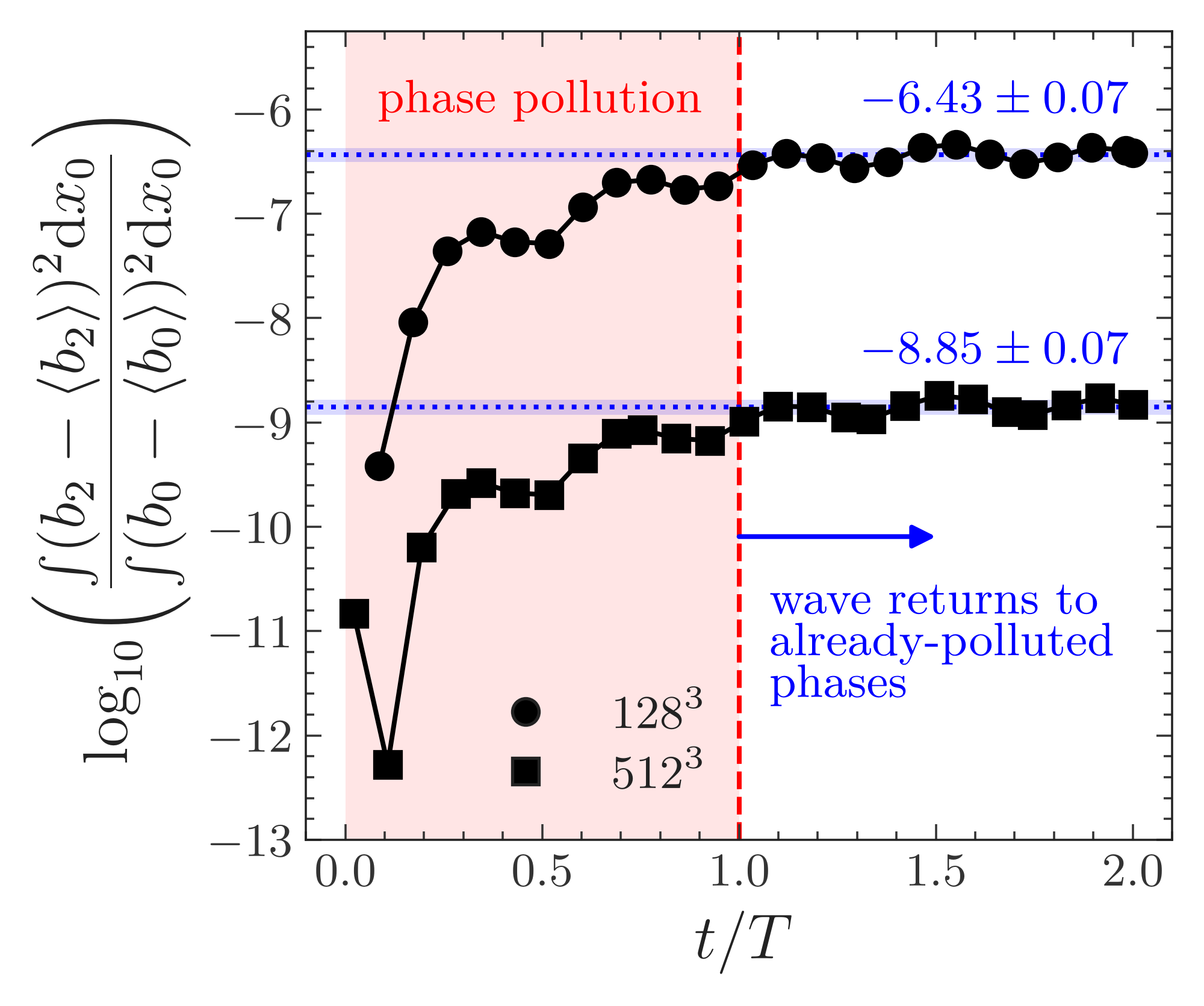}
    \caption{The time-evolving, line-integrated energy ratio of the out-of-plane component, $b_2$, to the component that primarily carries the signal, $b_0$, for the oblique slow magnetosonic wave. Both runs use our recommended scheme (Q26 compute, \citetalias{Balsara25b} averaging, and PPM-EP reconstruction), at $128^3$ (circles) and $512^3$ (squares). We report the statistically saturated ratio computed over the second wave period, $t/T \geq 1$, where $T$ is the wave period.}
    \label{fig:oblique-slow-wave}
\end{figure}

\Cref{fig:oblique-slow-wave} shows the evolution of $b_2$ compared to $b_0$ over the two wave periods, at both resolutions. Since $b_0$ is not constant in space, we do not compare the two fields directly, but rather the energy in their fluctuations (see the y-axis label). Moreover, because the fluctuation energy of $b_0$ is conserved to within $0.3\%$ at both resolutions, changes in the $b_2 / b_0$ ratio must directly follow from changes in $b_2$. With this in mind, we find that at both resolutions the ratio $b_2 / b_0$ grows over the first wave period ($t/T < 1$) as the wave sweeps over cells it has not yet sampled, polluting each with a small numerical deviation from the background state. The ratio saturates once the wave returns to those already-polluted phases at $t/T > 1$. Importantly, this saturated ratio reduces in magnitude as the resolution increases, and moreover remains many orders of magnitude below the wave signal. We therefore conclude that directional bias remains dynamically negligible in the solver.

\subsection{Stability of odd-even decoupling}
\label{sec:stress-tests:quirk}

Next we test a complementary regime to the previous test, where rather than a wave misaligned with the grid, we now consider a shock deliberately aligned with it. This is because a low-dissipation Riemann solver applied dimension-by-dimension (as our HLLD solver is) dissipates perturbations along a shock front less than those across it. When this is left unaddressed, even small perturbations at shock fronts can grow into corrugated, non-planar instabilities \citep{Quirk94a, Minoshima21a}. Fortunately, a solution exists: a shock-anisotropy correction \citep{Minoshima21a} (\sref{sec:methods:flux}), which we have extended our HLLD solver to include. Here we verify, using the \citet{Quirk94a} test problem, that our Riemann solver with this correction avoids forming this \emph{carbuncle} instability.

The test we use is a strong, planar shock initialised at $x_0 = 0.4$ within a domain $x_0, x_1 \in [0,1]$, with $\gamma = 5/3$, and run with $128^2$ cells. We use outflowing (Dirichlet) boundary conditions in $x_0$ and periodic boundary conditions in $x_1$ (and $x_2$; see \cref{note:quokka:3d}). We initialise the post-shock region, $x_0 \leq 0.4$, as:
\begin{align}
    \mVector{S}_\mathrm{p}^{[L; \mathrm{bg}]}
        = \begin{Bmatrix}
                \rho \\
                u_0 \\
                p
            \end{Bmatrix}
        = \begin{Bmatrix}
                3.692 \\
                -0.625 \\
                26.85
            \end{Bmatrix}
    ,
\end{align}
and the pre-shock region, $x_0 > 0.4$, as:
\begin{align}
    \mVector{S}_\mathrm{p}^{[R]}
        = \begin{Bmatrix}
                \rho \\
                u_0 \\
                p
            \end{Bmatrix}
        = \begin{Bmatrix}
                1 \\
                -5 \\
                0.6
            \end{Bmatrix}
    ,
\end{align}
with $\mVector{b} = \mVector{0}$ and $u_1 = u_2 = 0$; this corresponds with a pre-shock sonic Mach number $\mathcal{M} = 5$. We perturb the post-shock state at $x_0 = 0.4$, for every cell with an even $x_1$-index, as:
\begin{align}
    \mVector{S}_\mathrm{p}^{[L; \delta]}
        = \begin{Bmatrix}
                \delta\rho \\
                \delta u_0 \\
                \delta p
            \end{Bmatrix}
        = \begin{Bmatrix}
                -0.135 \\
                0.219 \\
                -1.31
            \end{Bmatrix}
    ,
\end{align}
where cells with an odd $x_1$-index remain unperturbed. The complete post-shock state is then $\mVector{S}_\mathrm{p}^{[L]} = \mVector{S}_\mathrm{p}^{[L; \mathrm{bg}]} + \mVector{S}_\mathrm{p}^{[L; \delta]}$.

\begin{figure}
    \centering
    \includegraphics[width=\linewidth]{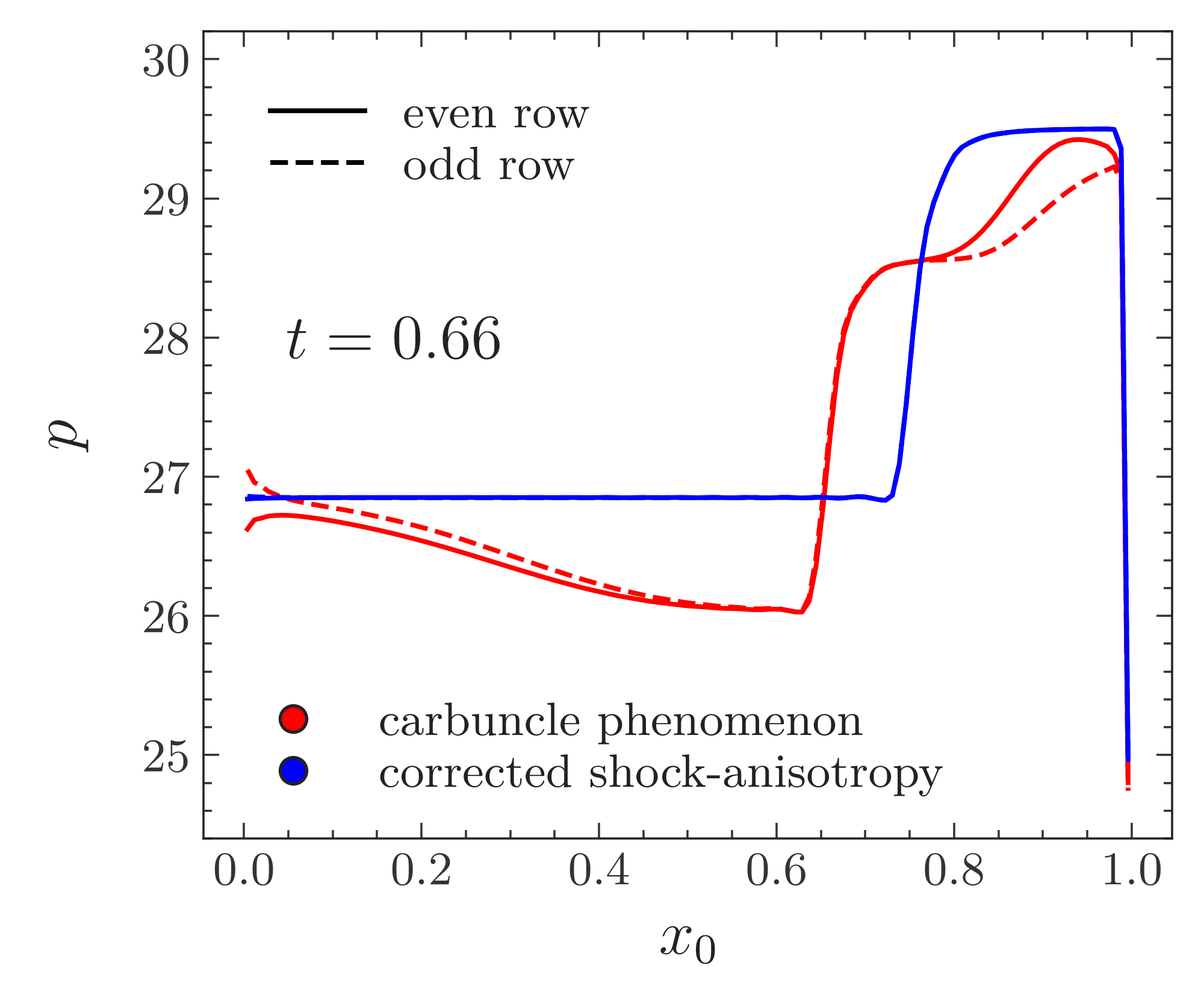}
    \caption{The \citet{Quirk94a} odd-even decoupling test (\sref{sec:stress-tests:quirk}), run with our recommended scheme (Q26 compute, \citetalias{Balsara25b} averaging, and PPM-EP reconstruction) at $128^2$ resolution, at $t \approx 0.66$. We show profiles of pressure along the $x_0$-axis for one of the perturbed rows (solid line; even $x_1$-index) and one unperturbed row (dashed line; odd $x_1$-index), with the \citet{Minoshima21a} shock-anisotropy correction disabled (red) and enabled (blue).}
    \label{fig:quirk-carbuncle}
\end{figure}

In \cref{fig:quirk-carbuncle} we compare our recommended scheme with and without the correction of \citet{Minoshima21a}. We plot pressure profiles at $t \approx 0.66$, along $x_0$ for both an even $x_1$-index (where the perturbation was applied; solid line) and an odd $x_1$-index (unperturbed; dashed line); we show the run without the correction in red, and the run with the correction in blue. Note that without the correction, the pressures in the even and odd rows drift apart, and the shock itself deforms. The run with the correction shows neither of these artefacts, and thus the \citet{Minoshima21a} shock-anisotropy correction is both necessary and effective at suppressing the carbuncle instability.

\subsection{Energy conservation at low Mach number}
\label{sec:stress-tests:balsara-vortex}

At low $\mathcal{M}$, approximate Riemann solvers recover the wrong scaling for pressure fluctuations, because the dissipation that keeps the solver upwind stable is set by the fast magnetosonic wave speed \citep{GuillardMurrone04a, watt2025mitigating}. This dissipation scales as $\bigO{\mathcal{M}}$, but the pressure fluctuations it acts on scale as $\bigO{\mathcal{M}^2}$. For flows in galaxies, $\mathcal{M} \gtrsim 1$, so this dissipation is reasonable, but in $\mathcal{M} \to 0$ flows, this dissipation is far too large, and can deform the flow \citep[see \eg][for a detailed analysis]{watt2025mitigating}. As with the carbuncle fix in \sref{sec:stress-tests:quirk}, \citet{Minoshima21a} provide a way to correct for this, through a low-$\mathcal{M}$ pressure rescaling in the Riemann solver. We have implemented this correction, and here validate it against the \citet{Balsara04a} vortex; we deliberately test at low resolution, since higher resolution is a way to resolve away this kind of issue.

We initialise a stable equilibrium vortex in a periodic domain spanning $x_0, x_1 \in [-5, 5]$, with $\gamma = 5/3$, and, to further stress the solver, advect the vortex diagonally along the grid,
\begin{align}
    \rho(\mVector{x})
        &= \rho_\mathrm{bg} \\
    \mVector{u}(\mVector{x})
        &= \mVector{u}_\mathrm{drift}
            + \mVector{u}_\mathrm{vortex} \\
    p(\mVector{x})
        &\scalebox{0.925}{$\displaystyle{}= p_\mathrm{bg} + \frac{1}{2} \rbrac{
                \delta b^2 (1 - r^2)
                - \rho_\mathrm{bg} (\delta u)^2
            } \exp\sbrac{1 - r^2}$} \\
    \mVector{b}(\mVector{x})
        &= \delta b \rbrac{
                x_0 \mVectorUnit{x}_1
                - x_1 \mVectorUnit{x}_0
            } \exp\sbrac{\frac{1 - r^2}{2}}
    ,
\end{align}
where,
\begin{align}
    \mVector{u}_\mathrm{drift}(\mVector{x})
        &= \frac{\delta u}{\sqrt{2}} \rbrac{
                \mVectorUnit{x}_0
                + \mVectorUnit{x}_1
            } \\
    \mVector{u}_\mathrm{vortex}(\mVector{x})
        &= \delta u \rbrac{
                x_0 \mVectorUnit{x}_1
                - x_1 \mVectorUnit{x}_0
            } \exp\sbrac{\frac{1 - r^2}{2}}
        ,
\end{align}
with $r^2 = x_0^2 + x_1^2$. We use a background state of $\rho_\mathrm{bg} = 1$ and $p_\mathrm{bg} = 1$, so $c_\mathrm{s} = \sqrt{\gamma p_\mathrm{bg} / \rho_\mathrm{bg}} = \sqrt{5/3}$, and use $\delta u = \mathcal{M} c_\mathrm{s}$ with $\mathcal{M} = \delta b = 0.01$.

Under these conditions, \ie~$p_\mathrm{bg} - \rho_\mathrm{bg} (\delta u)^2 / 2 > 0$, or equivalently $\mathcal{M} < \sqrt{2 / \gamma} = \sqrt{6/5}$, the vortex is an exact equilibrium solution to the ideal MHD equations, and thus the total kinetic, magnetic, and internal energy of the flow should remain constant. The extent to which these quantities are conserved is therefore a very strong diagnostic of numerically induced dissipation.

\begin{figure}
    \centering
    \includegraphics[width=\linewidth]{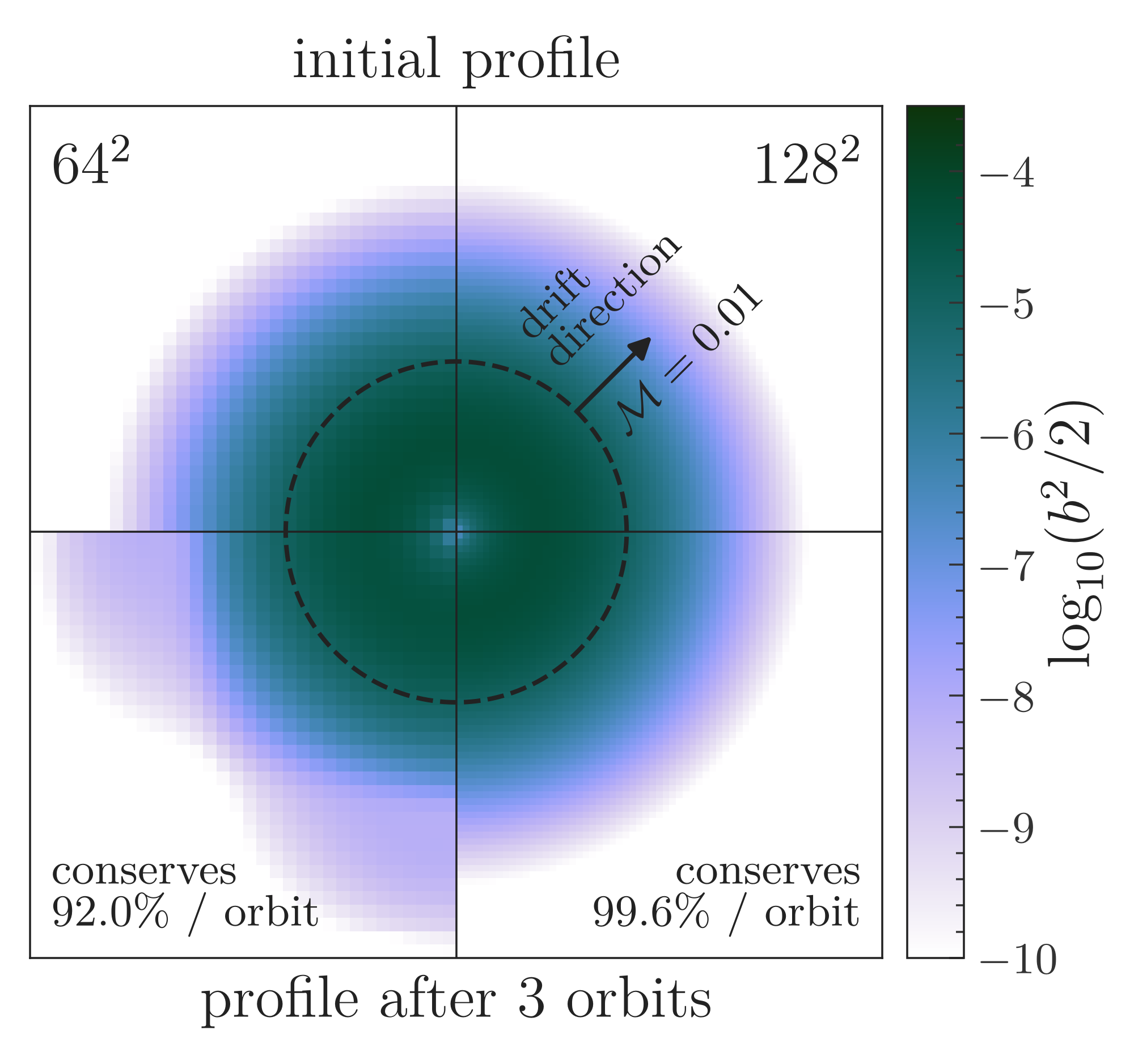}
    \caption{Slices of magnetic energy density in the \citet{Balsara04a} vortex (\sref{sec:stress-tests:balsara-vortex}), run with $\mathcal{M} = 0.01$ using our recommended scheme (Q26 compute, \citetalias{Balsara25b} averaging, and PPM-EP reconstruction). Each quadrant of the domain reveals a different solution: $64^2$ on the left and $128^2$ on the right, the initial condition on top and the final state after $3$ completed advection periods on the bottom. A reference circle (radius $=2$) shows the energy containing portion of the vortex; we indicate the measured fraction of magnetic energy retained per orbit.}
    \label{fig:balsara-vortex}
\end{figure}

We run the test at two resolutions, $64^2$ and $128^2$, to $t = 30\sqrt{2}/\delta u$, which is long enough for the vortex to advect diagonally across the periodic domain three times. At $64^2$, we find that the magnetic energy retains $92.0\%$ of its initial value per orbit, and at $128^2$ it retains $99.6\%$ per orbit; this is comparable to what \citet{watt2025mitigating} achieved using a custom MHD solver optimised specifically for this low-$\mathcal{M}$ regime. \Cref{fig:balsara-vortex} shows the vortex at both resolutions ($64^2$ on the left, and $128^2$ on the right), comparing the initial profile (top panels) to the final (bottom panels). We plot the $\log_{10}$ of the magnetic amplitude to reveal the warping at $64^2$, which is completely gone at $128^2$. Altogether, the vortex profiles retain both their amplitude and shape extremely well, confirming that the \citet{Minoshima21a} low-Mach pressure rescaling fix works as intended.

\subsection{3D shock robustness}
\label{sec:stress-tests:blast-wave}

Our next test is a 3D magnetised blast wave \citep{BalsaraSpicer99a}, with $\gamma = 5/3$ in a periodic domain $x_0,x_1,x_2 \in [-0.5,0.5]$. We initialise an overpressured spherical region, threaded by a strong background magnetic field oriented obliquely to the grid, to stress shock propagation across a range of angles to both the mesh and the field,
\begin{align}
    \rho(\mVector{x})
        &= 1 \\
    \mVector{u}(\mVector{x})
        &= 0 \\
    p(\mVector{x})
        &= \begin{cases}
                100
                    & , \qquad r < R \\
                1
                    & , \qquad r \geq R
            \end{cases} \\
    \mVector{b}(\mVector{x})
        &= \frac{10}{\sqrt{2}} \rbrac{\mVectorUnit{x}_0 + \mVectorUnit{x}_1}
    ,
\end{align}
where $r = \rbrac{x_0^2 + x_1^2 + x_2^2}^{1/2}$ and $R = 0.125$.

\begin{figure}
    \centering
    \includegraphics[width=\linewidth]{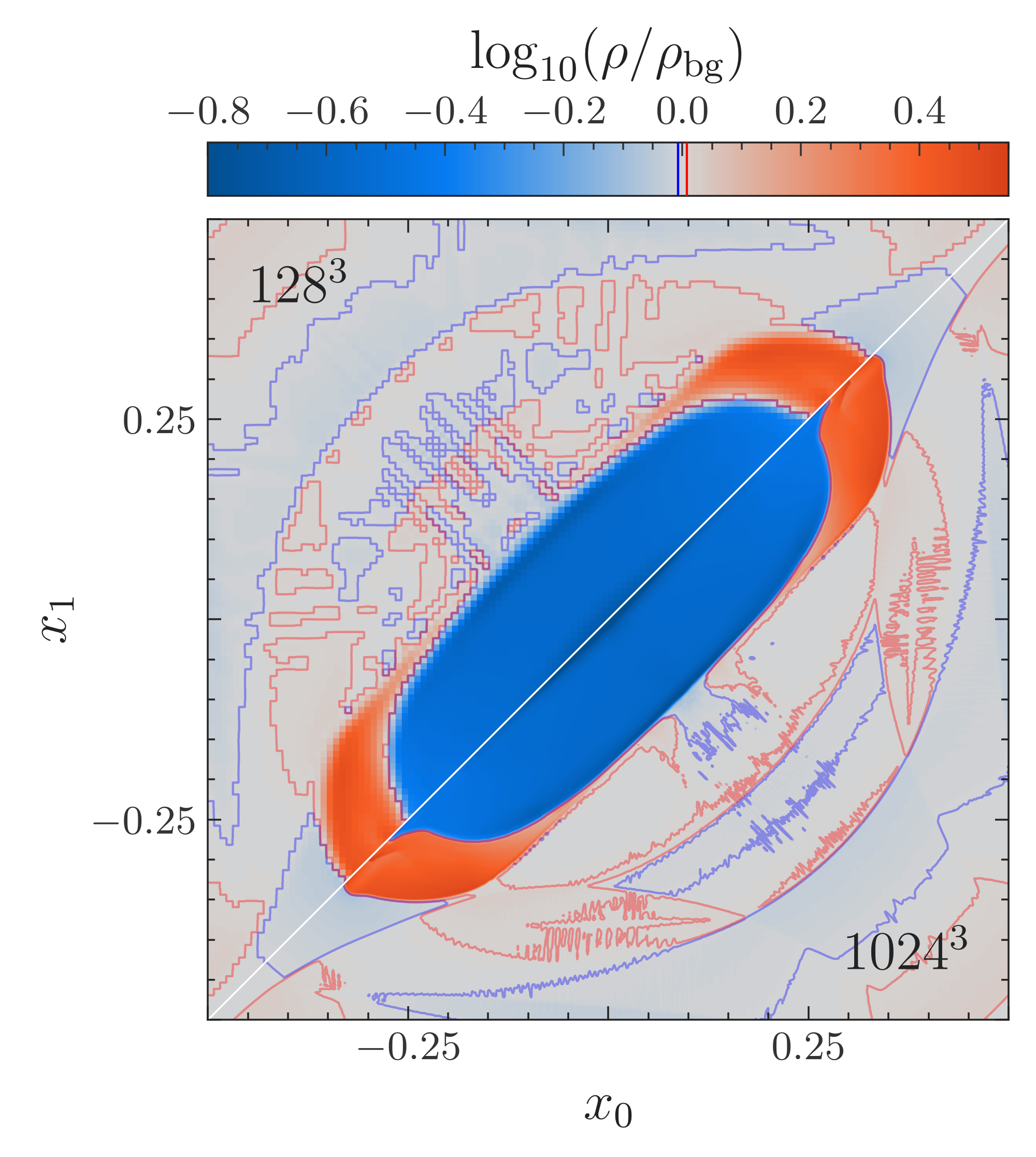}
    \caption{Density slices for the \citet{BalsaraSpicer99a} blast wave (\sref{sec:stress-tests:blast-wave}) through the mid-plane at $t = 0.05$, normalised by the ambient background value; run with our recommended scheme (Q26 compute, \citetalias{Balsara25b} averaging, and PPM-EP reconstruction). We compare $128^3$ (upper-left half) with $1024^3$ (in the lower-right), and show contours at $\log_{10}(\rho/\rho_\mathrm{bg}) = \pm 0.0075$.}
    \label{fig:blast-wave}
\end{figure}

We run this test to $t = 0.05$ both at $128^3$ and $1024^3$, and find that both runs remain stable, even in the $1024^3$ run where substantially sharper gradients are resolved. \Cref{fig:blast-wave} shows both resolutions in a single slice, split along the diagonal, with contours overlaid to trace the faint ($\lesssim 2\%$) density perturbations produced by fast magnetosonic waves. The blast has expanded freely along the diagonal background magnetic field and remains confined to it, forming an elongated under-dense cavity (blue region) with over-dense wings (red region) propagating outward. Both solutions stay symmetric about the diagonal and resolve similar structures, with $1024^3$ sharpening the faint wave fronts into thin, smooth arcs that $128^3$ can only trace at the grid scale.

\subsection{Resolved tearing-mediated reconnection}
\label{sec:stress-tests:current-sheet}

\begin{figure}
    \centering
    \includegraphics[width=\linewidth]{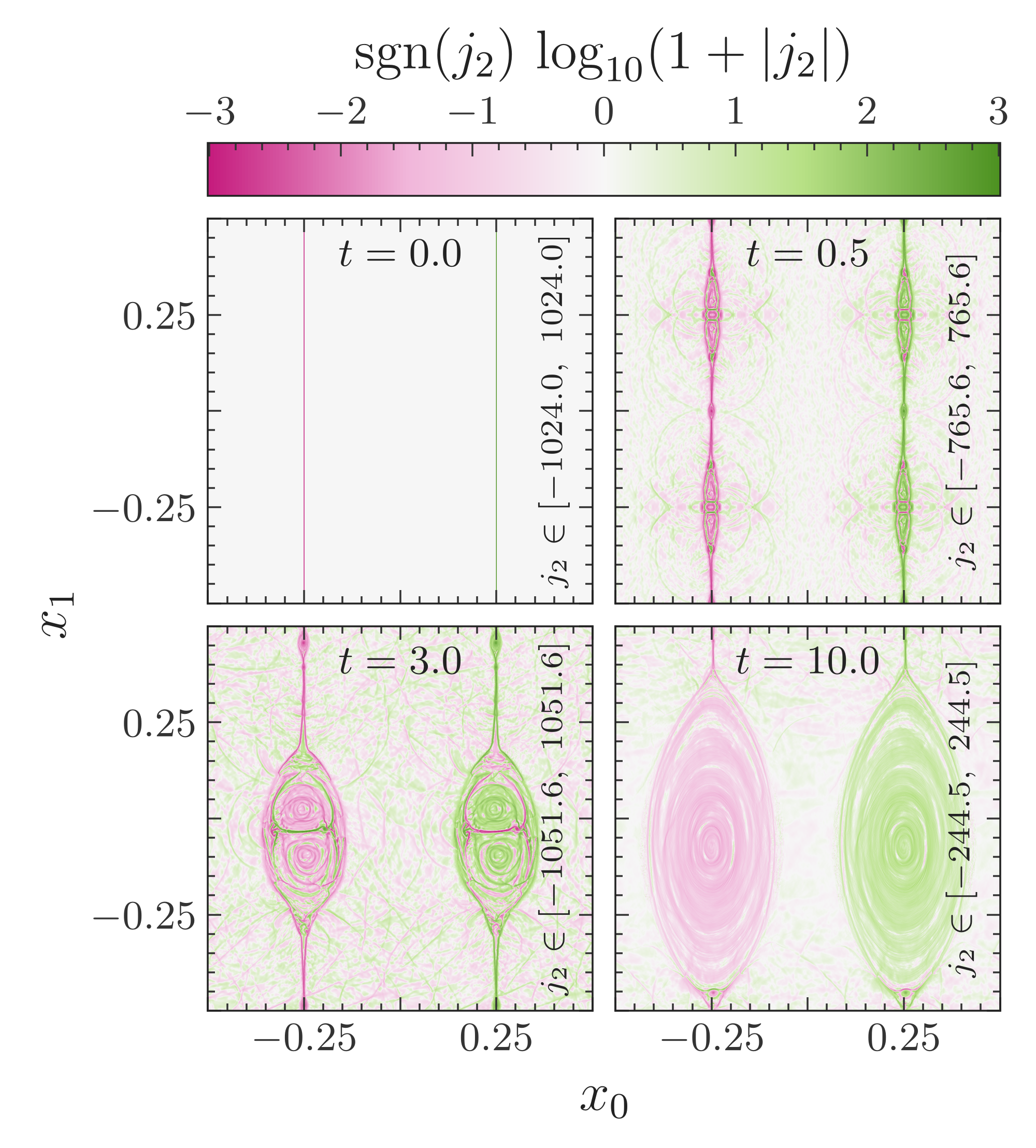}
    \caption{Evolution of the out-of-plane current density component $j_2 = \rbrac{\nabla \times \mVector{b}}_2$ for the \citet{HawleyStone95a} current sheet problem (\sref{sec:stress-tests:current-sheet}), run at $1024^2$ with our recommended scheme (Q26 compute, \citetalias{Balsara25b} averaging, and PPM-EP reconstruction). We show slices at $t = 0$, $0.5$, $3$, and $10$. Colours show the signed-logarithmic value of $j_2$; the annotation along the right edge of each panel shows the minimum and maximum $j_2$ in each snapshot.}
    \label{fig:current-sheet}
\end{figure}

\begin{figure*}
    \centering
    \includegraphics[width=\linewidth]{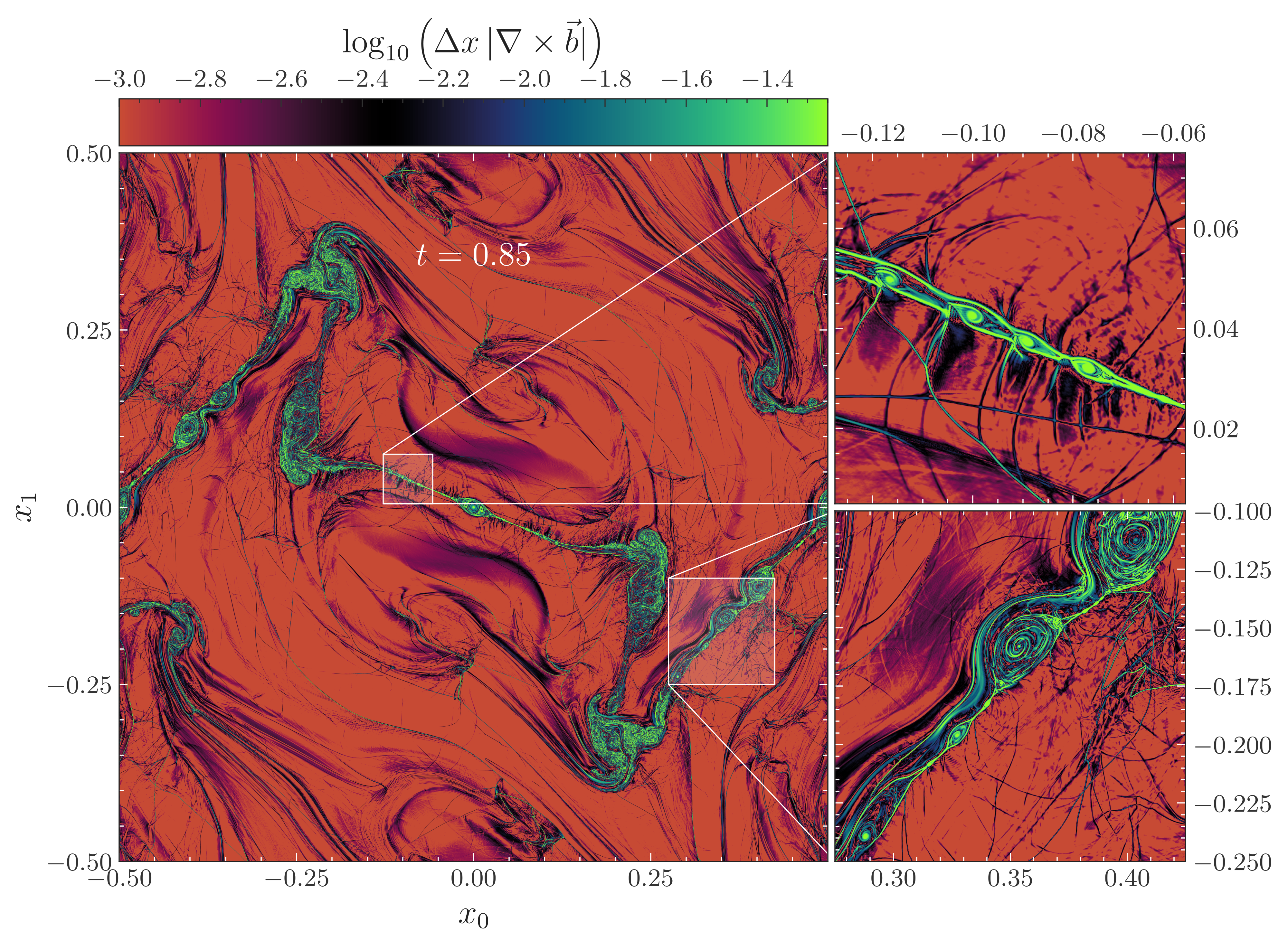}
    \caption{Current density magnitude, $\log_{10}\rbrac{\Delta x\, |\nabla\times\mVector{b}|}$, for the Orszag-Tang vortex (\sref{sec:scheme-comparison:orszag-tang:setup}) at $8192^2$ and $t=0.85$, run with our recommended scheme (Q26 compute, \citetalias{Balsara25b} averaging, and PPM-EP reconstruction). White boxes on the full-domain panel (left) mark the two regions showing zoomed insets (right).}
    \label{fig:orszag-tang:zoomins}
\end{figure*}

Next we demonstrate how our recommended scheme resolves tearing-unstable dynamics: first we show the plasmoid evolution in the current sheet test from \sref{sec:scheme-comparison:reconnection}, and then rerun the Orszag-Tang vortex from \sref{sec:scheme-comparison:orszag-tang} at higher resolution.

In \sref{sec:scheme-comparison:reconnection} we showed that our recommended scheme is one of only two EMF scheme combinations that remain stable for the \citet{HawleyStone95a} and \citet{Stone08a} current sheet problem with PPM-EP reconstruction at $1024^2$. We now return to the data from those runs and explore the dynamics that our recommended scheme resolves. \Cref{fig:current-sheet} shows the evolution of the out-of-plane current density across four snapshots. Initially, the magnetic field points along $x_1$ and reverses sign at $|x_0| = 0.25$, forming two thin, oppositely-signed current sheets (top-left panel). Along these reversals the velocity perturbation seeds a tearing instability that fragments sheets into plasmoids (top-right panel), which rapidly merge and grow as the flow evolves (bottom-left panel), until, by $t = 10$ (bottom-right panel), only a single, larger island remains at each of the two reconnection sites.

Turning to the Orszag-Tang vortex, tearing only starts to appear at $4096^2$ (\cref{fig:orszag-tang:reconstruction-schemes}), so we now repeat that setup at $8192^2$ to confirm that tearing modes are active at this resolution. \Cref{fig:orszag-tang:zoomins} shows a snapshot of current density for this run at $t = 0.85$, with two zoom-in regions highlighting the rich structures resolved on small scales: first, the current sheet along the vortex arm, which connects to the long-lived plasmoid in the centre, fragments into a chain of four plasmoids (top-right panel); further out, larger plasmoids sit intertwined with smaller ones (bottom-right panel). In both tests, our recommended scheme correctly resolves the tearing-mediated dynamics, with plasmoids that stay coherent as they advect along current sheets.

\subsection{Divergence preservation under adaptive mesh refinement}
\label{sec:stress-tests:field-loop}

\begin{figure*}
    \centering
    \includegraphics[width=0.95\linewidth]{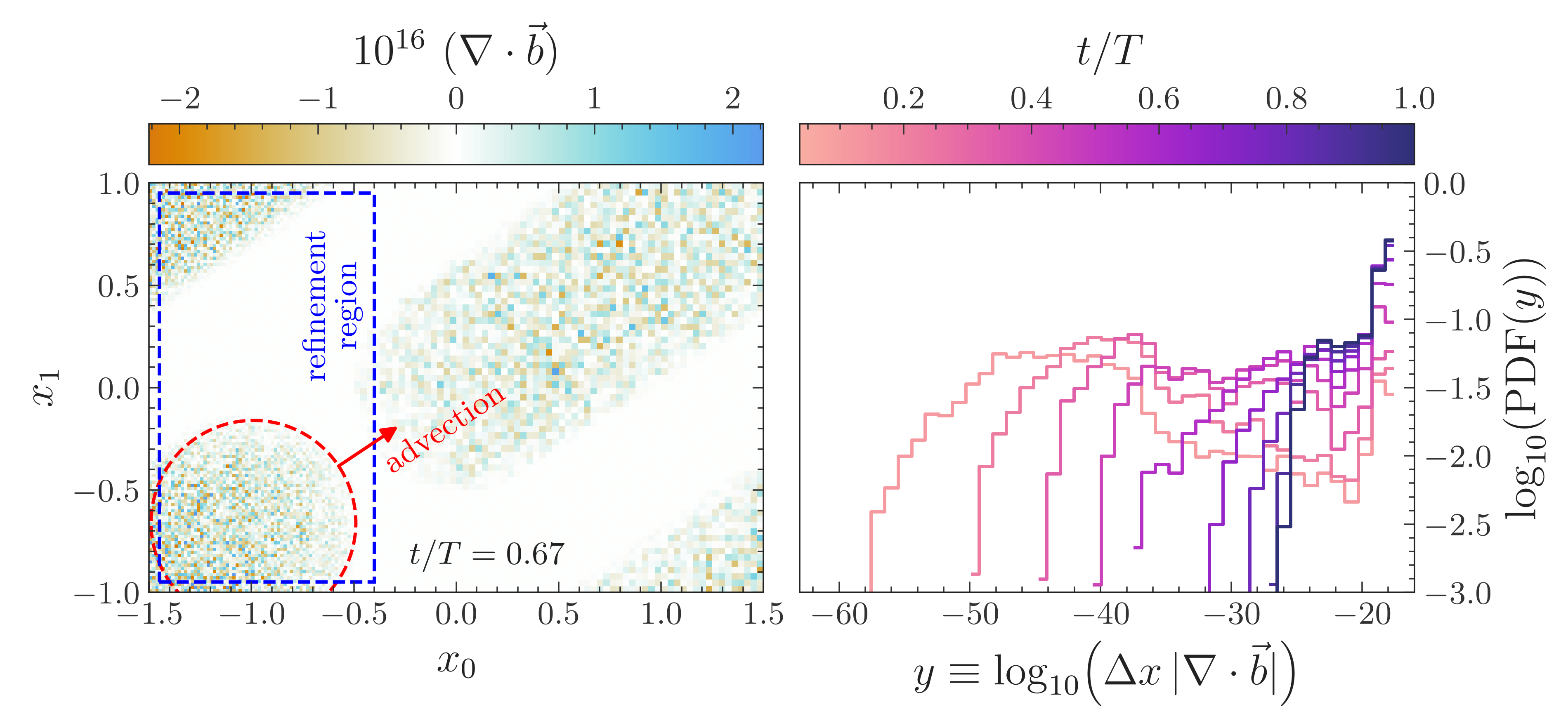}
    \caption{Field loop advection (\citet{Gardiner08a}; \sref{sec:stress-tests:field-loop}) with one level of static mesh refinement, run for a full period with our recommended scheme (Q26 compute, \citetalias{Balsara25b} averaging, and PPM-EP reconstruction). The left panel shows a slice of $\nabla\cdot\mVector{b}$, shown at the native resolution of each cell, at a time $t / T \approx 0.67$ where the loop is partially embedded within the refined region. The right panel shows the distribution of $y \equiv \log_{10}\!\left(\Delta x\, |\nabla\cdot\mVector{b}|\right)$, pooled over every cell in the AMR hierarchy at its own native $\Delta x$, evolving throughout the run, with colours showing time in units of the advection period $T$; we do not normalise by $|\mVector{b}|$, since it vanishes outside the loop and would make the ratio ill-defined there.}
    \label{fig:stress-tests:field-loop}
\end{figure*}

One of the most important constraints of any MHD simulation is satisfying $\nabla\cdot\mVector{b} = 0$. The CT formalism (\sref{sec:methods:mesh}) does so by construction, however, only up to numerical precision. In the presence of AMR (\sref{sec:methods:amr}), interpolation and EMF correction between coarse and fine AMR levels have the potential to create regions where spurious, finite-precision-level $\nabla\cdot\mVector{b}$ could, in principle, form. Here we confirm that our implementation maintains $\nabla\cdot\mVector{b} = 0$ to numerical precision, even with AMR.

Following \citet{Gardiner08a}, we set up a magnetic field loop advecting across a periodic domain spanning $x_0 \in [-1.5, 1.5]$, $x_1 \in [-1, 1]$, and $x_2 \in [-0.5, 0.5]$, with a refined region offset from the initial position of the loop (\cref{fig:stress-tests:field-loop} annotates its location). With $\gamma = 5/3$, we initialise,
\begin{align}
    \mVector{S}_\mathrm{p}
        = \begin{Bmatrix}
                \rho \\
                \mVector{u} \\
                p \\
                \mVector{b}
            \end{Bmatrix}
        = \begin{Bmatrix}
                1 \\
                \dfrac{1}{\sqrt{13}}\rbrac{3\,\mVectorUnit{x}_0 + 2\,\mVectorUnit{x}_1 + \mVectorUnit{x}_2} \\
                1 \\
                \dfrac{\delta b}{r}\rbrac{x_0\,\mVectorUnit{x}_1 - x_1\,\mVectorUnit{x}_0}
            \end{Bmatrix}
    ,
\end{align}
where $r = \rbrac{x_0^2 + x_1^2}^{1/2}$ and
\begin{align}
    \delta b
        = \begin{cases}
                10^{-3}
                    & , \qquad r < R \\
                0
                    & , \qquad r \geq R
            \end{cases}
    .
\end{align}
Since $\delta b \ll 1$ (magnetic pressure is far smaller than the background pressure), and $\mVector{u}$ is spatially uniform, the magnetic field loop is dynamically passive, and purely advects with the flow. We evolve this setup with radius $R = 0.5$ for one full advection period across the triply-periodic domain, $t / T = 1$. We choose to simulate at a low resolution of $96 \times 64 \times 32$ cells, and one additional AMR level, so that grid effects would be visually apparent.

\Cref{fig:stress-tests:field-loop} shows that no spurious $\nabla\cdot\mVector{b}$ is produced once the loop has entered the refined region, with $\nabla\cdot\mVector{b} \sim 10^{-16}$ (close to machine precision) both in the portion of the advection that took place on the base resolution (upper-right region of the slice panel) and once it crosses the refined region (bottom-left region). Beyond this single slice, we also show that the time-evolution of the distribution of $\nabla\cdot\mVector{b}$ values remains concentrated at the level of finite precision, rather than systematically shifting to increasingly larger values.

\subsection{Ohmic resistivity}
\label{sec:stress-tests:resistivity}

\begin{figure}
    \centering
    \includegraphics[width=0.95\linewidth]{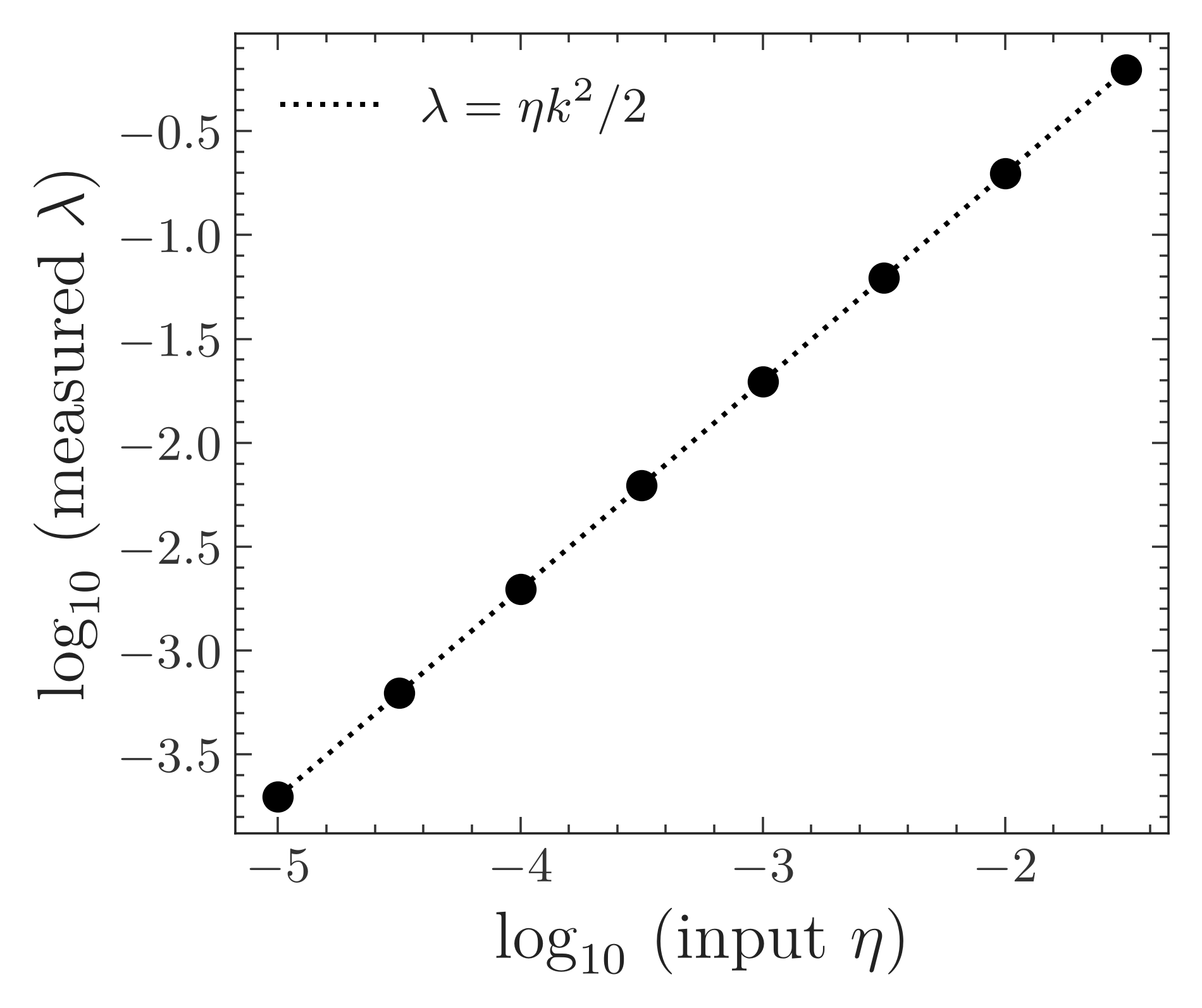}
    \caption{Measured decay rate of the resistively-damped Alfv\'en wave (\sref{sec:stress-tests:resistivity}) compared with the input resistivity $\eta$, run with our recommended scheme (Q26 compute, \citetalias{Balsara25b} averaging, and PPM-EP reconstruction), all at $N = 256$ cells. The dotted line shows the analytic prediction $\lambda = \eta k^2/2$.}
    \label{fig:resistive-alfven-decay}
\end{figure}

Having thoroughly validated our recommended scheme against ideal MHD, we now validate our implementation of constant Ohmic resistivity (\sref{sec:methods:resistivity}), which makes two additions to the ideal update: (i) a resistive correction to the EMF components (\cref{eqn:resistivity:emf-update}), and (ii) a corresponding Joule-heating term that is added to the total-energy flux. We validate these steps separately.

First, we check that a resistively-damped, linearly-polarised Alfv\'en wave decays as the analytic solution predicts. Substituting the resistive Ohm's law (\cref{eqn:resistivity:ohms-law}) into the induction equation (\cref{eqn:induction:continuous}) means that a wave that would ideally oscillate at $\omega_\mathrm{ideal} = c_\mathrm{A} k \cos\theta$ instead decays as $\exp\sbrac{-\lambda t}$, where
\begin{align}
    \lambda
        = \frac{\eta k^2}{2}
    , \label{eqn:resistivity:alfven-decay}
\end{align}
and oscillates at $\omega_\mathrm{real}$ (the real part of the complex frequency), where
\begin{align}
    \omega_\mathrm{real}
        = \sqrt{\omega_\mathrm{ideal}^2 - \lambda^2}
    . \label{eqn:resistivity:alfven-frequency}
\end{align}
We use the same linearly-polarised Alfv\'en wave setup as in \sref{sec:scheme-comparison:waves:alfven-linear}, run with the wave propagating grid-aligned, $\mVector{k}L/2\pi = \{1,0,0\}$, and a constant resistivity $\eta$ introduced. We run $8$ instances of this test with $N = 256$ cells throughout, varying $\eta \in [10^{-5}, 3.16 \times 10^{-2}]$ between each simulation, using $\mathrm{CFL} = 0.3$, consistent with the 3D stability requirement discussed in \sref{sec:methods:resistivity}. \Cref{fig:resistive-alfven-decay} shows that the measured decay rate matches the analytic $\lambda = \eta k^2/2$ (\cref{eqn:resistivity:alfven-decay}) across the full range, confirming that the resistive EMF correction performs as expected.

This test, however, cannot validate the Joule-heating term, since a periodic domain conserves total energy regardless of whether this term is implemented correctly, so a global energy check cannot distinguish a correct implementation from a broken, or missing, one. We separately confirm that the Joule-heating term is implemented correctly by checking, over a single time step, that the internal energy it deposits matches the $\eta\, j^2$ rate implied by \cref{eqn:resistivity:current-stencil} to machine precision.

\section{Conclusions}
\label{sec:conclusions}

In this paper we have implemented an MHD module, supporting both ideal and resistive (constant Ohmic) regimes, into the open-source code \quokka, extending its GPU-native radiation-hydrodynamic framework to magnetised flows. Our implementation couples to the existing Godunov finite-volume update of cell-centred state variables via a constrained transport (CT) update that stores magnetic fields on cell faces, using a staggered mesh
(see \cref{fig:schematic:grid}) that ensures $\nabla\cdot\mVector{b} = 0$ (to numerical precision) by construction. For the fluxes on these faces, we use an HLLD Riemann solver with stability and low-$\mathcal{M}$ pressure fixes based on \citet{Minoshima21a} (see \sref{sec:methods:flux}), and develop a first-order flux correction (FOFC) strategy that improves stability by detecting cells whose high-order update would produce an unphysical state and locally falling back to a more dissipative first-order update (see \sref{sec:methods:time-stepping}).

We divide our CT scheme (see \sref{sec:methods:emf}) up into two modular parts: (i) how the edge-centred EMFs are computed from neighbouring face- and cell-centred data (see \sref{sec:methods:emf:compute}), and (ii) how the four-quadrant estimates for the EMF around each edge are combined into a single upwinded value (see \sref{sec:methods:emf:averaging}). For the EMF compute step we implemented two recent, state-of-the-art EMF compute schemes produced by \citetalias{Felker18a} and \citetalias{Balsara25a} (see \cref{fig:schematic:fs18} and \ref{fig:schematic:b25}, respectively), and, in this paper, we also introduce a new scheme (Q26; see \cref{fig:schematic:q26}) which is based on the upwind CT approach of \citet{DelZanna07a}. Our new scheme computes EMFs from face-normal velocities constructed via a wavespeed-weighted average of the reconstructed left and right states at each face, using the same bounding wavespeeds as the Riemann solver, rather than reconstructing them separately from cell centres, which reduces the number of reconstruction operations and temporary variables required for this step. We also implemented two EMF averaging schemes, namely those of \citetalias{Londrillo04a} and \citetalias{Balsara25b}.

To determine which combination of these EMF compute and EMF averaging schemes best aligns with the performance goals of \quokka (see \sref{sec:intro:goals} for a discussion), we perform a rigorous and systematic comparison (\sref{sec:scheme-comparisons}) of the accuracy and efficiency of every combination of EMF compute and averaging together with three reconstruction methods: piecewise linear (PLM), piecewise parabolic (PPM), and extrema-preserving PPM (PPM-EP). We test all scheme combinations of EMF compute, EMF averaging, and reconstruction method against both linear and highly-non-linear problems, and compare not only accuracy but also GPU throughput and scaling. Based on our systematic comparison, we recommend our Q26 compute scheme coupled with \citetalias{Balsara25b} averaging and PPM-EP as the default scheme combination for \quokka, since it provides roughly $25$--$60\%$ higher GPU throughput than the alternatives we test (\sref{sec:scheme-comparison:speed}; \cref{fig:gpu-scaling}), resolves flow dynamics as accurately as those alternatives (\cref{fig:orszag-tang:emf-schemes}), and is one of the most stable scheme combinations at this accuracy in reconnection-dominated flows (\sref{sec:scheme-comparison:reconnection}). Finally, we further validate our recommended scheme against a broader suite of tests (\sref{sec:stress-tests}).

In future work we intend to expand these capabilities further into the non-ideal MHD regime, by adding treatments of ion-neutral drift and the Hall effect.

\section*{Acknowledgements}

We thank James R. Beattie and James Watt for many fruitful discussions that have greatly benefited this work.

N.~K. acknowledges financial support from the Australian Government via the Australian Government Research Training Program Fee-Offset Scholarship. This work was further supported by the Australian Research Council through award FL220100020 (to M.~R.~K.), and by computing resources supported by the Australian Government's National Collaborative Research Infrastructure Strategy (NCRIS), with access to Gadi at the National Computational Infrastructure and Setonix at the Pawsey Supercomputing Centre, through the National Computational Merit Allocation Scheme (award jh2). P.S.L. acknowledges the supports by the National Natural Science Foundation of China (NSFC) through grant No. 1241101426 and the National Key R\&D
Program of China (No.2022YFA1603101). We finally acknowledge access to the Marvin cluster, with support from the HPC@HRZ Team, and the High Performance Computing \& Analytics Lab at the University of Bonn.

This work made use of the \textsc{AMReX} framework \citep{Zhang19b}, on which \quokka~is built, and its turbulent forcing module, \textsc{TurbGen} \citep{Federrath2022_TurbGen}. We also relied on the following programming languages and software packages to analyse and produce visualisations of our simulation data: \textsc{C++} \citep{stroustrup2013cpp}, \textsc{python} along with \textsc{numpy} \citep{oliphant2006guide, van2011numpy, harris2020array}, \textsc{matplotlib} \citep{hunter2007matplotlib, bisong2019matplotlib}, and \textsc{cython} \citep{behnel2010cython, smith2015cython}. The exact MHD Riemann solver, \textsc{MHDRiemannSolver} \citep{Kriel26a}, was developed by N.~K. to compute reference solutions of the shock-tube test in \sref{sec:scheme-comparison:shocks:ryu-jones}.

\section*{Data Availability}

Most of the data underlying the results in this paper, along with all the scripts used to generate figures, are available on the GitHub repository: \url{https://github.com/AstroKriel/Kriel-quokka-mhd}. The remaining data is too large to host on GitHub and is available from the corresponding author on reasonable request. \quokka, including the MHD module presented in this paper and all of its capabilities, is open-source and available at \url{https://github.com/quokka-astro/quokka}.

\bibliographystyle{aasjournalv7}
\bibliography{header/refs}

\appendix
\section{A note on notation}
\label{sec:notation}

Throughout this paper we work with quantities that have many different variants, are stored on different stencils, and are evaluated at different times. To express all of this information consistently, we adopt the following conversions.

\subsection*{Field and index conventions}

We use lower-case symbols for scalar and vector fields, and upper-case symbols for collections (\eg~the state vector; see \cref{eqn:state:conserved}) and for rank-2 tensors. Spatial indices $i$, $j$, and $k$ denote mesh positions: integers refer to cell centres, and half-integers to faces or edges (see \cref{fig:schematic:grid}). Numbered subscripts indicate vector components aligned with coordinate axes, which correspond to the $x_0$, $x_1$, and $x_2$ directions (following the usual right-hand rule).

\subsection*{Brackets and superscripts}

Round brackets denote grouping, \eg~$a(b+c)$. Square brackets serve three roles: to group a set of arguments of a function or operator, \eg~$\exp\sbrac{1 - r^2}$; grouping the indices at which a quantity is sampled, \eg~$b_0\EvalAtH{i+1/2}{j}{k}$ (see \cref{fig:schematic:grid}); or, as a superscript, indicating a discrete case or instance of a quantity, either a time step, $\mVector{S}^{[n]}$ (see \cref{eqn:rk1:update,eqn:rk2:update}), or a specific reconstructed variant or wave type, $\varepsilon_2^{[LB]}$ and $\varepsilon_2^{[RT]}$ (see \cref{eqn:emf:ld04}). Curly brackets serve two roles: standard mathematical set membership, \eg~$S_\mathrm{I} \in \{L, R\}$ (see \sref{sec:methods:emf:compute}); or, as a subscript, a set of quantities that are jointly referenced or combined by the enclosing operator, \eg~$\bar{u}_{\cbrac{0,1}}$ for the average of $u_0$ and $u_1$ (see \cref{eqn:emf:fs18-average}). Superscripts are therefore used in two ways: (i) in square brackets, to indicate either a time step or a variant, and (ii) in the usual sense to denote powers (when not enclosed by square brackets).

\subsection*{Directional notation and interface labels}

Because the same procedure is applied independently along each direction, we describe the scheme through the flux across the $x_0$-oriented face; the corresponding expressions for the $x_1$- and $x_2$-faces follow by cyclic permutation of the $i$, $j$, and $k$ indices (see \cref{fig:schematic:grid}). For simplicity, we therefore describe all schemes in the local, 2D plane spanned by $x_0$ and $x_1$.

The interface states that are produced by reconstruction are labelled: left ($L$) and right ($R$), or top ($T$) and bottom ($B$). We denote side-based states by $S_\mathrm{I} \in \{L, R\}$ and $S_\mathrm{II} \in \{T, B\}$ (see \cref{fig:schematic:b-common}), and quadrant-based states either as single-sided labels $Q_\mathrm{I} \in \{T, R, B, L\}$ (see \cref{fig:schematic:q26}), or as corner combinations $Q_\mathrm{II} \in \{LB, RB, LT, RT\}$ (see \cref{fig:schematic:fs18}). When necessary, we explicitly indicate the order of reconstruction steps, \eg~$u_0^{[Q_\mathrm{II}: x_0 x_1]}$ and $u_0^{[Q_\mathrm{II}: x_1 x_0]}$.

\end{document}